\documentclass[prl,nofootinbib,twocolumn,superscriptaddress]{revtex4-2}

\usepackage[table,dvipsnames]{xcolor}
\usepackage[colorlinks,linktocpage]{hyperref}
\hypersetup{
	citecolor  = NavyBlue,
	linkcolor  = NavyBlue,
	urlcolor = NavyBlue
}

\usepackage{bbm}
\usepackage{amsfonts}
\usepackage{mathrsfs}
\usepackage{slashed}
\usepackage{bbold}
\usepackage{amsmath,amssymb}
\usepackage{epsfig}
\usepackage{graphicx}               
\usepackage{url}
\usepackage{soul}
\usepackage{pstricks}
\usepackage{multirow}

\newcommand\Tstrut{\rule{0pt}{3.5ex}}         
\newcommand\Bstrut{\rule[-2ex]{0pt}{0pt}}   

\def\ie{{\it i.e.}}
\def\eg{{\it e.g.}}

\def\vs{vs.}

\def\cf{{\it cf.}}

\usepackage[compat=1.1.0]{tikz-feynman}
\tikzfeynmanset{every blob = {/tikz/fill=gray!50, /tikz/minimum size=15pt}}
\tikzset{graviton/.style={decorate, decoration={snake, amplitude=.4mm, segment length=1.5mm, pre length=.5mm, post length=.5mm}, double}}

\def\C{{\textsc{c}}}
\def\c{$\ast$}
\def\cc{\c\,-}

\DeclareMathSymbol{\shortminus}{\mathbin}{AMSa}{"39}

\def\O{\mathcal{O}}
\def\S{\mathcal{S}}
\def\Q{\sigma}
\def\G{\mathcal{G}}
\def\K{\mathcal{K}}
\def\D{\mathcal{D}}
\def\x{\vec{x}}
\def\y{\vec{y}}
\def\z{\vec{z}}
\def\gh{\psi}
\def\l{\ell}
\def\lm1{{\ell\shortminus1}}
\def\ce{\gamma}

\def\*{\hspace{-1pt}}

\def\newpar{\vskip4pt}

\makeatletter 

\renewcommand\onecolumngrid{
	\do@columngrid{one}{\@ne}%
	\def\set@footnotewidth{\onecolumngrid}
	\def\footnoterule{\kern-6pt\hrule width 1.5in\kern6pt}%
}

\renewcommand\twocolumngrid{
	\def\footnoterule{
		\dimen@\skip\footins\divide\dimen@\thr@@
		\kern-\dimen@\hrule width.5in\kern\dimen@}
	\do@columngrid{mlt}{\tw@}
}%

\makeatother    

\begin{document}

\title{Structures of Neural Network Effective Theories}

\author{Ian Banta}
\affiliation{Department of Physics, University of California, Santa Barbara, CA 93106, USA}

\author{Tianji Cai}
\affiliation{Department of Physics, University of California, Santa Barbara, CA 93106, USA}

\author{Nathaniel Craig}
\affiliation{Department of Physics, University of California, Santa Barbara, CA 93106, USA}
\affiliation{Kavli Institute for Theoretical Physics, University of California, Santa Barbara, CA 93106, USA}

\author{Zhengkang Zhang}
\affiliation{Department of Physics, University of California, Santa Barbara, CA 93106, USA}
\affiliation{Kavli Institute for Theoretical Physics, University of California, Santa Barbara, CA 93106, USA}

\begin{abstract}
%
We develop a diagrammatic approach to effective field theories (EFTs) corresponding to deep neural networks at initialization, which dramatically simplifies computations of finite-width corrections to neuron statistics. 
The structures of EFT calculations make it transparent that a single condition governs criticality of all connected correlators of neuron preactivations.
Understanding of such EFTs may facilitate progress in both deep learning and field theory simulations.
\end{abstract}

\maketitle


\paragraph{Introduction}---\,
Machine learning (ML) has undergone a revolution in recent years, with applications ranging from image recognition and natural language processing, to self-driving cars and playing Go. 
Central to all these developments is the engineering of deep neural networks, a class of ML architectures consisting of multiple layers of artificial neurons. 
Such networks are apparently rather complex, with a deterring number of trainable parameters, which means practical applications have often been guided by expensive trial and error.
Nevertheless, extensive research is underway toward opening the black box.

That a theoretical understanding of such complex systems is possible has to do with the observation that a wide range of neural network architectures actually admit a simple limit: they reduce to Gaussian processes when the network width (number of neurons per layer) goes to infinity~\cite{Neal1996, Williams1996, lee2017deep, matthews2018gaussian, yang2019tensor, hanin2021random}, and evolve under gradient-based training as linear models governed by the neural tangent kernel~\cite{jacot2018neural, lee2019wide, yang2020tensor}. 
However, an infinitely-wide network neither exists in practice, nor provides an accurate model for deep learning. 
It is therefore crucial to understand finite-width effects, which have recently been studied by a variety of methods~\cite{antognini2019finite, hanin2019finite, huang2020dynamics, Dyer:2019uzd, Yaida:2019sjo, naveh2021predicting, seroussi2021separation, Aitken:2020tuu, Andreassen:2020cpx, zavatone2021asymptotics, naveh2021self, Roberts:2021fes, hanin2022correlation, Yaida:2022vsw}.

This line of research in ML theory has an intriguing synergy with theoretical physics~\cite{Roberts:2021lll}. 
In particular, it has been realized that neural networks have a natural correspondence with (statistical or quantum) field theories~\cite{schoenholz2017correspondence, cohen2021learning, Halverson:2020trp, Bachtis:2021xoh, Maiti:2021fpy, Erdmenger:2021sot, Erbin:2021kqf, Grosvenor:2021eol, Halverson:2021aot, Erbin:2022lls}. 
Infinite-width networks---which are Gaussian processes---correspond to free theories, while 
finite-width corrections in wide networks can be calculated perturbatively as in weakly-interacting theories. 
This allows for a systematically-improvable characterization of neural networks beyond the (very few) exactly-solvable special cases~\cite{zavatone2021exact, noci2021precise, hanin2022bayesian}.
Meanwhile, from an effective theory perspective~\cite{Roberts:2021fes}, information propagation through a deep neural network can be understood as a renormalization group (RG) flow. Examining scaling behaviors near RG fixed points reveals strategies to tune the network to criticality~\cite{raghu2017expressive, poole2016exponential, schoenholz2016deep}, which is crucial for mitigating the notorious exploding and vanishing gradient problems in practical applications.
In the reverse direction, this synergy also points to new opportunities to study field theories with neural networks~\cite{Halverson:2021aot}.

Inspired by recent progress, in this letter we further explore the structures of effective field theories (EFTs) corresponding to archetypical deep neural networks. 
To this end, we develop a novel diagrammatic formalism.\footnote{See also Refs.~\cite{Dyer:2019uzd, Aitken:2020tuu, Andreassen:2020cpx, cohen2021learning, Halverson:2020trp, Erbin:2021kqf, Grosvenor:2021eol, Maloney:2022cvb} for Feynman diagram-inspired approaches to ML.} 
Our approach largely builds on the frameworks of Refs.~\cite{Roberts:2021fes, hanin2022correlation}, which enable systematic calculations of finite-width corrections. 
The diagrammatic formalism dramatically simplifies these calculations, as we demonstrate by concisely reproducing known results in the main text and presenting further examples with new results in the Supplemental Material. 
Interestingly, the structures of diagrams in the RG analysis suggest that neural network EFTs are of a quite special type, where a single condition governs the critical tuning of all neuron correlators.
The study of these EFTs may lend new insights into both neural network properties and novel field-theoretic phenomena.

\newpar
\paragraph{EFT of deep neural networks}---\,
The archetype of deep neural networks, the multilayer perceptron, can be defined by a collection of neurons whose values $\phi_i^{(\l)}$ (called preactivations) are determined by the following operations given an input $\x\in \mathbb{R}^{n_0^{}}$:
\begin{eqnarray}
\phi_i^{(1)}(\x) &=& \sum_{j=1}^{n_0} W_{ij}^{(1)} x_j^{} +b_i^{(1)} \,,\nonumber\\
\phi_i^{(\l)}(\x) &=& \sum_{j=1}^{n_{\lm1}} W_{ij}^{(\l)} \Q\bigl(\phi_j^{(\lm1)}(\x)\bigr) +b_i^{(\l)} \quad (\l \ge 2)\,.
\label{eq:operations}
\end{eqnarray}
Here superscripts in parentheses label layers, subscripts $i,j$ label neurons within a layer (of which there are $n_\l$ at the $\l\,$th layer), and $\sigma(\phi)$ is the activation function  (common choices include $\tanh(\phi)$ or $\text{ReLU}(\phi)\equiv\max(0, \phi)$).
The weights $W_{ij}^{(\l)}$ and biases $b_i^{(\l)}$ ($\l = 1, \dots, L$) are the network parameters which are adjusted to minimize a loss function during training, such that the trained network can approximate the desired function.

The basic idea of an EFT of deep neural networks is to consider an ensemble of networks, where at initialization, $W_{ij}^{(\l)}$ and $b_i^{(\l)}$ are drawn independently from zero-mean Gaussian distributions with variances $C_W^{(\l)}/n_\lm1^{}$ and $C_b^{(\l)}$, respectively. 
The statistics of this ensemble encode both the typical behavior of neural networks initialized in this manner and how a particular network may fluctuate away from typicality. 
In the field theory language, these are captured by a Euclidean action, $\S[\phi] = -\log P(\phi)$, for all neuron preactivation fields $\phi_i^{(\l)}\*(\x)$, where $P(\phi)$ is the joint probability distribution. 
As we review in the Supplemental Material, at initialization the conditional probability distribution at each layer is Gaussian:
\begin{eqnarray}
&& P\bigl( \phi^{(\l)} \big| \phi^{(\lm1)} \bigr) = \bigl[\det \bigl(2\pi\G^{(\l)}\bigr)\bigr]^{\*-\*\frac{n_\l}{2}} \,e^{-\S_0^{(\l)}}\,, 
\label{eq:cond_prob}\\[8pt]
&&\S_0^{(\l)} \!= \!\int\! d\x_1^{} d\x_2^{} \,\frac{1}{2} \sum_{i=1}^{n_\l} \phi_i^{(\l)} \*(\x_1^{}) \bigl(\G^{(\l)}\bigr)^{\!-\*1}\!(\x_1^{}, \x_2^{}) \,\phi_i^{(\l)} \*(\x_2^{}) ,\quad\;
\label{eq:S_0}
\end{eqnarray}
where $\G^{(\l)} \*(\x_1^{}, \x_2^{}) = \frac{1}{n_\lm1^{}} \*\sum\limits_{j=1}^{n_{\lm1}} \G^{(\l)}_j\* (\x_1^{}, \x_2^{})$, with
\begin{eqnarray}
\G^{(\l)}_j\* (\x_1^{}, \x_2^{}) &=& C_b^{(\l)} + C_W^{(\l)}\,\Q\bigl(\phi^{(\lm1)}_j(\x_1^{})\bigr) \,\Q\bigl(\phi^{(\lm1)}_j(\x_2^{})\bigr) \nonumber\\[5pt]
&\equiv& C_b^{(\l)} + C_W^{(\l)}\, \Q_{j,\x_1^{}}^{(\lm1)} \,\Q_{j,\x_2^{}}^{(\lm1)}
\label{eq:G_j}
\end{eqnarray}
for $\l\ge 2$, and $\G^{(1)}_j\* (\x_1^{}, \x_2^{})= C_b^{(1)} + C_W^{(1)}\, x_{1j}^{} \,x_{2j}^{}$. 
We have taken the continuum limit in input space to better parallel field theory analyses. 
$\bigl(\G^{(\l)}\bigr)^{\!-\*1}$ is understood as the pseudoinverse when $\G^{(\l)}$ is not invertible.
We see that for $\l\ge 2$, $\G^{(\l)} \*(\x_1, \x_2)$ is an operator of the $(\l\*-\*1)\,$th-layer neurons, so Eq.~\eqref{eq:S_0} is actually an interacting theory with interlayer couplings. 
This also means the determinant in Eq.~\eqref{eq:cond_prob} is not a constant prefactor. 
To account for its effect, we introduce auxiliary anticommuting fields $\gh$, $\bar\gh$ which are analogs of ghosts and antighosts in the Faddeev-Popov procedure.
Including all layers, we have
\begin{equation}
e^{-\S[\phi]} = \int \!\D\gh \D\bar\gh\; e^{-\!\sum\limits_{\l=1}^{L} \bigl(\S_0^{(\l)}[\phi] + \S_\gh^{(\l)}[\phi, \gh, \bar\gh]\bigr)} \,,
\label{eq:exp-S}
\end{equation}
where $\S_0^{(\l)}$ is given by Eq.~\eqref{eq:S_0} above and
\begin{equation}
\S_\gh^{(\l)} = -\!\int\! d\x_1^{} d\x_2^{} \sum_{i'=1}^{n_\l/2} \bar\gh_{i'}^{(\l)} \*(\x_1^{}) \bigl(\G^{(\l)}\bigr)^{\!-\*1}\!(\x_1^{}, \x_2^{}) \,\gh_{i'}^{(\l)} \*(\x_2^{}) \,.
\label{eq:S_gh}
\end{equation}
The $\l\,$th-layer neurons interact with the $(\l\*-\*1)\,$th-layer and $(\l\*+\*1)\,$th-layer neurons via $\S_0^{(\l)}$ and $\S_0^{(\l+1)}$, respectively, while their associated ghosts have opposite-sign couplings to the $(\l\*-\*1)\,$th-layer neurons but do not couple to $(\l\*+\*1)\,$th-layer neurons. 
This means $\phi^{(\l)}$ and $\gh^{(\l)}$ loops cancel as far as their couplings to $\phi^{(\lm1)}$ are concerned, which must be the case since the network has directionality---neurons at a given layer cannot be affected by what happens at deeper layers.

\newpar
\paragraph{Neuron statistics from Feynman diagrams}---\,
We are interested in calculating neuron statistics, \ie\ connected correlators of neuron preactivation fields $\phi_i^{(\ell)}\*(\x)$ in the EFT above. 
More precisely, we would like to track the evolution of neuron correlators as a function of network layer $\ell$, which encodes how information is processed through a deep neural network and has an analogous form to RG flows in field theory. 
To this end, we develop an efficient diagrammatic framework to recursively determine $\ell$\,th-layer neuron correlators in terms of $(\ell\*-\*1)\,$th-layer neuron correlators.

Starting from the action Eq.~\eqref{eq:S_0}, we can derive the following Feynman rule (see Supplemental Material for details):
\begin{equation}
	\begin{tikzpicture}[baseline=(b)]
		\begin{feynman}
			\vertex[dot, minimum size=3pt, label = {below: {\scriptsize $\phi_i^{(\l)}\*(\x_1^{})$}}] (x1) {};
			\vertex[right = 20pt of x1] (v);
			\vertex[right = 20pt of v, dot, minimum size=3pt, label = {below: {\scriptsize $\phi_i^{(\l)}\*(\x_2^{})$}}] (x2){};
			\vertex[above = 2pt of v] (b){};
			\vertex[above = 17pt of v] (G){};
			\vertex[above = 20pt of v, label = {above: {\footnotesize $\frac{1}{n_\lm1}\, \G^{(\l)}_j\!(\x_1^{}, \x_2^{})$}}] (eq){};
			\diagram*{
				(x1) -- (v) -- (x2),
				(v) -- [graviton, thick] (G)
			};
		\end{feynman}
	\end{tikzpicture}
	\!\!=\!\!
	\begin{tikzpicture}[baseline=(b)]
		\begin{feynman}
			\vertex[dot, minimum size=3pt, label = {below: {\scriptsize $\phi_i^{(\l)}\*(\x_1^{})$}}] (x1) {};
			\vertex[right = 20pt of x1] (v);
			\vertex[right = 20pt of v, dot, minimum size=3pt, label = {below: {\scriptsize $\phi_i^{(\l)}\*(\x_2^{})$}}] (x2){};
			\vertex[above = 2pt of v] (b){};
			\vertex[above = 10pt of v, blob] (G){};
			\vertex[above = 20pt of v, label = {above: {\footnotesize $\frac{1}{n_\lm1}\bigl\langle \G^{(\l)}_j\!(\x_1^{}, \x_2^{})\bigr\rangle$}}] (eq){};
			\diagram*{
				(x1) -- (v) -- (x2),
				(v) -- [graviton, thick] (G)
			};
		\end{feynman}
	\end{tikzpicture}
	\!\!+\hspace{-7pt}
	\begin{tikzpicture}[baseline=(b)]
		\begin{feynman}
			\vertex[dot, minimum size=3pt, label = {below: {\scriptsize $\phi_i^{(\l)}\*(\x_1^{})$}}] (x1) {};
			\vertex[right = 20pt of x1] (v);
			\vertex[right = 20pt of v, dot, minimum size=3pt, label = {below: {\scriptsize $\phi_i^{(\l)}\*(\x_2^{})$}}] (x2){};
			\vertex[above = 2pt of v] (b){};
			\vertex[above = 18pt of v] (G){};
			\vertex[above = 20pt of v, label = {above: {\footnotesize \hspace{5pt}$\frac{C_W^{(\l)}}{n_{\lm1}} \,\Delta_{j}^{\*\*(\lm1)}\!(\x_1^{}, \x_2^{})$}}] (eq){};
			\diagram*{
				(x1) -- (v) -- (x2),
				(v) -- [photon] (G)
			};
		\end{feynman}
	\end{tikzpicture}
	.
	\label{eq:basic_rule}
\end{equation}
As indicated above, a blob means taking the expectation value of the operator (or product of operators) attached to it. 
Eq.~\eqref{eq:basic_rule} contains both what we normally call propagators and vertices: the first term on the right-hand side, when summed over $j$, is the full propagator (or two-point correlator) for $\phi_i^{(\ell)}$, 
\begin{equation}
\bigl\langle \phi^{(\l)}_{i_1^{}}\*(\x_1^{}) \,\phi^{(\l)}_{i_2^{}}\*(\x_2^{}) \bigr\rangle = \delta_{i_1^{} i_2^{}}\bigl\langle\G^{(\l)}\*(\x_1, \x_2)\bigr\rangle \,,
\label{eq:2-pt}
\end{equation}
while the second term is an interaction vertex between $\phi_i^{(\ell)}$ bilinears and operators built from $\phi_j^{(\lm1)}$:
\begin{equation}
	\Delta_{j}^{\!(\lm1)}(\x_1, \x_2) \equiv \Q_{j,\x_1}^{(\lm1)}\, \Q_{j,\x_2}^{(\lm1)} - \bigl\langle \Q_{j,\x_1}^{(\lm1)}\, \Q_{j,\x_2}^{(\lm1)} \bigr\rangle 
	\label{eq:Delta_def}
\end{equation}
for $\l\ge2$, and $\Delta_{j}^{\!(0)}(\x_1, \x_2)=0$.
From Eq.~\eqref{eq:basic_rule} it is clear that each $\phi^2\Delta$ vertex comes with a factor of $\frac{1}{n}$ (where $n$ collectively denotes $n_1^{}, \cdots, n_{L-1}^{}$). 
In the infinite-width limit, $n \to \infty$, the EFT is a free theory, whereas for large but finite $n$, we have a weakly-interacting theory where higher-point connected correlators can be perturbatively calculated as a $\frac{1}{n}$ expansion.

To see how this works, let us first take a closer look at the two-point correlator Eq.~\eqref{eq:2-pt} (which is automatically connected since we have normalized $\int\D\phi\, e^{-\S} = 1$, meaning the sum of vacuum bubbles vanishes). 
We can write it as an expansion in $\frac{1}{n}$:
\begin{equation}
\bigl\langle\G^{(\l)}\*(\x_1, \x_2)\bigr\rangle = \sum_{p=0}^{\infty} \frac{1}{n_\lm1^p}\, \K^{(\l)}_p \*(\x_1^{}, \x_2^{}) \,.
\label{eq:2-pt_exp}
\end{equation}
The leading-order (LO) term $\K_0^{(\l)}$ is known as the kernel; it is the propagator for $\phi_i^{(\ell)}$ in the free-theory limit $n\to\infty$. 
Evaluating $\bigl\langle\G^{(\l)}\*(\x_1, \x_2)\bigr\rangle$ in this limit amounts to using free-theory propagators $\K_0^{(\lm1)}$ for the previous-layer neurons $\phi_j^{(\lm1)}$ in the blob in Eq.~\eqref{eq:basic_rule}:
\begin{eqnarray}
\K^{(\l)}_0\!(\x_1^{}, \x_2^{}) &=& \sum_j \frac{1}{n_\lm1} 
\bigl\langle\G_j^{(\l)}\*(\x_1, \x_2)\bigr\rangle_{\K_0^{(\lm1)}}
\nonumber\\
&=& C_b^{(\l)} + C_W^{(\l)}\bigl\langle \Q_{\x_1^{}}^{} \Q_{\x_2^{}}^{} \bigr\rangle_{\K_0^{(\lm1)}} \,.
\label{eq:K_0_recursion}
\end{eqnarray}
Here subscript $\K_0^{(\lm1)}$\! means the expectation value is computed with the free-theory propagator $\K_0^{(\lm1)}$\! (\cf\ Eq.~\eqref{seq:bra_ket_K0} in the Supplemental Material). 
We have dropped both neuron and layer indices on $\Q$ because the $\K_0^{(\lm1)}$\! subscript already indicates the layer, and the expectation value is identical for all neurons in that layer.
One can further evaluate $\bigl\langle \Q_{\x_1^{}}^{} \Q_{\x_2^{}}^{} \bigr\rangle_{\K_0^{(\lm1)}}$ for specific choices of activation functions $\Q$, but we stay activation-agnostic for the present analysis.

Eq.~\eqref{eq:K_0_recursion} allows us to recursively determine $\K_0^{(\l)}$ from $\K_0^{(\lm1)}$, and has been well-known from studies of infinite-width networks. 
It may also be viewed as the RG flow of $\K_0^{}$, with ultraviolet boundary condition $\K_0^{(1)}(\x_1^{}, \x_2^{} )=C_b^{(1)} + \frac{C_W^{(1)}}{n_0} \,\x_1^{} \!\cdot\! \x_2^{}$. 
It is straightforward to extend the diagrammatic calculation to $\K_{p\ge1}^{}$. 
We present a simple derivation of the RG flow of $\K_1^{}$ in the Supplemental Material.

Next, consider the connected four-point correlator:
\begin{eqnarray}
&& \bigl\langle \phi^{(\l)}_{i_1^{}}\*(\x_1^{}) \,\phi^{(\l)}_{i_2^{}}\*(\x_2^{}) \,\phi^{(\l)}_{i_3^{}}\*(\x_3^{}) \,\phi^{(\l)}_{i_4^{}}\*(\x_4^{}) \bigr\rangle_\C \nonumber\\[2pt]
&=& \delta_{i_1^{} i_2^{}}\delta_{i_3^{} i_4^{}}\, \frac{1}{n_\lm1}\,V_4^{(\l)} \*(\x_1^{}, \x_2^{} ; \x_3^{}, \x_4^{}) 
+ \text{perms.},\quad
\label{eq:V4_def}
\end{eqnarray}
%
\noindent where
\begin{equation}
\frac{1}{n_\lm1}\,V_4^{(\l)} \*(\x_1^{}, \x_2^{} ; \x_3^{}, \x_4^{}) =
\, \sum_{j_1,j_2}
\begin{tikzpicture}[baseline=(b)]
\begin{feynman}
\vertex (l) {};
\vertex[below = 16pt of l, dot, minimum size=3pt, label = {below: {\footnotesize $\x_1^{}$}}] (x1) {};
\vertex[above = 16pt of l, dot, minimum size=3pt, label = {above: {\footnotesize $\x_2^{}$}}] (x2) {};
\vertex[right = 8pt of l, dot, minimum size=0pt] (v12) {};
\vertex[right = 20pt of v12, blob] (b) {};
\vertex[right = 20pt of b, dot, minimum size=0pt] (v34) {};
\vertex[right = 8pt of v34] (r) {};
\vertex[above = 16pt of r, dot, minimum size=3pt, label = {above: {\footnotesize $\x_3^{}$}}] (x3) {};
\vertex[below = 16pt of r, dot, minimum size=3pt, label = {below: {\footnotesize $\x_4^{}$}}] (x4) {};
\diagram*{
	(x1) -- (v12) -- (x2),
	(v12) -- [photon, edge label = {\scriptsize \;$\Delta_{j_1}$}, inner sep = 4pt] (b) -- [photon, edge label = {\scriptsize \,$\Delta_{j_2}$}, inner sep = 4pt] (v34), 
	(x3) -- (v34) -- (x4)
};
\end{feynman}
\end{tikzpicture}
.
\label{eq:V_4}
\end{equation}
From here on we label external legs only with input arguments $\x_1^{}, \x_2^{}$, etc.; it is clear that they represent $\l$\,th-layer neurons with pairwise-identical indices. 
We also omit ``$(\l\*-\*1)$'' superscripts on $\Delta_j^{(\lm1)}$ and drop the prefactor $\frac{C_W^{(\l)}}{n_{\lm1}}$ for compactness. 
Note that the blob in Eq.~\eqref{eq:V_4} is automatically connected because $\bigl\langle\Delta_{j}^{\*\*(\lm1)}\!(\x_1^{}, \x_2^{})\bigr\rangle = 0$ by definition.

To evaluate the diagram, we need to consider two cases, $j_1=j_2$ and $j_1\ne j_2$ 
For $j_1=j_2\equiv j$, the blob takes its free-theory value at LO:
\begin{eqnarray}
&&\sum_j
\begin{tikzpicture}[baseline=(b)]
\begin{feynman}
\vertex (l) {};
\vertex[below = 16pt of l, dot, minimum size=3pt, label = {below: {\footnotesize $\x_1^{}$}}] (x1) {};
\vertex[above = 16pt of l, dot, minimum size=3pt, label = {above: {\footnotesize $\x_2^{}$}}] (x2) {};
\vertex[right = 8pt of l, dot, minimum size=0pt] (v12) {};
\vertex[right = 20pt of v12, blob] (b) {};
\vertex[right = 20pt of b, dot, minimum size=0pt] (v34) {};
\vertex[right = 8pt of v34] (r) {};
\vertex[above = 16pt of r, dot, minimum size=3pt, label = {above: {\footnotesize $\x_3^{}$}}] (x3) {};
\vertex[below = 16pt of r, dot, minimum size=3pt, label = {below: {\footnotesize $\x_4^{}$}}] (x4) {};
\diagram*{
	(x1) -- (v12) -- (x2),
	(v12) -- [photon, edge label = {\scriptsize \;$\Delta_{j}$}, inner sep = 4pt] (b) -- [photon, edge label = {\scriptsize \,$\Delta_{j}$}, inner sep = 4pt] (v34), 
	(x3) -- (v34) -- (x4)
};
\end{feynman}
\end{tikzpicture}
\nonumber\\[5pt]
&=& \frac{\bigl(C_W^{(\l)}\bigr)^{\*2}\*}{n_\lm1} 
\bigl\langle \Delta (\x_1^{}, \x_2^{}) \Delta (\x_3^{}, \x_4^{}) \bigr\rangle_{\K^{(\lm1)}_0} 
\*+\O\Bigl(\frac{1}{n^2\*}\Bigr) 
\,.
\hspace{18pt}
\label{eq:4pt_1}
\end{eqnarray}
~\\
As in Eq.~\eqref{eq:K_0_recursion}, we have dropped the layer and neuron indices on $\Delta$. 
For $j_1\ne j_2$, free-theory propagators cannot connect $\Delta_{j_1}$ and $\Delta_{j_2}$, and the leading contribution is from inserting a connected four-point correlator of the $(\l\*-\*1)\,$th layer:
\begin{widetext}
\vskip-1pt
\begin{eqnarray}
&&\sum_{j_1, j_2}
\begin{tikzpicture}[baseline=(b)]
\begin{feynman}
\vertex (l) {};
\vertex[below = 16pt of l, dot, minimum size=3pt, label = {below: {\footnotesize $\x_1^{}$}}] (x1) {};
\vertex[above = 16pt of l, dot, minimum size=3pt, label = {above: {\footnotesize $\x_2^{}$}}] (x2) {};
\vertex[right = 8pt of l, dot, minimum size=0pt] (v12) {};
\vertex[right = 20pt of v12, blob] (b12) {};
\vertex[right = 16pt of b12, dot, minimum size=0pt] (w12) {};
\vertex[above = 1pt of w12, label = {above: {\scriptsize \hspace{-10pt} $\phi_{j_1}$}}] (w12u) {};
\vertex[below = 12pt of w12] (w12d) {};
\vertex[left = 10pt of w12d] (w12dl) {};
\vertex[right = 3pt of w12, blob, minimum size = 6pt] (b) {};
\vertex[right = 3pt of b, dot, minimum size=0pt] (w34) {};
\vertex[right = 16pt of w34, blob] (b34) {};
\vertex[above = 1pt of w34, label = {above: {\scriptsize \hspace{10pt} $\phi_{j_2}$}}] (w34u) {};
\vertex[below = 12pt of w34] (w34d) {};
\vertex[right = 10pt of w34d] (w34dr) {};
\vertex[right = 20pt of b34, dot, minimum size=0pt] (v34) {};
\vertex[right = 8pt of v34] (r) {};
\vertex[above = 16pt of r, dot, minimum size=3pt, label = {above: {\footnotesize $\x_3^{}$}}] (x3) {};
\vertex[below = 16pt of r, dot, minimum size=3pt, label = {below: {\footnotesize $\x_4^{}$}}] (x4) {};
\diagram*{
	(x1) -- (v12) -- (x2),
	(v12) -- [photon, edge label = {\scriptsize \;$\Delta_{j_1}$}, inner sep = 4pt] (b12) -- [quarter left] (w12) -- [quarter left] (b12),
	(b34) -- [quarter left] (w34) -- [quarter left] (b34) -- [photon, edge label = {\scriptsize \,$\Delta_{j_2}$}, inner sep = 4pt] (v34),
	(x3) -- (v34) -- (x4)
};
\draw [decoration={brace}, decorate] (w34dr) -- (w12dl)
node [pos=0.5, below = 1pt] {\scriptsize $\frac{1}{n_{\l\shortminus 2}\*}V_4^{(\lm1)}$};
\end{feynman}
\end{tikzpicture}
= \frac{\bigl(C_W^{(\l)}\bigr)^{\*2}}{4\,n_{\l\shortminus 2}} \prod_{\alpha=1}^4 \int \!d\y_\alpha^{} d\z_\alpha^{}\, \bigl( \K_0^{(\lm1)} \bigr)^{\!-1} \*(\y_\alpha^{}, \z_\alpha^{})
\, V_4^{(\lm1)} \!(\y_1^{}, \y_2^{} ; \y_3^{}, \y_4^{}) 
\nonumber\\[-16pt]
&&\hspace{150pt}
\Bigl\langle \Delta(\x_1^{}, \x_2^{})\, \phi(\z_1^{}) \,\phi(\z_2^{}) \Bigr\rangle_{\!\K^{(\lm1)}_0} 
\Bigl\langle \Delta(\x_3^{}, \x_4^{})\, \phi(\z_3^{}) \,\phi(\z_4^{}) \Bigr\rangle_{\!\K^{(\lm1)}_0} \*+\O\Bigl(\frac{1}{n^2\*}\Bigr) \nonumber\\[11pt]
&&= \frac{\bigl(C_W^{(\l)}\bigr)^{\*2}}{4\,n_{\l\shortminus 2}} \prod_{\alpha=1}^4 \int \! d\y_\alpha^{}\, V_4^{(\lm1)} \!(\y_1^{}, \y_2^{} ; \y_3^{}, \y_4^{}) 
\,\biggl\langle \frac{\delta^2 \Delta(\x_1^{}, \x_2^{})}{\delta \phi(\y_1^{}) \delta \phi(\y_2^{})} \biggr\rangle_{\!\!\K^{(\lm1)}_0}
\*\biggl\langle \frac{\delta^2 \Delta(\x_3^{}, \x_4^{})}{\delta \phi(\y_3^{}) \delta \phi(\y_4^{})} \biggr\rangle_{\!\!\K^{(\lm1)}_0} +\O\Bigl(\frac{1}{n^2\*}\Bigr)\,. \label{eq:4pt_2}
\end{eqnarray}
\vskip-1pt
\end{widetext}
In this diagram, internal solid lines denote $\phi_{j_1}^{(\lm1)}$, $\phi_{j_2}^{(\lm1)}$ propagators.
Exchanging the two $\phi_{j_1}^{(\lm1)}$ lines or the two $\phi_{j_2}^{(\lm1)}$ lines results in the same diagram, hence a symmetry factor $\frac{1}{2^2}=\frac{1}{4}$. 
The smaller blob at the center (together with the attached propagators) represents a connected four-point correlator of the $(\l\*-\*1)\,$th layer, $\frac{1}{n_{\l\shortminus 2}} V_4^{(\lm1)}$.
The larger blobs give rise to the correlators in the second line of Eq.~\eqref{eq:4pt_2}; they are automatically connected since $\bigl\langle\Delta_{j}^{\*\*(\lm1)}\!(\x_1^{}, \x_2^{})\bigr\rangle =\bigl\langle\phi_{j}^{\*(\lm1)}\!(\x)\bigr\rangle = 0$. 
A correlator $\bigl\langle \phi (\x) \dots\bigr\rangle$ by its standard definition includes the propagators $\K_0^{(\lm1)}(\x, \y)\dots$ (with $\y$ to be integrated over), so when we use correlators to build up diagrams, each internal propagator connecting two correlators (blobs) is counted twice. To avoid double-counting we thus insert an {\it inverse} propagator for each internal line in the diagram. 
This explains the factors of $\bigl( \K_0^{(\lm1)} \bigr)^{\!-1}$ in the first line of Eq.~\eqref{eq:4pt_2}, which effectively amputate the connected four-point correlator (or equivalently the larger blobs in the diagram). 
The final expression in Eq.~\eqref{eq:4pt_2} is obtained by Wick contraction, which yields factors of $\K_0^{(\lm1)}$ that cancel $\bigl( \K_0^{(\lm1)} \bigr)^{\!-1}$. 

Adding up Eqs.~\eqref{eq:4pt_1} and \eqref{eq:4pt_2} gives the final result for $V_4^{(\l)}$ in terms $V_4^{(\lm1)}$ and $\K_0^{(\lm1)}$, \ie\ the RG flow of $V_4$, which agrees with Refs.~\cite{Yaida:2019sjo,Roberts:2021fes}. 
Both equations are $\O(\frac{1}{n})$, so $V_4^{(\ell)}$ defined by Eq.~\eqref{eq:V4_def} is $\O(1)$.

The diagrammatic calculation extends straightforwardly to higher-point connected correlators, and provides a concise framework to systematically analyze finite-width effects in deep neural networks. 
In the Supplemental Material we present new results for the connected six-point and eight-point correlators as further examples.

The RG flow can also be formulated at the level of the EFT action. 
The idea is to consider a tower of EFTs, $\S_\text{eff}^{(\l)}$ ($\l=1,\dots,L$), obtained by integrating out the neurons and ghosts in all but the $\l\,$th layer. 
They take the form:
\begin{eqnarray}
\S_\text{eff}^{(\l)} &= & \int\!\* d\x_1^{} d\x_2^{} \, \bigl(\K_0^{(\l)} \*\*+\* \mu^{(\l)}\bigr)^{\!-\*1}\!(\x_1^{}, \x_2^{}) \nonumber\\
&&\hspace{20pt} \biggl[ \,\frac{1}{2}\,\phi_i^{(\l)} \*(\x_1^{}) \,\phi_i^{(\l)} \*(\x_2^{}) - \bar\gh_{i'}^{(\l)} \*(\x_1^{}) \,\gh_{i'}^{(\l)} \*(\x_2^{})\biggr] \nonumber\\[5pt]
&&-\int\!\* d\x_1^{} d\x_2^{} d\x_3^{} d\x_4^{} \,\lambda^{(\l)}\* (\x_1^{}, \x_2^{} ; \x_3^{}, \x_4^{}) \nonumber\\
&&\hspace{20pt} \biggl[ \,\frac{1}{8}\,\phi_i^{(\l)} \*(\x_1^{}) \,\phi_i^{(\l)} \*(\x_2^{}) \, \phi_j^{(\l)} \*(\x_3^{}) \,\phi_j^{(\l)} \*(\x_4^{}) \nonumber\\
&&\hspace{20pt} -\frac{1}{2}\,\phi_i^{(\l)} \*(\x_1^{}) \,\phi_i^{(\l)} \*(\x_2^{}) \, \bar\gh_{j'}^{(\l)} \*(\x_3^{}) \,\gh_{j'}^{(\l)} \*(\x_4^{}) \nonumber\\
&&\hspace{20pt} +\frac{1}{2}\, \bar\gh_{i'}^{(\l)} \*(\x_1^{}) \,\gh_{i'}^{(\l)} \*(\x_2^{}) \, \bar\gh_{j'}^{(\l)} \*(\x_3^{}) \,\gh_{j'}^{(\l)} \*(\x_4^{}) \biggr]\nonumber\\[5pt]
&&+\,\cdots
\label{eq:S_eff}
\end{eqnarray}
where summation over repeated indices is assumed. 
The EFT couplings $\mu^{(\l)}\*, \lambda^{(\l)} \*\sim\* \O\bigl(\frac{1}{n}\bigr)$ can be determined from the connected correlators, so their RG flows directly follow from those of the latter discussed above. 
For example, matching the connected four-point correlator relates $\lambda^{(\l)}$ to $V_4^{(\l)}$ and $\K^{(\l)}_0$:
\begin{eqnarray}
&&\frac{1}{n_\lm1}\,V_4^{(\l)} \*(\x_1^{}, \x_2^{} ; \x_3^{}, \x_4^{}) = 
\begin{tikzpicture}[baseline=(m)]
	\begin{feynman}
		\vertex (l) {};
		\vertex[below = 16pt of l, dot, minimum size=3pt, label = {below: {\footnotesize $\x_1^{}$}}] (x1) {};
		\vertex[above = 16pt of l, dot, minimum size=3pt, label = {above: {\footnotesize $\x_2^{}$}}] (x2) {};
		\vertex[right = 8pt of l, dot, minimum size=0pt] (v12) {};
		\vertex[right = 2pt of v12] (m) {};
		\vertex[right = 2pt of m, dot, minimum size=0pt] (v34) {};
		\vertex[right = 8pt of v34] (r) {};
		\vertex[above = 16pt of r, dot, minimum size=3pt, label = {above: {\footnotesize $\x_3^{}$}}] (x3) {};
		\vertex[below = 16pt of r, dot, minimum size=3pt, label = {below: {\footnotesize $\x_4^{}$}}] (x4) {};
		\diagram*{
			(x1) -- (v12) -- (x2),
			(x3) -- (v34) -- (x4)
		};
	\draw[fill = black] (m) ellipse (3pt and 1pt);
	\end{feynman}
\end{tikzpicture}
+\O \Bigl(\frac{1}{n^{\*2\*}}\Bigr)
\nonumber\\[-2pt]
=&& \;\prod_{\alpha=1}^4 \*\int\*\* d\y_\alpha^{} \,\*\K_0^{(\l)}\!(\x_\alpha^{}, \y_\alpha^{}) \,\*\lambda^{(\l)}\! (\y_1^{}, \y_2^{} ; \y_3^{}, \y_4^{}) 
+\O \Bigl(\frac{1}{n^{\*2\*}}\Bigr),\hspace{20pt}
\label{eq:V4_EFT}
\end{eqnarray}
where we use an elongated vertex to indicate pairing of the four arguments of $\lambda^{(\l)}$. 
For the two-point correlator, the calculation involves the following diagrams:
\begin{equation}
\begin{tikzpicture}[baseline=(x1)]
\begin{feynman}
\vertex[dot, minimum size = 3pt] (x1) {};
\vertex[right = 30pt of x1, dot, minimum size = 3pt] (x2) {};
\vertex[below = 15pt of x1] (x1d) {};
\vertex[left = 5pt of x1d] (x1dl) {};
\vertex[below = 15pt of x2] (x2d) {};
\vertex[right = 5pt of x2d] (x2dr) {};
\diagram*{
	(x1) -- (x2)
};
\draw [decoration={brace}, decorate] (x2dr) -- (x1dl) node [pos=0.5, below = 1pt] {\footnotesize $\;\;\K_0^{(\l)}$};
\end{feynman}
\end{tikzpicture}
+
\begin{tikzpicture}[baseline=(x1)]
	\begin{feynman}
		\vertex[dot, minimum size = 3pt] (x1) {};
		\vertex[right = 30pt of x1, dot, minimum size = 3pt] (x2) {};
		\vertex[below = 15pt of x1] (x1d) {};
		\vertex[left = 5pt of x1d] (x1dl) {};
		\vertex[below = 15pt of x2] (x2d) {};
		\vertex[right = 64pt of x2d] (x2dr) {};
		\diagram*{
			(x1) -- [insertion={[size=2pt]0.5}, edge label' = {\scriptsize $\;\;\mu^{(\l)}$}]  (x2)
		};
	\draw [decoration={brace}, decorate] (x2dr) -- (x1dl) node [pos=0.5, below = 1pt] {\footnotesize $\;\;\K_1^{(\l)}$};
\end{feynman}
\end{tikzpicture}
\hspace{-60pt}\,+\;\;
\begin{tikzpicture}[baseline=(x1)]
	\begin{feynman}
		\vertex[dot, minimum size=3pt] (x1) {};
		\vertex[right = 15pt of x1, dot, minimum size = 0pt, label = {below: {\scriptsize $\;\;\lambda^{\*(\l)}$}}] (v) {};
		\vertex[above = 15pt of v, dot, minimum size = 0pt] (t) {};
		\vertex[right = 30pt of x1, dot, minimum size = 3pt] (x2) {};
		\diagram*{
			(x1) -- (v) -- [half left] (t) -- [half left] (v) -- (x2)
		};
		\draw[fill = black] (v) ellipse (3pt and 1pt);
	\end{feynman}
\end{tikzpicture}
\;\;+\;\cdots
\label{eq:2-pt_EFT}
\end{equation}
The alternative pairing of legs at the quartic vertex in the last diagram results in an $\O\bigl(\frac{n}{n}\bigr)$ contribution, which, however, is canceled by diagrams with ghost loops due to the opposite-sign coupling:
\begin{equation}
	\sum_j\;\,
	\begin{tikzpicture}[baseline=(vu)]
		\begin{feynman}
			\vertex[dot, minimum size=3pt] (x1) {};
			\vertex[right = 15pt of x1, dot, minimum size = 0pt] (vd) {};
			\vertex[right = 30pt of x1, dot, minimum size = 3pt] (x2) {};
			\vertex[above = 2pt of vd, dot, minimum size = 0pt] (v) {};
			\vertex[above = 2pt of v, dot, minimum size = 0pt] (vu) {};
			\vertex[above = 15pt of vu, dot, minimum size = 0pt, label = {above: {\scriptsize $\phi_j$}}] (t) {};
			\diagram*{
				(x1) -- (vd) -- (x2),
				(vu) -- [half left] (t) -- [half left] (vu)
			};
			\draw[fill = black] (v) ellipse (1pt and 3pt);
		\end{feynman}
	\end{tikzpicture}
	\;\;+\;
	\sum_{j'}\;\,
	\begin{tikzpicture}[baseline=(vu)]
		\begin{feynman}
			\vertex[dot, minimum size=3pt] (x1) {};
			\vertex[right = 15pt of x1, dot, minimum size = 0pt] (vd) {};
			\vertex[right = 30pt of x1, dot, minimum size = 3pt] (x2) {};
			\vertex[above = 2pt of vd, dot, minimum size = 0pt] (v) {};
			\vertex[above = 2pt of v, dot, minimum size = 0pt] (vu) {};
			\vertex[above = 15pt of vu, dot, minimum size = 0pt, label = {above: {\scriptsize $\gh_{j'}$}}] (t) {};
			\diagram*{
				(x1) -- (vd) -- (x2),
				(vu) -- [ghost, half left] (t) -- [ghost, half left] (vu)
			};
			\draw[fill = black] (v) ellipse (1pt and 3pt);
		\end{feynman}
	\end{tikzpicture}
	\;=\, 0 
\end{equation}
Similar cancellations also explain the exclusion of $\O\bigl(\frac{n}{n^{\*2\*}}\bigr)$ loop diagrams in the calculation of the connected four-point correlator in Eq.~\eqref{eq:V4_EFT}.\footnote{We can reproduce the effective theory of Ref.~\cite{Roberts:2021fes} by integrating out the ghosts from Eq.~\eqref{eq:S_eff}. Calculations within this ghostless effective theory give rise to $\O\bigl(\frac{n_\lm1}{n_\l}\bigr)$ terms, necessitating either working in the regime $n_\l^{}\gg n_\lm1 \gg 1$ or marginalizing the action over all but an $\O(1)$ number of neurons in the $(\l\*-\*1)\,$th layer to have a perturbative $\frac{1}{n}$ expansion. Retaining the ghosts avoids such subtleties, rendering $\mu^{(\l)}$ genuinely $\O\bigl(\frac{1}{n}\bigr)$ and the difference between $\lambda^{(\l)}$ and the (amputated) connected four-point correlator genuinely $\O\bigl(\frac{1}{n^{\*2}}\bigr)$.}

\newpar
\paragraph{Structures of RG flow and criticality}---\,
%
The RG analysis of neuron statistics is highly relevant for the critical tuning of deep neural networks. 
The necessity of tuning has long been appreciated in practical applications of deep learning, especially in the context of mitigating the infamous exploding and vanishing gradient problems which make it difficult to train deep networks given finite machine precision. 
In the EFT framework, this is related to the fact that generic choices of hyperparameters $C_b^{(\l)}, C_W^{(\l)}$ lead to exponential scaling of neuron correlators under RG. 
Taming the exponential behaviors requires tuning the network to criticality by judiciously setting these hyperparameters~\cite{raghu2017expressive, poole2016exponential, schoenholz2016deep}. 
At the kernel level, the criticality analysis of Ref.~\cite{Roberts:2021fes} reveals two prominent universality classes which networks with a variety of activation functions fall into: scale-invariant (including \eg\ ReLU) and $\K^\star=0$ (including \eg\ tanh). 
In each case, $\K_0^{(\l)}$ flows toward a nontrivial fixed point as $\l$ increases; crucially, the scaling near the fixed point is power-law rather than exponential, which allows information to propagate through the layers so the network can learn nontrivial features from data.

While previous criticality analyses have mostly focused on the two-point correlator, it is important to also consider higher-point correlators because they encode fluctuations across the ensemble. 
In other words, it is not sufficient to require the networks are well-behaved on average, but the scaling behavior of each network must be close to the average.
At first sight, criticality seems to impose more constraints than the number of tunable hyperparameters, if we require power-law scaling of all higher-point correlators at arbitrary input points. 
However, as we will show, the structures of RG flow, manifest in the diagrammatic formulation, are such that tuning $\K_0^{}$ to criticality actually ensures power-law scaling of all higher-point connected correlators near the fixed point.

Let us start with the two-point correlator. 
The asymptotic scaling behavior (exponential \vs\ power-law) can be inferred from the following question: upon an infinitesimal variation at the $(\l\*-\*1)\,$th layer,
\begin{equation}
\bigl\langle\G^{(\lm1)}\*(\x_1^{}, \x_2^{})\bigr\rangle \to \bigl\langle\G^{(\lm1)}\*(\x_1^{}, \x_2^{})\bigr\rangle + \delta \bigl\langle\G^{(\lm1)}\*(\x_1^{}, \x_2^{})\bigr\rangle\,,
\end{equation}
how does $\bigl\langle\G^{(\l)}\*(\x_1^{}, \x_2^{})\bigr\rangle$ change? 
Diagrammatically, this can be calculated as follows:
\begin{equation}
\begin{tikzpicture}[baseline=(b.base)]
\begin{feynman}
\vertex[dot, minimum size=3pt] (x1) {};
\vertex[right = 20pt of x1, blob, minimum size = 8pt] (b) {{\scriptsize $\delta$}};
\vertex[right = 20pt of b, dot, minimum size = 3pt] (x2) {};
\diagram*{
	(x1) -- (b) -- (x2)
};
\end{feynman}
\end{tikzpicture}
\;\;\;= \;\;\sum_j\;\;
\begin{tikzpicture}[baseline=(b)]
\begin{feynman}
\vertex[dot, minimum size=3pt] (x1) {};
\vertex[right = 20pt of x1] (v);
\vertex[right = 20pt of v, dot, minimum size = 3pt] (x2) {};
\vertex[above = 5pt of v] (b) {};
\vertex[above = 12pt of v, blob] (G) {};
\vertex[above = 20pt of G, dot, minimum size = 0pt] (w) {};
\diagram*{
	(x1) -- (v) -- (x2),
	(v) -- [photon, edge label' = {\scriptsize \,$\Delta_j$}] (G) -- [quarter left] (w) -- [quarter left] (G)
};
\vertex[above = 18pt of G, blob, minimum size = 8pt] (d) {{\scriptsize $\delta$}};
\vertex[right = 14pt of G, label = {above: {\scriptsize $\phi_j$}}] (Gr) {};
\end{feynman}
\end{tikzpicture}
\,,
\label{eq:dG_diag}
\end{equation}
where a blob labeled ``$\delta$'' denotes the variation of the (two-point) correlator. 
The result is
\begin{eqnarray}
\delta \bigl\langle\G^{(\l)}\*(\x_1^{}, \x_2^{})\bigr\rangle &=& \int\! d\y_1^{} d\y_2^{} 
\,\chi^{(\l)}\*(\x_1^{}, \x_2^{}; \y_1^{}, \y_2^{}) \nonumber\\[2pt]
&&\hspace{40pt} \delta \bigl\langle\G^{(\lm1)}\*(\y_1^{}, \y_2^{})\bigr\rangle \,,
\end{eqnarray}
or, equivalently:
\begin{equation}
\frac{\delta \bigl\langle\G^{(\l)}\*(\x_1^{}, \x_2^{})\bigr\rangle}{\delta \bigl\langle\G^{(\lm1)}\*(\y_1^{}, \y_2^{})\bigr\rangle} \,=\, \chi^{(\l)}\*(\x_1^{}, \x_2^{}; \y_1^{}, \y_2^{}) \,, 
\end{equation}
where
\begin{eqnarray}
&&\chi^{(\l)}\*(\x_1^{}, \x_2^{}; \y_1^{}, \y_2^{}) \nonumber\\[2pt]
&=& \frac{C_W^{(\l)}}{2}\! \int \!d\z_1^{} d\z_2^{} \,\bigl(\K_0^{(\lm1)}\bigr)^{\!-1}\* (\y_1^{}, \z_1^{}) \,\bigl(\K_0^{(\lm1)}\bigr)^{\!-1}\* (\y_2^{}, \z_2^{}) 
\nonumber\\
&&\hspace{40pt}\Bigl\langle \Delta (\x_1^{}, \x_2^{}) \,\phi(\z_1^{}) \,\phi(\z_2^{}) \Bigr\rangle_{\K^{(\lm1)}_0} +\O\Bigl(\frac{1}{n}\Bigr)
\nonumber\\
&=& \frac{C_W^{(\l)}}{2} \biggl\langle \frac{\delta^2 \Delta(\x_1^{}, \x_2^{})}{\delta \phi(\y_1^{}) \delta \phi(\y_2^{})} \biggr\rangle_{\!\!\K^{(\lm1)}_0} +\O\Bigl(\frac{1}{n}\Bigr) \,.
\label{eq:chi}
\end{eqnarray}
We can clearly see the structural similarity to Eq.~\eqref{eq:4pt_2} above. 
Ultimately, the same subdiagram enters both equations, and we can write:
\begin{equation}
\chi^{(\l)}\*(\x_1^{}, \x_2^{}; \y_1^{}, \y_2^{}) = \sum_j
\begin{tikzpicture}[baseline=(b12)]
\begin{feynman}
\vertex (l) {};
\vertex[below = 16pt of l, dot, minimum size = 3pt, label = {left: {\footnotesize $\x_1^{}$}}] (x1) {};
\vertex[above = 16pt of l, dot, minimum size = 3pt, label = {left: {\footnotesize $\x_2^{}$}}] (x2) {};
\vertex[right = 8pt of l, dot, minimum size = 0pt] (v12) {};
\vertex[right = 20pt of v12, blob] (b12) {};
\vertex[right = 16pt of b12, dot, minimum size = 0pt] (r) {};
\vertex[below = 16pt of r, dot, minimum size = 0pt, label = {right: {\footnotesize $\phi_j (\y_1^{})$}}] (y1) {};
\vertex[above = 16pt of r, dot, minimum size = 0pt, label = {right: {\footnotesize $\phi_j (\y_2^{})$}}] (y2) {};
\diagram*{
	(x1) -- (v12) -- (x2),
	(v12) -- [photon, edge label = {\scriptsize \,$\Delta_j$}, inner sep = 4pt] (b12),
	(y1) -- (b12) -- (y2)
};
\end{feynman}
\end{tikzpicture}
,
\end{equation}
where the $\phi_j^{(\lm1)}$ legs are amputated, and an exchange symmetry between them in the full diagram is assumed (hence a symmetry factor $\frac{1}{2}$ in Eq.~\eqref{eq:chi}).

The same pattern persists for higher-point connected correlators. 
For the connected four-point correlator, an infinitesimal variation of $V_4^{(\lm1)}\*(\x_1^{}, \x_2^{}; \x_3^{}, \x_4^{})$ results in a change in $V_4^{(\l)}\*(\x_1^{}, \x_2^{}; \x_3^{}, \x_4^{})$: 
\vspace{8pt}
\begin{equation}
\begin{tikzpicture}[baseline=(d.base)]
\begin{feynman}
\vertex (l) {};
\vertex[below = 16pt of l, dot, minimum size=3pt] (x1) {};
\vertex[above = 16pt of l, dot, minimum size=3pt] (x2) {};
\vertex[right = 8pt of l, dot, minimum size=0pt] (v12) {};
\vertex[right = 8pt of v12, dot, minimum size=0pt] (v34) {};
\vertex[right = 8pt of v34] (r) {};
\vertex[above = 16pt of r, dot, minimum size=3pt] (x3) {};
\vertex[below = 16pt of r, dot, minimum size=3pt] (x4) {};
\diagram*{
	(x1) -- (v12) -- (x2),
	(x3) -- (v34) -- (x4)
};
\vertex[right = 4pt of v12, blob, minimum size = 8pt] (d) {{\scriptsize $\delta$}};
\end{feynman}
\end{tikzpicture}
\;=\; \sum_{j_1,j_2}
\begin{tikzpicture}[baseline=(d.base)]
\begin{feynman}
\vertex (l) {};
\vertex[below = 16pt of l, dot, minimum size=3pt] (x1) {};
\vertex[above = 16pt of l, dot, minimum size=3pt] (x2) {};
\vertex[right = 8pt of l, dot, minimum size=0pt] (v12) {};
\vertex[right = 20pt of v12, blob] (b12) {};
\vertex[right = 16pt of b12, dot, minimum size=0pt] (w12) {};
\vertex[above = 1pt of w12, label = {above: {\scriptsize \hspace{-10pt} $\phi_{j_1}$}}] (w12u) {};
\vertex[right = 8pt of w12, dot, minimum size=0pt] (w34) {};
\vertex[above = 1pt of w34, label = {above: {\scriptsize \hspace{10pt} $\phi_{j_2}$}}] (w34u) {};
\vertex[right = 16pt of w34, blob] (b34) {};
\vertex[right = 20pt of b34, dot, minimum size=0pt] (v34) {};
\vertex[right = 8pt of v34] (r) {};
\vertex[above = 16pt of r, dot, minimum size=3pt] (x3) {};
\vertex[below = 16pt of r, dot, minimum size=3pt] (x4) {};
\diagram*{
	(x1) -- (v12) -- (x2),
	(v12) -- [photon, edge label = {\scriptsize \;$\Delta_{j_1}$}, inner sep = 4pt] (b12) -- [quarter left] (w12) -- [quarter left] (b12),
	(b34) -- [quarter left] (w34) -- [quarter left] (b34) -- [photon, edge label = {\scriptsize \,$\Delta_{j_2}$}, inner sep = 4pt] (v34),
	(x3) -- (v34) -- (x4)
};
\vertex[right = 4pt of w12, blob, minimum size = 8pt] (d) {{\scriptsize $\delta$}};
\end{feynman}
\end{tikzpicture}
\;\;,
\label{eq:dV4_diag}
\vspace{8pt}
\end{equation}
and we find:
\begin{eqnarray}
&&\frac{n_{\l\shortminus 2}}{n_\lm1} \frac{\delta V_4^{(\l)} \*\bigl(\x_1^{}, \x_2^{} ; \x_3^{}, \x_4^{}\bigr)}{\delta V_4^{(\lm1)} \*\bigl(\y_1^{}, \y_2^{} ; \y_3^{}, \y_4^{}\bigr)} \nonumber\\
&=& \frac{1}{2}\Bigl[ \chi^{(\l)}\*\bigl(\x_1^{}, \x_2^{}; \y_1^{}, \y_2^{}\bigr) \, \chi^{(\l)}\*\bigl(\x_3^{}, \x_4^{}; \y_3^{}, \y_4^{}\bigr) \nonumber\\
&&\hspace{10pt}+\, \chi^{(\l)}\*\bigl(\x_1^{}, \x_2^{}; \y_3^{}, \y_4^{}\bigr) \, \chi^{(\l)}\*\bigl(\x_3^{}, \x_4^{}; \y_1^{}, \y_2^{}\bigr) \Bigr] \,,
\label{eq:dV4}
\end{eqnarray}
where the symmetrized form arises because $V_4^{(\l)} \*\bigl(\x_1^{}, \x_2^{} ; \x_3^{}, \x_4^{}\bigr) = V_4^{(\l)} \*\bigl(\x_3^{}, \x_4^{} ; \x_1^{}, \x_2^{}\bigr)$.

Generally, the connected $2k$-point correlator is defined by
\begin{eqnarray}
&& \bigl\langle \phi^{(\l)}_{i_1^{}}\*(\x_1^{}) \,\dots \, \phi^{(\l)}_{i_{2k}^{}}\*(\x_{2k}^{}) \bigr\rangle_\C \nonumber\\[2pt]
&=& \delta_{i_1^{} i_2^{}} \*\cdots\, \delta_{i_{2k\shortminus 1}^{} i_{2k}^{}}\, \frac{1}{n_\lm1^{k\shortminus 1}}\,V_{2k}^{(\l)} \*\bigl(\x_1^{}, \x_2^{} ; \dots ; \x_{2k\shortminus 1}^{}, \x_{2k}^{}\bigr) \nonumber\\
&&+\, \text{perms.},
\label{eq:V2k_def}
\end{eqnarray}
where $V_{2k}^{(\l)}$ can be shown to be $\O(1)$~\cite{hanin2022correlation}. 
Its variation follows from:
\begin{equation}
\begin{tikzpicture}[baseline=(d.base)]
\begin{feynman}
\vertex (d) {};
\vertex[left = 4pt of d, dot, minimum size = 0pt] (v12) {};
\vertex[left = 16pt of v12] (l) {};
\vertex[below = 8pt of l, dot, minimum size = 3pt] (x1) {};
\vertex[above = 8pt of l, dot, minimum size = 3pt] (x2) {};
\vertex[above = 4pt of d, dot, minimum size = 0pt] (v34) {};
\vertex[above = 16pt of v34] (u) {};
\vertex[left = 8pt of u, dot, minimum size = 3pt] (x3) {};
\vertex[right = 8pt of u, dot, minimum size = 3pt] (x4) {};
\vertex[right = 4pt of d, dot, minimum size = 0pt] (v56) {};
\vertex[right = 16pt of v56, dot, minimum size = 0pt] (r) {};
\vertex[above = 8pt of r, dot, minimum size = 3pt] (x5) {};
\vertex[below = 8pt of r, dot, minimum size = 3pt] (x6) {};
\diagram*{
	(x1) -- (v12) -- (x2),
	(x3) -- (v34) -- (x4),
	(x5) -- (v56) -- (x6)
};
\vertex[right = 0pt of d, blob, minimum size = 8pt] (dd) {{\scriptsize $\delta$}};
\vertex[below = 15pt of d, dot, minimum size = 1pt] (d2) {};
\vertex[left = 8pt of d2] (d2l) {};
\vertex[above = 2pt of d2l, dot, minimum size = 1pt] (d1) {};
\vertex[right = 8pt of d2] (d2r) {};
\vertex[above = 2pt of d2r, dot, minimum size = 1pt] (d3) {};
\end{feynman}
\end{tikzpicture}
\;\;\,=\; \sum_{j_1, \dots, j_k}
\begin{tikzpicture}[baseline=(d.base)]
\begin{feynman}
\vertex (d) {};
\vertex[left = 4pt of d, dot, minimum size = 0pt] (w12) {};
\vertex[below = 1pt of w12, label = {below: {\scriptsize \hspace{-10pt} $\phi_{j_2}$}}] (w12d) {};
\vertex[left = 16pt of w12, blob] (b12) {};
\vertex[left = 20pt of b12, dot, minimum size = 0pt] (v12) {};
\vertex[left = 8pt of v12] (l) {};
\vertex[below = 16pt of l, dot, minimum size = 3pt] (x1) {};
\vertex[above = 16pt of l, dot, minimum size = 3pt] (x2) {};
\vertex[above = 4pt of d, dot, minimum size = 0pt] (w34) {};
\vertex[right = 13pt of w34, label = {above: {\scriptsize $\phi_{j_1}$}}, inner sep = -1pt] (w34r) {};
\vertex[above = 16pt of w34, blob] (b34) {};
\vertex[above = 20pt of b34, dot, minimum size = 0pt] (v34) {};
\vertex[above = 8pt of v34] (u) {};
\vertex[left = 16pt of u, dot, minimum size = 3pt] (x3) {};
\vertex[right = 16pt of u, dot, minimum size = 3pt] (x4) {};
\vertex[right = 4pt of d, dot, minimum size = 0pt] (w56) {};
\vertex[below = 1pt of w56, label = {below: {\scriptsize \hspace{12pt} $\phi_{j_k}$}}] (w56d) {};
\vertex[right = 16pt of w56, blob] (b56) {};
\vertex[right = 20pt of b56, dot, minimum size = 0pt] (v56) {};
\vertex[right = 8 pt of v56, dot, minimum size = 0pt] (r) {};
\vertex[above = 16pt of r, dot, minimum size = 3pt] (x5) {};
\vertex[below = 16pt of r, dot, minimum size = 3pt] (x6) {};
\diagram*{
	(x1) -- (v12) -- (x2),
	(x3) -- (v34) -- (x4),
	(x5) -- (v56) -- (x6),
	(v12) -- [photon, edge label' = {\scriptsize \;$\Delta_{j_2}$}, inner sep = 4pt] (b12) -- [quarter left] (w12) -- [quarter left] (b12),
	(v34) -- [photon, edge label = {\scriptsize $\Delta_{j_1}$}, inner sep = 4pt] (b34) -- [quarter left] (w34) -- [quarter left] (b34),
	(v56) -- [photon, edge label = {\scriptsize \,$\Delta_{j_k}$}, inner sep = 4pt] (b56) -- [quarter left] (w56) -- [quarter left] (b56)
};
\vertex[right = 0pt of d, blob, minimum size = 8pt] (dd) {{\scriptsize $\delta$}};
\vertex[below = 25pt of d, dot, minimum size = 1pt] (d2) {};
\vertex[left = 10pt of d2] (d2l) {};
\vertex[above = 3pt of d2l, dot, minimum size = 1pt] (d1) {};
\vertex[right = 10pt of d2] (d2r) {};
\vertex[above = 3pt of d2r, dot, minimum size = 1pt] (d3) {};
\end{feynman}
\end{tikzpicture}
\;\;,\label{eq:dV2k_diag}
\end{equation}
and we have
\begin{eqnarray}
&&\biggl(\frac{n_{\l\shortminus 2}}{n_\lm1}\biggr)^{\!\* k\shortminus 1} \* \frac{\delta V_{2k}^{(\l)} \*\bigl(\x_1^{}, \x_2^{} ; \dots ; \x_{2k\shortminus 1}^{}, \x_{2k}^{}\bigr)}{\delta V_{2k}^{(\lm1)} \*\bigl(\y_1^{}, \y_2^{} ; \dots ; \y_{2k\shortminus 1}^{}, \y_{2k}^{}\bigr)} \nonumber\\
&=& \text{sym.} \Biggl[\prod_{k'\*=1}^{k} \chi^{(\l)}\*\bigl(\x_{2k'\*\shortminus 1}^{}, \x_{2k'}^{}; \y_{2k'\*\shortminus 1}^{}, \y_{2k'}^{}\bigr) \Biggr] \,,
\end{eqnarray}
where ``sym.''\ means symmetrizing the expression in the same way as in Eq.~\eqref{eq:dV4}.

The quantity $\chi^{(\l)}\*(\x_1^{}, \x_2^{}; \y_1^{}, \y_2^{})$ in the equations above is a generalization of the parallel and perpendicular susceptibilities, $\chi_\parallel^{}$ and $\chi_\perp^{}$, introduced in Ref.~\cite{Roberts:2021fes} when analyzing the special case of two nearby inputs. 
In the nearby-inputs limit, tuning the network to criticality means adjusting the hyperparameters $C_W^{(\l)}$, $C_b^{(\l)}$ such that the kernel recursion Eq.~\eqref{eq:K_0_recursion} has a fixed point $\K^\star$ where $\chi_\parallel^{} = \chi_\perp^{} =1$. 
In the Supplemental Material, we show that, at least for the scale-invariant and $\K^\star=0$ universality classes, this tuning actually implies a stronger condition is satisfied (at LO in $\frac{1}{n}$):
\begin{eqnarray}
&&\chi^{(\l)}\*(\x_1^{}, \x_2^{}; \y_1^{}, \y_2^{}) \Bigr|_{\K_0^{\*(\lm1)} \*= \,\K^\star} \nonumber\\
&=& \frac{1}{2} \Bigl[\delta(\x_1^{} - \y_1^{}) \,\delta(\x_2^{} - \y_2^{}) \*+\*\delta(\x_1^{} - \y_2^{}) \,\delta(\x_2^{} - \y_1^{}) \Bigr] .\qquad
\label{eq:crit}
\end{eqnarray}
Eq.~\eqref{eq:crit} ensures perturbations around the fixed point stay constant through the layers, not just for the two-point correlator, $\delta \bigl\langle\G^{(\l)}\*(\x_1^{}, \x_2^{})\bigr\rangle = \delta \bigl\langle\G^{(\lm1)}\*(\x_1^{}, \x_2^{})\bigr\rangle$, but for the entire tower of higher-point connected correlators, $\frac{1}{n_\lm1^{k\shortminus 1}}\, \delta V_{2k}^{(\l)} \*\bigl(\x_1^{}, \dots, \x_{2k}^{}\bigr) = \frac{1}{n_{\l\shortminus 2}^{k\shortminus 1}}\, \delta V_{2k}^{(\lm1)} \*\bigl(\x_1^{}, \dots, \x_{2k}^{}\bigr)$. 
This in turn implies that for all of them, RG flow toward the fixed point is power-law instead of exponential, once the single condition Eq.~\eqref{eq:crit} is satisfied.
The discussion above makes it transparent that power-law scaling of higher-point connected correlators at criticality (previously observed in Refs.~\cite{Roberts:2021fes, hanin2022correlation} up to eight-point level in the degenerate-input limit) has its roots in the structures of EFT interactions, as manifested by the common structure shared by the diagrams in Eqs.~\eqref{eq:dG_diag}, \eqref{eq:dV4_diag} and \eqref{eq:dV2k_diag}.

\newpar
\paragraph{Summary and outlook}---\,
%
In this letter, we introduced a diagrammatic formalism that significantly simplifies perturbative calculations of finite-width effects in EFTs corresponding to archetypical deep neural networks. 
The concise reproduction of known results and derivation of new results highlights the efficiency of the diagrammatic approach, while the incorporation of ghosts vastly simplifies $\frac{1}{n}$ counting in the EFT action. 
Our analysis also made transparent the structures of such EFTs which underlie the success of critical tuning in deep neural networks. 
In fact, a universal diagrammatic structure emerges in the RG analysis of all higher-point connected correlators of neuron preactivations, which means criticality (\ie\ power-law as opposed to exponential scaling) of all the neuron statistics at initialization is governed by a single condition, Eq.~\eqref{eq:crit}. 

From the deep learning point of view, an obvious next step is to extend the diagrammatic formalism to incorporate gradient-based training and simplify perturbative calculations involving the neural tangent kernel~\cite{jacot2018neural, lee2019wide} and its differentials~\cite{hanin2019finite, huang2020dynamics, Dyer:2019uzd, Roberts:2021fes}. 
From the fundamental physics point of view, we are hopeful that much more can be learned from the intimate connection between neural networks and field theories. 
Understanding the structures of EFTs corresponding to other neural network architectures (\eg\ recurrent neural networks~\cite{Grosvenor:2021eol, segadlo2021unified} and transformers~\cite{Dinan:2023aah}) will allow us to gain further insights into this connection and potentially point to novel ML architecture designs for simulating field theories.

\begin{acknowledgments}
\newpar
\paragraph{Acknowledgments}---\,
We are particularly grateful to Sho Yaida for helpful conversations throughout the course of this work.
We thank Hannah Day, Marat Freytsis, Boris Hanin, Yonatan Kahn, and Anindita Maiti for useful discussions and comments on a preliminary draft, and Guy Gur-Ari for related discussions. 
Feynman diagrams in this work were drawn using \texttt{tikz-feynman}~\cite{Ellis:2016jkw}. 
This work was supported in part by the U.S.\ Department of Energy under the grant DE-SC0011702. 
This work was performed in part at the Aspen Center for Physics, supported by the National Science Foundation under Grant No.~NSF PHY-2210452, and the Kavli Institute for Theoretical Physics, supported by the National Science Foundation under Grant No.~NSF PHY-1748958.
\end{acknowledgments}

\bibliography{ml}

\onecolumngrid
\newpage

\section{Supplemental Material}
\renewcommand{\theequation}{A.\arabic{equation}}
\setcounter{equation}{0}

\subsection{EFT action}

In this section, we detail the intermediate steps that lead to the EFT action for multilayer perceptrons, given by Eq.~\eqref{eq:exp-S}. 
The initial steps of the derivation closely follow Ch.~4 of Ref.~\cite{Roberts:2021fes}. 
We start with the operations in Eq.~\eqref{eq:operations} which recursively determine $\l\,$th-layer preactivations from $(\l\*-\*1)\,$th-layer preactivations:
\begin{equation}
\phi_i^{(\l)}(\x) = \sum_{j=1}^{n_{\lm1}} W_{ij}^{(\l)} \Q\bigl(\phi_j^{(\lm1)}(\x)\bigr) +b_i^{(\l)} \,,
\label{seq:operations}
\end{equation}
where it is understood that $\Q\bigl(\phi_j^{(\lm1)}(\x)\bigr)$ should be replaced by $x_j$ when $\l=1$. 
Considering an ensemble of networks where the weights $W_{ij}^{(\l)}$ and biases $b_i^{(\l)}$ are drawn independently from some probability distributions, $P_W^{(\l)}$ and $P_b^{(\l)}$, at initialization, we can write down the following conditional probability distribution (treating $\x$ as discrete for the moment):
\begin{equation}
P\bigl( \phi^{(\l)} \big| \phi^{(\lm1)} \bigr) = \prod_{i,j} \int dW_{ij}\, P_W^{(\l)}\bigl(W_{ij}\bigr) \prod_i \int db_i \, P_b^{(\l)} (b_i) \,\prod_{i,\x} 
\delta\Bigl( \phi_i^{(\l)}(\x) - \sum_{j=1}^{n_{\lm1}} W_{ij}\, \Q\bigl(\phi_j^{(\lm1)}(\x)\bigr) -b_i \Bigr) \,.
\label{seq:cond_prob}
\end{equation}
The integrals over weights and biases can be performed analytically when $P_W^{(\l)}$ and $P_b^{(\l)}$ are Gaussian. 
Concretely, assuming they are zero-mean Gaussians with variances $C_W^{(\l)}/n_\lm1$ and $C_b^{(\l)}$, respectively:
\begin{equation}
P_W^{(\l)}(W) = \frac{1}{\sqrt{2\pi C_W^{(\l)}/n_\lm1}} \exp \Biggl(-\frac{W^2}{2C_W^{(\l)}/n_\lm1}\Biggr) \,,\qquad\quad
P_b^{(\l)}(b) = \frac{1}{\sqrt{2\pi C_b^{(\l)}}} \exp \Biggl(-\frac{b^2}{2C_b^{(\l)}}\Biggr) \,,
\label{seq:PWPb}
\end{equation}
and rewriting the delta functions in Eq.~\eqref{seq:cond_prob} as
\begin{equation}
\delta\Bigl( \phi_i^{(\l)}(\x) - \sum_{j=1}^{n_{\lm1}} W_{ij} \Q\bigl(\phi_j^{(\lm1)}(\x)\bigr) -b_i \Bigr) =
\int\frac{d\Lambda_i(\x)}{2\pi} \,\exp \biggl[i \,\Lambda_i(\x) \Bigl( \phi_i^{(\l)}(\x) - \sum_{j=1}^{n_{\lm1}} W_{ij} \Q\bigl(\phi_j^{(\lm1)}(\x)\bigr) -b_i \Bigr) \biggr]
\end{equation}
we can complete the squares and integrate out $W_{ij}$, $b_i$. 
Finally integrating over the auxiliary variables $\Lambda_i(\x)$ (where the integrand is again Gaussian) and taking the continuum limit, we obtain Eq.~\eqref{eq:cond_prob} in the main text, reprinted here for convenience:
\begin{equation}
P\bigl( \phi^{(\l)} \big| \phi^{(\lm1)} \bigr) = 
\Bigl[\det \Bigl(2\pi\G^{(\l)}\Bigr)\Bigr]^{\*-\*\frac{n_\l}{2}} \exp \Biggl[ - \!\int\! d\x_1^{} d\x_2^{} \,\frac{1}{2} \sum_{i=1}^{n_\l} \phi_i^{(\l)} \*(\x_1^{}) \bigl(\G^{(\l)}\bigr)^{\!-\*1}\!(\x_1^{}, \x_2^{}) \,\phi_i^{(\l)} \*(\x_2^{})\Biggr] \,,
\label{seq:cond_prob_result}
\end{equation}
where
\begin{equation}
\G^{(\l)} (\x_1^{}, \x_2^{}) \equiv \frac{1}{n_\lm1}\sum_{j=1}^{n_\lm1} \G^{(\l)}_j (\x_1^{}, \x_2^{}) \,,\qquad\quad
\G^{(\l)}_j (\x_1^{}, \x_2^{}) \equiv C_b^{(\l)} + C_W^{(\l)}\, \Q_{j,\x_1}^{(\lm1)}\, \Q_{j,\x_2}^{(\lm1)} = \G^{(\l)}_j (\x_2^{}, \x_1^{})  \,,
\end{equation}
with the understanding that $\Q_{j,\x_1}^{(\lm1)} \equiv \Q\bigl(\phi_j^{(\lm1)}(\x_1)\bigr)$ should be replaced by $x_{1j}^{}$ ($j$\,th component of $\x_1$), and similarly for $\Q_{j,\x_2}^{(\lm1)}$, when $\l=1$. 
The (pseudo)inverse $\bigl(\G^{(\l)}\bigr)^{\!-\*1}$ satisfies:
\begin{equation}
\int d\x_1^{} d\x_2^{}\, \G^{(\l)} (\x_1^{}, \x_1^{\,\prime}) \,\bigl(\G^{(\l)}\bigr)^{\!-\*1}(\x_1^{}, \x_2^{}) \,\G^{(\l)} (\x_2^{}, \x_2^{\,\prime}) = \G(\x_1^{\,\prime}, \x_2^{\,\prime}) \,.
\end{equation}

Motivated by the superficial analogy of the steps above with the Faddeev-Popov procedure (where Eq.~\eqref{seq:operations} plays the role of ``choosing a gauge''), we next rewrite the functional determinant as a path integral over a set of anticommuting fields, \ie\ ghosts and antighosts (a procedure also known as supersymmetrization in other contexts):
\begin{equation}
\Bigl[\det \Bigl(2\pi\G^{(\l)}\Bigr)\Bigr]^{\*-\*\frac{n_\l}{2}} = 
\int \D\gh \D\bar\gh \,\exp \Biggl[ \,\sum_{i'=1}^{n_\l/2} \bar\gh_{i'}^{(\l)} \*(\x_1^{}) \bigl(\G^{(\l)}\bigr)^{\!-\*1}\!(\x_1^{}, \x_2^{}) \,\gh_{i'}^{(\l)} \*(\x_2^{}) \Biggr] \,.
\end{equation}
Accounting for all layers $\l = 1, \dots, L$ simply amounts to multiplying the conditional probability distributions:
\begin{equation}
e^{-\S} = P \bigl(\phi^{(1)}, \dots, \phi^{(L)}\bigr) = P\bigl(\phi^{(1)}\bigr) \,P\bigl(\phi^{(2)} | \phi^{(1)}\bigr) \,\dots \, P\bigl(\phi^{(L)} | \phi^{(L\shortminus 1)}\bigr) \,,
\end{equation}
and we arrive at Eq.~\eqref{eq:exp-S} in the main text.
Note that the total number of ghosts and antighosts is the same as the number of neurons, $\sum\limits_{\l=1}^{L} n_\l^{}$. 
So while the randomly-initialized weights and biases introduce independent  stochasticity at each layer, there are no physical ``field'' degrees of freedom with independent fluctuations. (The analogous statement in the Faddeev-Popov procedure is that considering a family of gauges does not introduce new physical degrees of freedom into the gauge theory.)

\subsection{Feynman rule and general procedure of diagrammatic calculation}

To show that the Feynman rule in Eq.~\eqref{eq:basic_rule} is the correct one, we first note that if $\phi_j^{(\lm1)}(\x)$ were classical background fields, we would simply have a free theory for $\phi_i^{(\l)}(\x)$ with propagator $\G^{(\l)}\*(\x_1, \x_2)$. 
In this case, the two-point correlator is given by
\begin{equation}
\bigl\langle \phi^{(\l)}_{i_1^{}}\*(\x_1^{}) \,\phi^{(\l)}_{i_2^{}}\*(\x_2^{}) \bigr\rangle = \delta_{i_1^{} i_2^{}} \,\G^{(\l)}\*(\x_1^{}, \x_2^{}) \,,
\label{seq:2-pt_classical}
\end{equation}
and, by simple Wick contraction, the four-point correlator is given by
\begin{equation}
\bigl\langle \phi^{(\l)}_{i_1^{}}\*(\x_1^{}) \,\phi^{(\l)}_{i_2^{}}\*(\x_2^{}) \,\phi^{(\l)}_{i_3^{}}\*(\x_3^{}) \,\phi^{(\l)}_{i_4^{}}\*(\x_4^{}) \bigr\rangle
= \delta_{i_1^{} i_2^{}} \delta_{i_3^{} i_4^{}} \,\G^{(\l)}\*(\x_1^{}, \x_2^{})\, \G^{(\l)}\*(\x_3^{}, \x_4^{}) + \text{perms.}
\label{seq:4-pt_classical}
\end{equation}
In reality, $\phi_j^{(\lm1)}(\x)$ are not classical background fields but exhibit statistical fluctuations. 
This means we should take the ensemble average of the expressions on the right-hand sides of Eqs.~\eqref{seq:2-pt_classical} and \eqref{seq:4-pt_classical}, which can be represented diagrammatically as:
\vspace{-10pt}
\begin{eqnarray}
\bigl\langle \phi^{(\l)}_{i_1^{}}\*(\x_1^{}) \,\phi^{(\l)}_{i_2^{}}\*(\x_2^{}) \bigr\rangle &=& \delta_{i_1^{} i_2^{}} \bigl\langle \G^{(\l)}\*(\x_1, \x_2) \bigr\rangle
= \delta_{i_1^{} i_2^{}} \,\sum_j
\begin{tikzpicture}[baseline=(b)]
\begin{feynman}
\vertex[dot, minimum size=3pt, label = {below: {\scriptsize $\phi_i^{(\l)}\*(\x_1^{})$}}] (x1) {};
\vertex[right = 20pt of x1] (v);
\vertex[right = 20pt of v, dot, minimum size=3pt, label = {below: {\scriptsize $\phi_i^{(\l)}\*(\x_2^{})$}}] (x2){};
\vertex[above = 2pt of v] (b){};
\vertex[above = 12pt of v, blob] (G){};
\vertex[above = 20pt of v, label = {above: {\footnotesize $\frac{1}{n_\lm1}\bigl\langle \G^{(\l)}_j\!(\x_1^{}, \x_2^{})\bigr\rangle$}}] (eq){};
\diagram*{
	(x1) -- (v) -- (x2),
	(v) -- [graviton, thick] (G)
};
\end{feynman}
\end{tikzpicture}
\,,
\label{seq:2-pt}
\\[5pt]
\bigl\langle \phi^{(\l)}_{i_1^{}}\*(\x_1^{}) \,\phi^{(\l)}_{i_2^{}}\*(\x_2^{}) \,\phi^{(\l)}_{i_3^{}}\*(\x_3^{}) \,\phi^{(\l)}_{i_4^{}}\*(\x_4^{}) \bigr\rangle
&=& \delta_{i_1^{} i_2^{}} \delta_{i_3^{} i_4^{}} \bigl\langle \G^{(\l)}\*(\x_1, \x_2)\, \G^{(\l)}\*(\x_3, \x_4) \bigr\rangle + \text{perms.} 
\nonumber\\[5pt]
&=&
\delta_{i_1^{} i_2^{}} \delta_{i_3^{} i_4^{}} \,\sum_{j_1,j_2}
\begin{tikzpicture}[baseline=(b)]
\begin{feynman}
\vertex (l) {};
\vertex[below = 16pt of l, dot, minimum size=3pt, label = {below: {\footnotesize $\x_1^{}$}}] (x1) {};
\vertex[above = 16pt of l, dot, minimum size=3pt, label = {above: {\footnotesize $\x_2^{}$}}] (x2) {};
\vertex[right = 8pt of l, dot, minimum size=0pt] (v12) {};
\vertex[right = 20pt of v12, blob] (b) {};
\vertex[right = 20pt of b, dot, minimum size=0pt] (v34) {};
\vertex[right = 8pt of v34] (r) {};
\vertex[above = 16pt of r, dot, minimum size=3pt, label = {above: {\footnotesize $\x_3^{}$}}] (x3) {};
\vertex[below = 16pt of r, dot, minimum size=3pt, label = {below: {\footnotesize $\x_4^{}$}}] (x4) {};
\diagram*{
	(x1) -- (v12) -- (x2),
	(v12) -- [graviton, thick, edge label = {\scriptsize $\;\; \G_{j_1}^{(\l)}$}, inner sep = 4pt] (b) -- [graviton, thick, edge label = {\scriptsize $\;\G_{j_2}^{(\l)} $}, inner sep = 4pt] (v34), 
	(x3) -- (v34) -- (x4)
};
\end{feynman}
\end{tikzpicture}
+ \text{perms.} 
\label{seq:4-pt}
\end{eqnarray}
The same discussion applies to higher-point correlators. 
The end result amounts to simply stipulating the rule on the left-hand side of Eq.~\eqref{eq:basic_rule} 
and requiring that each double wavy line should connect to a blob (\ie\ it cannot be an external line). 
Furthermore, since we are mostly interested in connected correlators, it is convenient to decompose $\G^{(\l)}_j\!(\x_1^{}, \x_2^{})$ into an expectation value piece and a fluctuating piece as on the right-hand side of Eq.~\eqref{eq:basic_rule}; the former is automatically disconnected from the rest of a diagram, so only the latter explicitly enters our calculations. 
One further simplification due to this decomposition follows from the fact that the fluctuation piece is tadpole-free:
\begin{equation}
\begin{tikzpicture}[baseline=(b)]
\begin{feynman}
\vertex[dot, minimum size=3pt, label = {below: {\footnotesize $\x_1^{}$}}] (x1) {};
\vertex[right = 20pt of x1] (v);
\vertex[right = 20pt of v, dot, minimum size=3pt, label = {below: {\footnotesize $\x_2^{}$}}] (x2){};
\vertex[above = 2pt of v] (b){};
\vertex[above = 3pt of v, label = {right: {\footnotesize $\!\!\Delta_j$}}] (j){};
\vertex[above = 12pt of v, blob] (G){};
\vertex[above = 20pt of v] (eq){};
\diagram*{
	(x1) -- (v) -- (x2),
	(v) -- [photon] (G)
};
\end{feynman}
\end{tikzpicture}
= \frac{C_W^{(\l)}}{n_{\lm1}} \,
\bigl\langle \Delta_{j}^{\*\*(\lm1)}\!(\x_1^{}, \x_2^{}) \bigr\rangle = 0 \,.
\label{seq:tadpole_free}
\end{equation}
This means the blobs in Eq.~\eqref{eq:V_4} and Eq.~\eqref{seq:V_6} below, involving two and three $\Delta_{j}^{\*\*(\lm1)}$'s, respectively, are automatically connected, since any way of disconnecting the blobs would result in a tadpole.

We would like to note that the way we use Feynman diagrams in neural network EFT calculations is perhaps slightly different from what one is used to in other contexts. 
Usually one would derive Feynman rules for propagators and interaction vertices, and use them to build diagrams from which one can calculate correlators in terms of parameters of the theory. 
In the present case, however, our goal is to derive RG flows, which are {\it relations} between correlators. 
The general strategy here is to first write $\l$\,th-layer $\phi$ correlators in terms of $(\l\*-\*1)$\,th-layer $\Delta$ correlators, \ie\ expectation values of (products of) $\Delta_j^{(\lm1)}$'s, as summarized by the Feynman rule Eq.~\eqref{eq:basic_rule} and exemplified in Eqs.~\eqref{seq:2-pt} and \eqref{seq:4-pt} above (upon replacing $\G_j^{(\l)}$'s by $\Delta_j^{(\lm1)}$'s to isolate the connected contribution), and then calculate these $(\l\*-\*1)$\,th-layer $\Delta$ correlators in terms of $(\l\*-\*1)$\,th-layer $\phi$ correlators. 
In the second step, if a $\Delta$ correlator involves identical neuron indices (\eg\ in Eq.~\eqref{eq:4pt_1}), it simply takes its free-theory value expressed in terms of free propagators (\ie\ two-point  $\phi$ correlators) at LO; if distinct neuron indices are involved, we need to insert mixed $\Delta$\,-\,$\phi$ correlators (\eg\ the larger blobs in Eq.~\eqref{eq:4pt_2}) to bridge the $\Delta$'s and four- or higher-point $\phi$ correlators. 
In either case, we can express the result in terms of free-theory expectation values of $(\l\*-\*1)\,$th-layer single neuron operators:
\begin{equation}
\Bigl\langle \mathcal{O} \bigl(\phi_i^{(\lm1)}(\x_1^{}) ,  \phi_i^{(\lm1)}(\x_2^{}) , \dots \bigr) \Bigr\rangle_{\K_0^{(\lm1)}} \equiv \frac{\int \D\phi \,\mathcal{O}\bigl(\phi(\x_1^{}), \phi(\x_2^{}), \dots \bigr) \,e^{-\int d\y_1^{} d\y_2^{} \,\frac{1}{2}\,\phi(\y_1^{})\, \bigl(\K_0^{(\lm1)}\bigr)^{-1}(\y_1^{}, \y_2^{}) \,\phi(\y_2^{})}}{\int \D\phi \,e^{-\int d\y_1^{} d\y_2^{} \,\frac{1}{2}\,\phi(\y_1^{})\, \bigl(\K_0^{(\lm1)}\bigr)^{-1}(\y_1^{}, \y_2^{}) \,\phi(\y_2^{})}} \,,
\label{seq:bra_ket_K0}
\end{equation}
where $\mathcal{O}$ represents a product of $\Delta$'s and $\phi$'s (with the exception of the LO two-point correlator $\K_0$ where $\mathcal{O} = \Q_{\x_1^{}} \Q_{\x_2^{}}$; see Eq.~\eqref{eq:K_0_recursion}). 
By Wick contractions we can then rewrite these expectation values in terms of those of functional derivatives of $\Delta$'s (as in \eg\ Eq.~\eqref{eq:4pt_2}). 

To systematically implement this general procedure, it is convenient to introduce the following \cc blob notation:
\begin{equation}
\begin{tikzpicture}[baseline=(bd.base)]
\begin{feynman}
\vertex[blob] (b) {\c};
\vertex[below = 2pt of b] (bd) {};
\vertex[left = 14pt of b] (bl) {};
\vertex[below = 14pt of bl, dot, minimum size = 0pt] (v12) {};
\vertex[below = 14pt of v12, dot, minimum size = 3pt, label = {below: {\scriptsize $\x_1^{}$}}] (x1) {};
\vertex[left = 14pt of v12, dot, minimum size = 3pt, label = {left: {\scriptsize $\x_2^{}$}}] (x2) {};
\vertex[above = 14pt of bl, dot, minimum size = 0pt] (v34) {};
\vertex[left = 14pt of v34, dot, minimum size = 3pt, label = {left: {\scriptsize $\x_{2m\shortminus 1}^{}$}}] (x3) {};
\vertex[above = 14pt of v34, dot, minimum size = 3pt, label = {above: {\scriptsize $\x_{2m}^{}$}}] (x4) {};
\vertex[right = 25pt of b] (br) {};
\vertex[above = 25pt of br] (y1) {\hspace{15pt}{\footnotesize $\phi_j(y_1^{})$}};
\vertex[below = 25pt of br] (y2) {\hspace{15pt}{\footnotesize $\phi_j(y_{2r}^{})$}};
\vertex[left = 26pt of b, dot, minimum size = 1pt] (dl2) {};
\vertex[right = 2pt of dl2] (dl2r) {};
\vertex[below = 6pt of dl2r, dot, minimum size = 1pt] (dl1) {};
\vertex[above = 6pt of dl2r, dot, minimum size = 1pt] (dl3) {};
\vertex[right = 26pt of b, dot, minimum size = 1pt] (dr2) {};
\vertex[left = 2pt of dr2] (dr2l) {};
\vertex[above = 6pt of dr2l, dot, minimum size = 1pt] (dr1) {};
\vertex[below = 6pt of dr2l, dot, minimum size = 1pt] (dr3) {};
\diagram*{
	(x1) -- (v12) -- (x2),
	(x3) -- (v34) -- (x4),
	(v12) -- [photon, edge label' = {\scriptsize $\Delta_j$}, inner sep = 1pt] (b) -- [photon, edge label' = {\scriptsize $\Delta_j$}, inner sep = 1pt] (v34),
	(y1) -- (b) -- (y2)
};
\end{feynman}
\end{tikzpicture}
\equiv\;
\begin{tikzpicture}[baseline=(bd.base)]
\begin{feynman}
\vertex[blob] (b) {};
\vertex[below = 2pt of b] (bd) {};
\vertex[left = 14pt of b] (bl) {};
\vertex[below = 14pt of bl, dot, minimum size = 0pt] (v12) {};
\vertex[below = 14pt of v12, dot, minimum size = 3pt, label = {below: {\scriptsize $\x_1^{}$}}] (x1) {};
\vertex[left = 14pt of v12, dot, minimum size = 3pt, label = {left: {\scriptsize $\x_2^{}$}}] (x2) {};
\vertex[above = 14pt of bl, dot, minimum size = 0pt] (v34) {};
\vertex[left = 14pt of v34, dot, minimum size = 3pt, label = {left: {\scriptsize $\x_{2m\shortminus 1}^{}$}}] (x3) {};
\vertex[above = 14pt of v34, dot, minimum size = 3pt, label = {above: {\scriptsize $\x_{2m}^{}$}}] (x4) {};
\vertex[right = 25pt of b] (br) {};
\vertex[above = 25pt of br] (y1) {\hspace{15pt}{\footnotesize $\phi_j(y_1^{})$}};
\vertex[below = 25pt of br] (y2) {\hspace{15pt}{\footnotesize $\phi_j(y_{2r}^{})$}};
\vertex[left = 26pt of b, dot, minimum size = 1pt] (dl2) {};
\vertex[right = 2pt of dl2] (dl2r) {};
\vertex[below = 6pt of dl2r, dot, minimum size = 1pt] (dl1) {};
\vertex[above = 6pt of dl2r, dot, minimum size = 1pt] (dl3) {};
\vertex[right = 26pt of b, dot, minimum size = 1pt] (dr2) {};
\vertex[left = 2pt of dr2] (dr2l) {};
\vertex[above = 6pt of dr2l, dot, minimum size = 1pt] (dr1) {};
\vertex[below = 6pt of dr2l, dot, minimum size = 1pt] (dr3) {};
\diagram*{
	(x1) -- (v12) -- (x2),
	(x3) -- (v34) -- (x4),
	(v12) -- [photon, edge label' = {\scriptsize $\Delta_j$}, inner sep = 1pt] (b) -- [photon, edge label' = {\scriptsize $\Delta_j$}, inner sep = 1pt] (v34),
	(y1) -- (b) -- (y2)
};
\end{feynman}
\end{tikzpicture}
- \;\;
\begin{pmatrix}
\text{diagrams where the $\phi^{2r}\Delta^m$ blob becomes} \\
\text{disconnected due to contractions among $\phi$'s}
\end{pmatrix} \,,
\label{seq:ast_blob}
\end{equation}
where all the $m$ $\Delta$'s and $2r$ $\phi$'s carry the same neuron index $j$. 
The diagrams being subtracted off in Eq.~\eqref{seq:ast_blob} are those where the $\phi^{2r}\Delta^m$ blob becomes disconnected due to Wick contractions between any number of pairs of $\phi$ legs.
In the main text, we only encountered the $m=2, r=0$ and $m=r=1$ cases when calculating the connected four-point correlator. 
In these cases, there is no distinction between \cc blobs, full blobs and connected blobs, because disconnecting those blobs in any way would give zero due to the tadpole-free condition Eq.~\eqref{seq:tadpole_free}. 
For general $m, r$, though, we have to keep in mind that Wick-contracting the $\phi$'s may not be the only way to disconnect the $\phi^{2r}\Delta^m$ blob; nor does disconnecting a $\phi^{2r}\Delta^m$ blob in a diagram necessarily make the full diagram disconnected. 
Nevertheless, we will see in the examples in the next section that the use of \cc blobs conveniently organizes the derivation of RG flows of connected $\phi$ correlators and neatly takes care of subtleties regarding double-counting.
The \cc blobs also admit simple expressions: requiring that each $\phi$ must be Wick contracted with one of the $\Delta$'s, we arrive at the following general formula at LO in $\frac{1}{n}$:
\begin{equation}
\begin{tikzpicture}[baseline=(bd.base)]
\begin{feynman}
\vertex[blob] (b) {\c};
\vertex[below = 2pt of b] (bd) {};
\vertex[left = 14pt of b] (bl) {};
\vertex[below = 14pt of bl, dot, minimum size = 0pt] (v12) {};
\vertex[below = 14pt of v12, dot, minimum size = 3pt, label = {below: {\scriptsize $\x_1^{}$}}] (x1) {};
\vertex[left = 14pt of v12, dot, minimum size = 3pt, label = {left: {\scriptsize $\x_2^{}$}}] (x2) {};
\vertex[above = 14pt of bl, dot, minimum size = 0pt] (v34) {};
\vertex[left = 14pt of v34, dot, minimum size = 3pt, label = {left: {\scriptsize $\x_{2m\shortminus 1}^{}$}}] (x3) {};
\vertex[above = 14pt of v34, dot, minimum size = 3pt, label = {above: {\scriptsize $\x_{2m}^{}$}}] (x4) {};
\vertex[right = 25pt of b] (br) {};
\vertex[above = 25pt of br] (y1) {\hspace{15pt}{\footnotesize $\phi_j(y_1^{})$}};
\vertex[below = 25pt of br] (y2) {\hspace{15pt}{\footnotesize $\phi_j(y_{2r}^{})$}};
\vertex[left = 26pt of b, dot, minimum size = 1pt] (dl2) {};
\vertex[right = 2pt of dl2] (dl2r) {};
\vertex[below = 6pt of dl2r, dot, minimum size = 1pt] (dl1) {};
\vertex[above = 6pt of dl2r, dot, minimum size = 1pt] (dl3) {};
\vertex[right = 26pt of b, dot, minimum size = 1pt] (dr2) {};
\vertex[left = 2pt of dr2] (dr2l) {};
\vertex[above = 6pt of dr2l, dot, minimum size = 1pt] (dr1) {};
\vertex[below = 6pt of dr2l, dot, minimum size = 1pt] (dr3) {};
\diagram*{
	(x1) -- (v12) -- (x2),
	(x3) -- (v34) -- (x4),
	(v12) -- [photon, edge label' = {\scriptsize $\Delta_j$}, inner sep = 1pt] (b) -- [photon, edge label' = {\scriptsize $\Delta_j$}, inner sep = 1pt] (v34),
	(y1) -- (b) -- (y2)
};
\end{feynman}
\end{tikzpicture}
=
\biggl(\frac{C_W^{(\l)}}{n_\lm1}\biggr)^{\*m} \*
\int \* \prod_{\alpha=1}^{2r} d\z_\alpha^{} \, \K_0^{(\lm1)}\* (\y_\alpha^{}, \z_\alpha^{}) \,\biggl\langle \frac{\delta^{2r} \bigl(\Delta(\x_1^{}, \x_2^{}) \,\cdots\, \Delta(\x_{2m\shortminus 1}^{}, \x_{2m}^{}) \bigr)}{\delta \phi(\z_1^{}) \,\cdots\, \delta \phi(\z_{2r}^{})} \biggr\rangle_{\!\!\K^{(\lm1)}_0} \,.
\label{seq:ast_blob_result}
\end{equation}
As explained below Eq.~\eqref{eq:4pt_2} in the main text, when calculating a full diagram we would always convolve the expression for the subdiagram in Eq.~\eqref{seq:ast_blob_result} with inverse propagators associated with the $\phi$ legs (which would become internal lines in the full diagram), or, in other words, we would amputate the $\phi$ legs in Eq.~\eqref{seq:ast_blob_result}. 
This would leave us with just the expectation value on the right-hand side of Eq.~\eqref{seq:ast_blob_result}. 
In what follows we will use \cc blobs to build up diagrams, although they coincide with full blobs (in which case the ``\c'' label is redundant) when $r=0$ or $m=r=1$.

We finally remark on the comparison of our results with Ref.~\cite{Roberts:2021fes}, which also presented the two-point correlator up to NLO and connected four-point correlator at LO. 
The results in Ref.~\cite{Roberts:2021fes} are also written in terms of free-theory expectation values of single neuron operators, but the operators are products of $\Q$'s and $\phi$'s whereas our results are written in terms of derivatives of products of $\Delta$'s. 
To compare the results, one can use the definition Eq.~\eqref{eq:Delta_def} to rewrite our results in terms of derivatives of $\Q$'s, and use Wick contraction to reduce the results in Ref.~\cite{Roberts:2021fes} to the same expressions. 
For example, the expectation value $\bigl\langle \sigma_{\alpha_1}^{} \sigma_{\alpha_2}^{} (z_{\beta_1}^{} z_{\beta_2}^{} -g_{\beta_1\beta_2}^{}) \bigr\rangle_g$ in the notation of Ref.~\cite{Roberts:2021fes}, which corresponds to $\bigl\langle \Q(\x_1^{}) \,\Q(\x_2^{}) \bigl(\phi(\y_1^{}) \,\phi(\y_2^{}) -\K_0^{(\lm1)}(\y_1^{}, y_2^{})\bigr)\bigr\rangle_{\K_0^{(\lm1)}}$ in the notation here, can be further evaluated by Wick contracting each of the two $\phi$'s with $\Q(\x_1^{}) \,\Q(\x_2^{})$; the other possible Wick contraction---contracting the two $\phi$'s with each other---yields a term that cancels against the $\K_0^{(\lm1)}$ term. 
As a result,
\begin{equation}
\bigl\langle \Q(\x_1^{}) \,\Q(\x_2^{}) \bigl(\phi(\y_1^{}) \,\phi(\y_2^{}) -\K_0^{(\lm1)}(\y_1^{}, y_2^{})\bigr)\bigr\rangle_{\K_0^{(\lm1)}} = \int d\z_1^{} d\z_2^{} \,\K_0^{(\lm1)} (\y_1^{}, \z_1^{}) \, \K_0^{(\lm1)} (\y_2^{}, \z_2^{}) \,\biggl\langle \frac{\delta^2 \bigl(\Q(\x_1^{}) \Q(\x_2^{}) \bigr)}{\delta \phi(\z_1^{}) \,\delta \phi(\z_2^{})} \biggr\rangle_{\!\!\K^{(\lm1)}_0} \,,
\end{equation}
which, up to prefactors, is the same as the expression on the right-hand side of Eq.~\eqref{seq:ast_blob_result} with $m=r=1$ upon substituting in Eq.~\eqref{eq:Delta_def} (the last term in Eq.~\eqref{eq:Delta_def}, as an expectation value itself, has vanishing functional derivatives).

\subsection{Further demonstration of diagrammatic calculations}

We now present three additional calculations to further demonstrate our diagrammatic approach: the two-point correlator at next-to-leading order (NLO), connected six-point correlator at LO and connected eight-point correlator at LO, which are of $\O\bigl(\frac{1}{n}\bigr)$, $\O\bigl(\frac{1}{n^2}\bigr)$ and $\O\bigl(\frac{1}{n^3}\bigr)$, respectively. 
The two-point correlator at NLO was previously computed in Ref.~\cite{Roberts:2021fes}, and we reproduce their result. 
To the best of our knowledge, the results for the connected six-point and eight-point correlators are new; as an independent check, we have also calculated them using the algebraic approach of Ref.~\cite{hanin2022correlation} and found full agreement with our diagrammatic results.\footnote{Ref.~\cite{hanin2022correlation} presented the connected six-point and eight-point correlators in the degenerate input limit. Our results agree with Ref.~\cite{hanin2022correlation} in that limit upon correcting some errors in arXiv v2 of the latter. We thank Boris Hanin for correspondence on this point.}

\subsubsection{Two-point correlator at NLO}

Let us first consider the calculation of two-point correlator at NLO, \ie\ the $p=1$ term of the series in Eq.~\eqref{eq:2-pt_exp}, $\frac{1}{n_\lm1} \K_1^{(\l)}\!(\x_1^{}, \x_2^{})$. 
There are two sources of $\O\bigl(\frac{1}{n}\bigr)$ corrections to the RG flow of two-point correlator. 
First, we can have one of the $(\l\*-\*1)\,$th-layer propagators take its NLO piece:
\begin{equation}
\sum_j
\begin{tikzpicture}[baseline=(b)]
\begin{feynman}
\vertex[dot, minimum size=3pt, label = {below: {\footnotesize $\x_1^{}$}}] (x1) {};
\vertex[right = 20pt of x1] (v);
\vertex[right = 20pt of v, dot, minimum size = 3pt, label = {below: {\footnotesize $\x_2^{}$}}] (x2) {};
\vertex[above = 5pt of v] (b) {};
\vertex[above = 3pt of v, label = {right: \!\*{\scriptsize $\Delta_j$}}] (j){};
\vertex[above = 12pt of v, blob] (G) {};
\vertex[above = 22pt of v, dot, minimum size = 0pt] (g) {};
\vertex[above = 18pt of g, dot, minimum size = 0pt] (w) {};
\diagram*{
	(x1) -- (v) -- (x2),
	(v) -- [photon] (G) -- (g) -- [half left] (w) -- [half left] (g)
};
\vertex[above = 0pt of G, blob] (K0) {\c};
\vertex[above = 8pt of w] (K1) {{\scriptsize $\frac{1}{n_{\l\shortminus 2}} \K_1^{(\lm1)}$}};
\vertex[right = 15pt of G, label = {above: {\scriptsize $\phi_j$}}] (Gr) {};
\end{feynman}
\end{tikzpicture}
= \frac{C_W^{(\l)}}{2\, n_{\l\shortminus 2}} \int \* d\y_1^{} \, d\y_2^{} \; \K_1^{(\lm1)} \*(\y_1^{}, \y_2^{}) \, \biggl\langle \frac{\delta^2 \Delta(\x_1^{}, \x_2^{})}{\delta \phi(\y_1^{}) \, \delta \phi(\y_2^{})} \biggr\rangle_{\!\!\K^{(\lm1)}_0}
\!+\O\Bigl(\frac{1}{n^2}\Bigr) \,,
\label{seq:K1_K}
\end{equation}
where we have used Eq.~\eqref{seq:ast_blob_result} with $m=r=1$, and the symmetry factor $\frac{1}{2}$ comes from exchanging the two $\phi_j^{}$ legs attached to the \cc blob. 
The sum over $j$ yields a factor of $n_\lm1$ which cancels the $\frac{1}{n_\lm1}$ factor from the $\phi^2\Delta$ vertex. 

The other contribution comes from inserting a connected four-point correlator of the $(\l\*-\*1)\,$th layer:
\begin{equation}
\sum_j
\begin{tikzpicture}[baseline=(b)]
\begin{feynman}
\vertex[dot, minimum size = 3pt, label = {below: {\footnotesize $\x_1^{}$}}] (x1) {};
\vertex[right = 20pt of x1] (v);
\vertex[right = 20pt of v, dot, minimum size = 3pt, label = {below: {\footnotesize $\x_2^{}$}}] (x2) {};
\vertex[above = 5pt of v] (b) {};
\vertex[above = 3pt of v, label = {right: \!{\scriptsize $\Delta_j$}}] (j){};
\vertex[above = 12pt of v, blob] (G) {};
\vertex[left = 2pt of G, dot, minimum size = 0pt] (Gl) {};
\vertex[right = 2pt of G, dot, minimum size = 0pt] (Gr) {};
\vertex[above = 4pt of G, dot, minimum size = 0pt] (Gu) {};
\vertex[left = 2pt of Gu, dot, minimum size = 0pt] (Gul) {};
\vertex[right = 2pt of Gu, dot, minimum size = 0pt] (Gur) {};
\vertex[above = 20pt of G, blob, minimum size = 6pt] (V4) {};
\vertex[left = 3pt of V4, dot, minimum size = 0pt] (V4l) {};
\vertex[right = 3pt of V4, dot, minimum size = 0pt] (V4r) {};
\diagram*{
	(x1) -- (v) --	(x2),
	(v) -- [photon] (G) -- (Gl) -- [half left] (V4l) -- [quarter right] (Gul) -- (Gur) -- [quarter right] (V4r) -- [half left] (Gr),
};
\vertex[above = 0pt of G, blob] (K0) {\c};
\vertex[above = 10pt of V4] (VV4) {{\scriptsize $\frac{1}{n_{\l\shortminus 2}} V_4^{(\lm1)}$}};
\vertex[right = 18pt of G, label = {above: {\scriptsize $\phi_j$}}] (Gr) {};
\end{feynman}
\end{tikzpicture}
= \frac{C_W^{(\l)}}{8\,n_{\l\shortminus 2}} \int \*\prod_{\alpha=1}^4 d\y_\alpha^{} \; V_4^{(\lm1)}\* (\y_1^{}, \y_2^{}; \y_3^{}, \y_4^{}) \, \biggl\langle \frac{\delta^4 \Delta(\x_1^{}, \x_2^{})}{\delta \phi(\y_1^{}) \,\delta \phi(\y_2^{}) \,\delta \phi(\y_3^{}) \,\delta \phi(\y_4^{})} \biggr\rangle_{\!\!\K^{(\lm1)}_0}
\!+\O\Bigl(\frac{1}{n^2}\Bigr) \,,
\label{seq:K1_V}
\end{equation}
where we have used Eq.~\eqref{seq:ast_blob_result} with $m=1, r=2$, and the symmetry factor $\frac{1}{2^3} = \frac{1}{8}$ comes from exchanging $\y_1^{} \leftrightarrow \y_2^{}$, $\y_3^{} \leftrightarrow \y_4^{}$ and $(\y_1^{} ,\y_2^{}) \leftrightarrow (\y_3^{} ,\y_4^{})$ among the four $\phi_j^{}$ legs attached to $V_4^{(\lm1)}\* (\y_1^{}, \y_2^{}; \y_3^{}, \y_4^{})$.
It is worth noting that given the definition of \cc blob in Eq.~\eqref{seq:ast_blob}, the diagram in Eq.~\eqref{seq:K1_V} does not include contributions where two of the four $\phi_j$ legs are Wick contracted:
\begin{equation}
\sum_j\;
\begin{tikzpicture}[baseline=(b)]
\begin{feynman}
\vertex[dot, minimum size = 3pt, label = {below: {\footnotesize $\x_1^{}$}}] (x1) {};
\vertex[right = 20pt of x1] (v);
\vertex[right = 20pt of v, dot, minimum size = 3pt, label = {below: {\footnotesize $\x_2^{}$}}] (x2) {};
\vertex[above = 5pt of v] (b) {};
\vertex[above = 3pt of v, label = {right: \!\!\! {\scriptsize $\Delta_j$}}] (j){};
\vertex[above = 12pt of v, blob] (G) {};
\vertex[above = 18pt of G, blob, minimum size = 6pt] (V4) {};
\vertex[above = 0pt of V4, dot, minimum size = 0pt] (v4) {};
\vertex[left = 3pt of V4, dot, minimum size = 0pt] (V4l) {};
\vertex[right = 3pt of V4, dot, minimum size = 0pt] (V4r) {};
\vertex[above = 18pt of V4, dot, minimum size = 0pt] (t) {};
\diagram*{
	(x1) -- (v) -- (x2),
	(v) -- [photon] (G) -- [quarter left] (V4l), 
	(G) -- [quarter right] (V4r),
	(v4) -- [half left] (t) -- [half left] (v4)
};
\vertex[above = 0pt of V4, blob, minimum size = 6pt] (VV4) {};
\vertex[right = 15pt of G, label = {above: {\scriptsize $\phi_j$}}] (Gr) {};
\end{feynman}
\end{tikzpicture}
\;+\,
\begin{tikzpicture}[baseline=(b)]
\begin{feynman}
\vertex[dot, minimum size = 3pt, label = {below: {\footnotesize $\x_1^{}$}}] (x1) {};
\vertex[right = 20pt of x1] (v);
\vertex[right = 20pt of v, dot, minimum size = 3pt, label = {below: {\footnotesize $\x_2^{}$}}] (x2) {};
\vertex[above = 5pt of v] (b) {};
\vertex[above = 3pt of v, label = {right: \!\!\! {\scriptsize $\Delta_j$}}] (j){};
\vertex[above = 12pt of v, blob] (G) {};
\vertex[above = 18pt of G, blob, minimum size = 6pt] (V4) {};
\vertex[below = 3pt of V4, dot, minimum size = 0pt] (V4d) {};
\vertex[above = 3pt of V4, dot, minimum size = 0pt] (V4u) {};
\vertex[above = 18pt of V4, dot, minimum size = 0pt] (t) {};
\diagram*{
	(x1) -- (v) -- (x2),
	(v) -- [photon] (G) -- [quarter left] (V4d) -- [quarter left] (G),
	(V4u) -- [half left] (t) -- [half left] (V4u)
};
\vertex[right = 15pt of G, label = {above: {\scriptsize $\phi_j$}}] (Gr) {};
\end{feynman}
\end{tikzpicture}
\;\,.
\label{seq:K1_subtract}
\end{equation}
Contracting $\phi_j^{}$ legs like this disconnects the $\phi^4_j\Delta_j^{}$ subdiagram while the full diagram remains connected. 
On the other hand, these contributions were in fact already included in Eq.~\eqref{seq:K1_K}, because the upper parts of these diagrams (involving the $(\l\*-\*1)\,$th-layer connected four-point correlator) are simply NLO corrections to the $\phi_j^{(\lm1)}$ propagator. 
Quite generally, our definition of \cc blob in Eq.~\eqref{seq:ast_blob} conveniently avoids double-counting of such diagrams.

Adding up both contributions discussed above, we obtain the RG flow of the NLO two-point correlator:
\begin{equation}
\frac{1}{n_\lm1} \frac{}{}\K_1^{(\l)}\!(\x_1, \x_2) = \text{Eq.~\eqref{seq:K1_K}} + \text{Eq.~\eqref{seq:K1_V}} \,.
\end{equation}
which can be used to recursively determine $\K_1^{(\l)}$ from $\K_1^{(\lm1)}$, $V_4^{(\lm1)}$ and $\K_0^{(\lm1)}$.

\subsubsection{Connected six-point correlator}

We next demonstrate the derivation of RG flow of the connected six-point correlator at LO. 
Similarly to Eq.~\eqref{eq:V_4} for the connected four-point correlator, we have
\begin{equation}
\frac{1}{n_\lm1^2}\,V_6^{(\l)} \*(\x_1^{}, \x_2^{} ; \x_3^{}, \x_4^{} ; \x_5^{}, \x_6^{}) 
= \sum_{j_1,j_2,j_3} 
\begin{tikzpicture}[baseline=(b.base)]
\begin{feynman}
\vertex[blob] (b) {};
\vertex[left = 20pt of b, dot, minimum size = 0pt] (v12) {};
\vertex[left = 8pt of v12] (l) {};
\vertex[below = 16pt of l, dot, minimum size = 3pt, label = {left: {\footnotesize $\x_1^{}$}}] (x1) {};
\vertex[above = 16pt of l, dot, minimum size = 3pt, label = {left: {\footnotesize $\x_2^{}$}}] (x2) {};
\vertex[above = 20pt of b, dot, minimum size = 0pt] (v34) {};
\vertex[above = 8pt of v34] (u) {};
\vertex[left = 16pt of u, dot, minimum size = 3pt, label = {above: {\footnotesize $\x_3^{}$}}] (x3) {};
\vertex[right = 16pt of u, dot, minimum size = 3pt, label = {above: {\footnotesize $\x_4^{}$}}] (x4) {};
\vertex[right = 20pt of b, dot, minimum size = 0pt] (v56) {};
\vertex[right = 8 pt of v56, dot, minimum size = 0pt] (r) {};
\vertex[above = 16pt of r, dot, minimum size = 3pt, label = {right: {\footnotesize $\x_5^{}$}}] (x5) {};
\vertex[below = 16pt of r, dot, minimum size = 3pt, label = {right: {\footnotesize $\x_6^{}$}}] (x6) {};
\diagram*{
	(x1) -- (v12) -- (x2),
	(x3) -- (v34) -- (x4),
	(x5) -- (v56) -- (x6),
	(v12) -- [photon, edge label' = {\scriptsize\, $\Delta_{j_1}$}] (b) -- [photon, edge label' = {\scriptsize $\Delta_{j_2}$}] (v34),
	(v56) -- [photon, edge label = {\scriptsize \,$\Delta_{j_3}$}] (b)
};
\end{feynman}
\end{tikzpicture}
.
\label{seq:V_6}
\end{equation}

We need to consider three cases. 
First, if $j_1, j_2, j_3$ are all equal, $j_1=j_2=j_3\equiv j$, we can simply use free-theory propagators $\K_0^{(\lm1)}$ to connect the $\phi_j$ fields contained in $\Delta_j$ to obtain the LO result:
\begin{equation}
	\sum_j
\begin{tikzpicture}[baseline=(b.base)]
\begin{feynman}
\vertex[blob] (b) {\c};
\vertex[left = 20pt of b, dot, minimum size = 0pt] (v12) {};
\vertex[left = 8pt of v12] (l) {};
\vertex[below = 16pt of l, dot, minimum size = 3pt, label = {left: {\footnotesize $\x_1^{}$}}] (x1) {};
\vertex[above = 16pt of l, dot, minimum size = 3pt, label = {left: {\footnotesize $\x_2^{}$}}] (x2) {};
\vertex[above = 20pt of b, dot, minimum size = 0pt] (v34) {};
\vertex[above = 8pt of v34] (u) {};
\vertex[left = 16pt of u, dot, minimum size = 3pt, label = {above: {\footnotesize $\x_3^{}$}}] (x3) {};
\vertex[right = 16pt of u, dot, minimum size = 3pt, label = {above: {\footnotesize $\x_4^{}$}}] (x4) {};
\vertex[right = 20pt of b, dot, minimum size = 0pt] (v56) {};
\vertex[right = 8 pt of v56, dot, minimum size = 0pt] (r) {};
\vertex[above = 16pt of r, dot, minimum size = 3pt, label = {right: {\footnotesize $\x_5^{}$}}] (x5) {};
\vertex[below = 16pt of r, dot, minimum size = 3pt, label = {right: {\footnotesize $\x_6^{}$}}] (x6) {};
\diagram*{
	(x1) -- (v12) -- (x2),
	(x3) -- (v34) -- (x4),
	(x5) -- (v56) -- (x6),
	(v12) -- [photon, edge label' = {\scriptsize $\Delta_j$}] (b) -- [photon, edge label' = {\scriptsize $\Delta_j$}] (v34),
	(v56) -- [photon, edge label = {\scriptsize $\Delta_j$}] (b)
};
\end{feynman}
\end{tikzpicture}
=\,\frac{\bigl(C_W^{(\l)}\bigr)^{\*3}}{n_\lm1^2} \,
\Bigl\langle \Delta(\x_1^{}, \x_2^{}) \,\Delta (\x_3^{}, \x_4^{}) \,\Delta (\x_5^{}, \x_6^{}) \Bigr\rangle_{\K^{(\lm1)}_0} 
\*+\O\Bigl(\frac{1}{n^3}\Bigr) \,,
\label{seq:V6_1}
\end{equation}
where the sum over $j$ yields a factor of $n_\lm1$ which cancels one of the three factors of $\frac{1}{n_\lm1}$ from $\phi^2\Delta$ vertices and renders the result $\O\bigl(\frac{1}{n^2}\bigr)$. 

Second, if $j_1, j_2, j_3$ take two distinct values, we need to use a connected four-point correlator at the $(\l\*-\*1)\,$th layer to connect neurons with distinct indices (while still using free propagators to connect neurons with identical indices):
\begin{eqnarray}
&&\sum_{j_1, j_2}\;
\begin{tikzpicture}[baseline=(K0.base)]
\begin{feynman}
\vertex[blob] (K0) {\c};
\vertex[left = 20pt of K0, dot, minimum size = 0pt] (v12) {};
\vertex[left = 8pt of v12] (l) {};
\vertex[below = 16pt of l, dot, minimum size = 3pt, label = {left: {\footnotesize $\x_1^{}$}}] (x1) {};
\vertex[above = 16pt of l, dot, minimum size = 3pt, label = {left: {\footnotesize $\x_2^{}$}}] (x2) {};
\vertex[right = 20pt of K0, dot, minimum size = 0pt] (v56) {};
\vertex[right = 8pt of v56, dot, minimum size = 0pt] (r) {};
\vertex[above = 16pt of r, dot, minimum size = 3pt, label = {right: {\footnotesize $\x_5^{}$}}] (x5) {};
\vertex[below = 16pt of r, dot, minimum size = 3pt, label = {right: {\footnotesize $\x_6^{}$}}] (x6) {};
\vertex[above = 18pt of K0, blob, minimum size = 6pt] (V4) {};
\vertex[above = 3pt of V4, dot, minimum size = 0pt] (V4u) {};
\vertex[below = 3pt of V4, dot, minimum size = 0pt] (V4d) {};
\vertex[above = 18pt of V4, blob] (K0t) {\c};
\vertex[above = 20pt of K0t, dot, minimum size = 0pt] (v34) {};
\vertex[above = 8pt of v34] (u) {};
\vertex[left = 16pt of u, dot, minimum size = 3pt, label = {above: {\footnotesize $\x_3^{}$}}] (x3) {};
\vertex[right = 16pt of u, dot, minimum size = 3pt, label = {above: {\footnotesize $\x_4^{}$}}] (x4) {};
\diagram*{
	(x1) -- (v12) -- (x2),
	(x3) -- (v34) -- (x4),
	(x5) -- (v56) -- (x6),
	(v12) -- [photon, edge label' = {\scriptsize \;$\Delta_{j_1}$}] (K0) -- [quarter left] (V4d) -- [quarter left] (K0) -- [photon, edge label' = {\scriptsize \;$\Delta_{j_1}$}] (v56),
	(v34) -- [photon, edge label = {\scriptsize $\Delta_{j_2}$}] (K0t) -- [quarter left] (V4u) -- [quarter left] (K0t)
};
\vertex[right = 14pt of K0, label = {above: {\scriptsize $\phi_{j_1}$}}] (K0r) {};
\vertex[right = 14pt of K0t, label = {below: {\scriptsize $\phi_{j_2}$}}] (K0tr) {};
\end{feynman}
\end{tikzpicture}
+\;\text{perms.}\, =\, \frac{\bigl(C_W^{(\l)}\bigr)^{\*3}}{4\,n_\lm1 n_{\l\shortminus 2}} \int \*\prod_{\alpha=1}^4 d\y_\alpha^{} \; V_4^{(\lm1)}\* (\y_1^{}, \y_2^{}; \y_3^{}, \y_4^{}) \nonumber\\[10pt]
&&
\hspace{20pt}
\Biggl[ 
\biggl\langle \frac{\delta^2 \bigl( \Delta(\x_1^{}, \x_2^{}) \, \Delta(\x_5^{}, \x_6^{}) \bigr)}{\delta \phi(\y_1^{}) \,\delta \phi(\y_2^{}) } \biggr\rangle_{\!\!\K^{(\lm1)}_0}
\biggl\langle \frac{\delta^2 \Delta(\x_3^{}, \x_4^{})}{\delta \phi(\y_3^{}) \,\delta \phi(\y_4^{}) } \biggr\rangle_{\!\!\K^{(\lm1)}_0}
\!+\, \biggl\langle \frac{\delta^2 \bigl( \Delta(\x_1^{}, \x_2^{}) \, \Delta(\x_3^{}, \x_4^{}) \bigr)}{\delta \phi(\y_1^{}) \,\delta \phi(\y_2^{}) } \biggr\rangle_{\!\!\K^{(\lm1)}_0}
\biggl\langle \frac{\delta^2 \Delta(\x_5^{}, \x_6^{})}{\delta \phi(\y_3^{}) \,\delta \phi(\y_4^{}) } \biggr\rangle_{\!\!\K^{(\lm1)}_0}
\nonumber\\[10pt]
&&
\hspace{20pt}
\;\;+\, \biggl\langle \frac{\delta^2 \bigl( \Delta(\x_3^{}, \x_4^{}) \, \Delta(\x_5^{}, \x_6^{}) \bigr)}{\delta \phi(\y_1^{}) \,\delta \phi(\y_2^{}) } \biggr\rangle_{\!\!\K^{(\lm1)}_0}
\biggl\langle \frac{\delta^2 \Delta(\x_1^{}, \x_2^{})}{\delta \phi(\y_3^{}) \,\delta \phi(\y_4^{}) } \biggr\rangle_{\!\!\K^{(\lm1)}_0}
\Biggr]
+\O\Bigl(\frac{1}{n^3}\Bigr) \,,
\label{seq:V6_2}
\end{eqnarray}
The diagram is automatically connected as a result of our \cc blob definition Eq.~\eqref{seq:ast_blob}. 
To arrive at the expression in Eq.~\eqref{seq:V6_2}, we have used Eq.~\eqref{seq:ast_blob_result} with $(m,r)=(2, 1)$ and $(1,1)$ for the two \cc blobs, respectively, and the symmetry factor $\frac{1}{2^2} = \frac{1}{4}$ comes from exchanging $\y_1^{} \leftrightarrow \y_2^{}$ and $\y_3^{} \leftrightarrow \y_4^{}$ among the four $\phi_j^{}$ legs attached to $V_4^{(\lm1)}\* (\y_1^{}, \y_2^{}; \y_3^{}, \y_4^{})$. 
Compared to the first contribution in Eq.~\eqref{seq:V6_1}, here the $j$ sum yields an additional factor of $n_\lm1$ while the connected four-point correlator inserted carries a factor of $\frac{1}{n_{\l \shortminus 2}}$, so the end result is again $\O\bigl(\frac{1}{n^2}\bigr)$. 
Note that Eq.~\eqref{seq:V6_2} holds regardless of whether $j_1=j_2$ terms are included in the sum (the same is true for Eq.~\eqref{eq:4pt_2} in the main text). 

Finally, if $j_1, j_2, j_3$ are all distinct, we must use either a connected six-point correlator or two connected four-point correlators to connect the $(\l\*-\*1)\,$th-layer neurons. 
In the former case we have
\begin{eqnarray}
\sum_{j_1, j_2, j_3}
\begin{tikzpicture}[baseline=(b12.base)]
\begin{feynman}
\vertex[blob, minimum size = 6pt] (d) {};
\vertex[left = 3pt of d, dot, minimum size = 0pt] (w12) {};
\vertex[left = 20pt of d, blob] (b12) {\c};
\vertex[left = 20pt of b12, dot, minimum size = 0pt] (v12) {};
\vertex[left = 8pt of v12] (l) {};
\vertex[below = 16pt of l, dot, minimum size = 3pt, label = {left: {\footnotesize $\x_1^{}$}}] (x1) {};
\vertex[above = 16pt of l, dot, minimum size = 3pt, label = {left: {\footnotesize $\x_2^{}$}}] (x2) {};
\vertex[above = 3pt of d, dot, minimum size = 0pt] (w34) {};
\vertex[above = 20pt of d, blob] (b34) {\c};
\vertex[above = 20pt of b34, dot, minimum size = 0pt] (v34) {};
\vertex[above = 8pt of v34] (u) {};
\vertex[left = 16pt of u, dot, minimum size = 3pt, label = {above: {\footnotesize $\x_3^{}$}}] (x3) {};
\vertex[right = 16pt of u, dot, minimum size = 3pt, label = {above: {\footnotesize $\x_4^{}$}}] (x4) {};
\vertex[right = 3pt of d, dot, minimum size = 0pt] (w56) {};
\vertex[right = 20pt of d, blob] (b56) {\c};
\vertex[right = 20pt of b56, dot, minimum size = 0pt] (v56) {};
\vertex[right = 8 pt of v56, dot, minimum size = 0pt] (r) {};
\vertex[above = 16pt of r, dot, minimum size = 3pt, label = {right: {\footnotesize $\x_5^{}$}}] (x5) {};
\vertex[below = 16pt of r, dot, minimum size = 3pt, label = {right: {\footnotesize $\x_6^{}$}}] (x6) {};
\diagram*{
	(x1) -- (v12) -- (x2),
	(x3) -- (v34) -- (x4),
	(x5) -- (v56) -- (x6),
	(v12) -- [photon, edge label' = {\scriptsize \;$\Delta_{j_1}$}] (b12) -- [quarter left] (w12) -- [quarter left] (b12),
	(v34) -- [photon, edge label = {\scriptsize $\Delta_{j_2}$}] (b34) -- [quarter left] (w34) -- [quarter left] (b34),
	(v56) -- [photon, edge label = {\scriptsize \;$\Delta_{j_3}$}] (b56) -- [quarter left] (w56) -- [quarter left] (b56)
};
\vertex[right = 14pt of b12, label = {below: {\scriptsize $\phi_{j_1}$}}] (b12r) {};
\vertex[below = 8pt of b34, label = {right: {\scriptsize $\phi_{j_2}$}}] (b12r) {};
\vertex[left = 10pt of b56, label = {below: {\scriptsize $\phi_{j_3}$}}] (b12r) {};
\end{feynman}
\end{tikzpicture}
&=&\, \frac{\bigl(C_W^{(\l)}\bigr)^{\*3}}{8\,n_{\l\shortminus 2}^2}
\prod_{\alpha=1}^6 \int d\y_\alpha^{} \; V_6^{(\lm1)}\* (\y_1^{}, \y_2^{}; \y_3^{}, \y_4^{}; \y_5^{}, \y_6^{})
\nonumber\\[10pt]
&&\hspace{-60pt}
\biggl\langle \frac{\delta^2 \Delta(\x_1^{}, \x_2^{})}{\delta \phi(\y_1^{}) \, \delta \phi(\y_2^{})} \biggr\rangle_{\!\!\K^{(\lm1)}_0}
\biggl\langle \frac{\delta^2 \Delta(\x_3^{}, \x_4^{})}{\delta \phi(\y_3^{}) \, \delta \phi(\y_4^{})} \biggr\rangle_{\!\!\K^{(\lm1)}_0}
\biggl\langle \frac{\delta^2 \Delta(\x_5^{}, \x_6^{})}{\delta \phi(\y_5^{}) \, \delta \phi(\y_6^{})} \biggr\rangle_{\!\!\K^{(\lm1)}_0} 
\!+\O\Bigl(\frac{1}{n^3}\Bigr) \,,
\label{seq:V6_3a}
\end{eqnarray}
while in the latter case we have
\begin{eqnarray}
\sum_{j_1, j_2, j_3} 
\begin{tikzpicture}[baseline=(b12.base)]
\begin{feynman}
\vertex[blob] (b) {\c};
\vertex[left = 18pt of b, blob, minimum size = 6pt] (V4a) {};
\vertex[left = 3pt of V4a, dot, minimum size = 0pt] (w12) {};
\vertex[right = 3pt of V4a, dot, minimum size = 0pt] (u12) {};
\vertex[left = 18pt of V4a, blob] (b12) {\c};
\vertex[left = 20pt of b12, dot, minimum size = 0pt] (v12) {};
\vertex[left = 8pt of v12] (l) {};
\vertex[below = 16pt of l, dot, minimum size = 3pt, label = {left: {\footnotesize $\x_1^{}$}}] (x1) {};
\vertex[above = 16pt of l, dot, minimum size = 3pt, label = {left: {\footnotesize $\x_2^{}$}}] (x2) {};
\vertex[above = 20pt of b, dot, minimum size = 0pt] (v34) {};
\vertex[above = 8pt of v34] (u) {};
\vertex[left = 16pt of u, dot, minimum size = 3pt, label = {above: {\footnotesize $\x_3^{}$}}] (x3) {};
\vertex[right = 16pt of u, dot, minimum size = 3pt, label = {above: {\footnotesize $\x_4^{}$}}] (x4) {};
\vertex[right = 18pt of b, blob, minimum size = 6pt] (V4b) {};
\vertex[right = 3pt of V4b, dot, minimum size = 0pt] (w56) {};
\vertex[left = 3pt of V4b, dot, minimum size = 0pt] (u56) {};
\vertex[right = 18pt of V4b, blob] (b56) {\c};
\vertex[right = 20pt of b56, dot, minimum size = 0pt] (v56) {};
\vertex[right = 8 pt of v56, dot, minimum size = 0pt] (r) {};
\vertex[above = 16pt of r, dot, minimum size = 3pt, label = {right: {\footnotesize $\x_5^{}$}}] (x5) {};
\vertex[below = 16pt of r, dot, minimum size = 3pt, label = {right: {\footnotesize $\x_6^{}$}}] (x6) {};
\diagram*{
	(x1) -- (v12) -- (x2),
	(x3) -- (v34) -- (x4),
	(x5) -- (v56) -- (x6),
	(v12) -- [photon, edge label' = {\scriptsize \;$\Delta_{j_1}$}] (b12) -- [quarter left] (w12) -- [quarter left] (b12),
	(v34) -- [photon, edge label = {\scriptsize $\Delta_{j_2}$}] (b),
	(v56) -- [photon, edge label = {\scriptsize \;$\Delta_{j_3}$}] (b56) -- [quarter left] (w56) -- [quarter left] (b56),
	(b) -- [quarter left] (u12) -- [quarter left] (b) -- [quarter left] (u56) -- [quarter left] (b)
};
\vertex[right = 13pt of b12, label = {below: {\scriptsize $\phi_{j_1}$}}] (b12r) {};
\vertex[left = 8pt of b, label = {below: {\scriptsize $\phi_{j_2}$}}] (bl) {};
\vertex[right = 13pt of b, label = {below: {\scriptsize $\phi_{j_2}$}}] (br) {};
\vertex[left = 8pt of b56, label = {below: {\scriptsize $\phi_{j_3}$}}] (b56l) {};
\end{feynman}
\end{tikzpicture}
+\;\text{perms.}\, &=&\, \frac{\bigl(C_W^{(\l)}\bigr)^{\*3}}{16\, n_{\l\shortminus 2}^2} \int \*\prod_{\alpha=1}^8 d\y_\alpha^{} \; V_4^{(\lm1)}\* (\y_1^{}, \y_2^{}; \y_3^{}, \y_4^{}) \,V_4^{(\lm1)}\* (\y_5^{}, \y_6^{}; \y_7^{}, \y_8^{}) 
\nonumber\\[10pt]
&&\hspace{-180pt}
\Biggl[
\biggl\langle \frac{\delta^4 \Delta(\x_3^{}, \x_4^{})}{\delta \phi(\y_3^{}) \, \delta \phi(\y_4^{}) \, \delta \phi(\y_7^{}) \, \delta \phi(\y_8^{})} \biggr\rangle_{\!\!\K^{(\lm1)}_0}
\biggl\langle \frac{\delta^2 \Delta(\x_1^{}, \x_2^{})}{\delta \phi(\y_1^{}) \, \delta \phi(\y_2^{})} \biggr\rangle_{\!\!\K^{(\lm1)}_0}
\biggl\langle \frac{\delta^2 \Delta(\x_5^{}, \x_6^{})}{\delta \phi(\y_5^{}) \, \delta \phi(\y_6^{})} \biggr\rangle_{\!\!\K^{(\lm1)}_0} 
\nonumber\\[10pt]
&&\hspace{-184pt}
+\,
\biggl\langle \frac{\delta^4 \Delta(\x_1^{}, \x_2^{})}{\delta \phi(\y_3^{}) \, \delta \phi(\y_4^{}) \, \delta \phi(\y_7^{}) \, \delta \phi(\y_8^{})} \biggr\rangle_{\!\!\K^{(\lm1)}_0}
\biggl\langle \frac{\delta^2 \Delta(\x_3^{}, \x_4^{})}{\delta \phi(\y_1^{}) \, \delta \phi(\y_2^{})} \biggr\rangle_{\!\!\K^{(\lm1)}_0}
\biggl\langle \frac{\delta^2 \Delta(\x_5^{}, \x_6^{})}{\delta \phi(\y_5^{}) \, \delta \phi(\y_6^{})} \biggr\rangle_{\!\!\K^{(\lm1)}_0} 
\nonumber\\[10pt]
&&\hspace{-184pt}
+\,
\biggl\langle \frac{\delta^4 \Delta(\x_5^{}, \x_6^{})}{\delta \phi(\y_3^{}) \, \delta \phi(\y_4^{}) \, \delta \phi(\y_7^{}) \, \delta \phi(\y_8^{})} \biggr\rangle_{\!\!\K^{(\lm1)}_0}
\biggl\langle \frac{\delta^2 \Delta(\x_1^{}, \x_2^{})}{\delta \phi(\y_1^{}) \, \delta \phi(\y_2^{})} \biggr\rangle_{\!\!\K^{(\lm1)}_0}
\biggl\langle \frac{\delta^2 \Delta(\x_3^{}, \x_4^{})}{\delta \phi(\y_5^{}) \, \delta \phi(\y_6^{})} \biggr\rangle_{\!\!\K^{(\lm1)}_0} 
\Biggr]
+\O\Bigl(\frac{1}{n^3}\Bigr) \,.
\label{seq:V6_3b}
\end{eqnarray}
Symmetry factors and $\frac{1}{n}$ counting should be obvious at this point. 
Note that with the use of \cc blobs, Eq.~\eqref{seq:V6_3b} excludes both disconnected diagrams (which we should obviously exclude), and also the following connected diagram obtained by contracting two of the four $\phi_{j_2}$ legs (hence disconnecting the $\phi_{j_2}^4\Delta_{j_2}^{}$ blob):
\begin{equation}
\sum_{j_1, j_2, j_3}
\begin{tikzpicture}[baseline=(b12.base)]
\begin{feynman}
\vertex[] (c) {};
\vertex[left = 10pt of c, blob, minimum size = 6pt] (V4a) {};
\vertex[left = 3pt of V4a, dot, minimum size = 0pt] (w12) {};
\vertex[right = 3pt of V4a, dot, minimum size = 0pt] (u12) {};
\vertex[left = 18pt of V4a, blob] (b12) {\c};
\vertex[left = 20pt of b12, dot, minimum size = 0pt] (v12) {};
\vertex[left = 8pt of v12] (l) {};
\vertex[below = 16pt of l, dot, minimum size = 3pt, label = {left: {\footnotesize $\x_1^{}$}}] (x1) {};
\vertex[above = 16pt of l, dot, minimum size = 3pt, label = {left: {\footnotesize $\x_1^{}$}}] (x2) {};
\vertex[above = 12pt of c, dot, minimum size = 0pt] (b34d) {};
\vertex[above = 16pt of c, blob] (b34) {\c};
\vertex[above = 20pt of b34, dot, minimum size = 0pt] (v34) {};
\vertex[above = 8pt of v34] (u) {};
\vertex[left = 16pt of u, dot, minimum size = 3pt, label = {above: {\footnotesize $\x_3^{}$}}] (x3) {};
\vertex[right = 16pt of u, dot, minimum size = 3pt, label = {above: {\footnotesize $\x_4^{}$}}] (x4) {};
\vertex[right = 10pt of c, blob, minimum size = 6pt] (V4b) {};
\vertex[right = 3pt of V4b, dot, minimum size = 0pt] (w56) {};
\vertex[left = 3pt of V4b, dot, minimum size = 0pt] (u56) {};
\vertex[right = 18pt of V4b, blob] (b56) {\c};
\vertex[right = 20pt of b56, dot, minimum size = 0pt] (v56) {};
\vertex[right = 8 pt of v56, dot, minimum size = 0pt] (r) {};
\vertex[above = 16pt of r, dot, minimum size = 3pt, label = {right: {\footnotesize $\x_5^{}$}}] (x5) {};
\vertex[below = 16pt of r, dot, minimum size = 3pt, label = {right: {\footnotesize $\x_6^{}$}}] (x6) {};
\diagram*{
	(x1) -- (v12) -- (x2),
	(x3) -- (v34) -- (x4),
	(x5) -- (v56) -- (x6),
	(v12) -- [photon, edge label' = {\scriptsize \;$\Delta_{j_1}$}] (b12) -- [quarter left] (w12) -- [quarter left] (b12),
	(v34) -- [photon, edge label = {\scriptsize $\Delta_{j_2}$}] (b34),
	(v56) -- [photon, edge label = {\scriptsize \;$\Delta_{j_3}$}] (b56) -- [quarter left] (w56) -- [quarter left] (b56),
	(u12) -- [quarter left] (b34d) -- [quarter left] (u56) -- [quarter left] (u12)
};
\vertex[above = 0pt of b34, blob] (bb34) {\c};
\vertex[right = 13pt of b12, label = {below: {\scriptsize $\phi_{j_1}$}}] (b12r) {};
\vertex[below = 15pt of b34, label = {below: {\scriptsize \;\;$\phi_{j_2}$}}] (b34dd) {};
\vertex[left = 8pt of b56, label = {below: {\scriptsize $\phi_{j_3}$}}] (b56l) {};
\end{feynman}
\end{tikzpicture}
+\;\text{perms.}
\label{seq:V6_subtract}
\end{equation}
On the other hand, this contribution was already accounted for in Eq.~\eqref{seq:V6_3a}. 
So just as in the calculation of $\K_1^{(\l)}$ in the previous subsection (see discussion around Eq.~\eqref{seq:K1_subtract}), the use of \cc blobs neatly avoids double-counting.

To summarize, the RG flow of the connected six-point correlator at LO is obtained from just four sets of Feynman diagrams discussed above:
\begin{equation}
\frac{1}{n_\lm1^2} V_6^{(\l)} \*(\x_1^{}, \x_2^{} ; \x_3^{}, \x_4^{} ; \x_5^{}, \x_6^{}) = 
\,\text{Eq.~\eqref{seq:V6_1}} + \, \text{Eq.~\eqref{seq:V6_2}} + \, \text{Eq.~\eqref{seq:V6_3a}} + \,\text{Eq.~\eqref{seq:V6_3b}}\,.
\end{equation}
This allows us to recursively determine $V_6^{(\l)}$ from $V_6^{(\lm1)}$, $V_4^{(\lm1)}$ and $\K_0^{(\lm1)}$.

\subsubsection{Connected eight-point correlator}

Finally, we present a summary of the connected eight-point correlator calculation in Table.~\ref{tab:8pt}. 
As in the case of $V_6$ above, we organize the calculation by the number of distinct $j$ indices that are summed over. 
For simplicity we drop the summation over $j_1, j_2, \dots$ as well as all labels in the diagrams; it should be clear at this point that at LO all $\Delta$'s and $\phi$'s attached to the same \cc blob share the same neuron index, whereas the smaller $V_{2k}^{(\lm1)}$ blobs can connect different neuron indices. 
A new feature worth noting is that starting at the eight-point level we need to subtract off disconnected diagrams that are not excluded in the definition of \cc blob in Eq.~\eqref{seq:ast_blob}. 
Meanwhile, the use of \cc blobs still saves us from double-counting in the same way as discussed around Eqs.~\eqref{seq:K1_subtract} and \eqref{seq:V6_subtract}. 
The $\frac{1}{n}$ counting is transparent from the connected diagram in each row of the table: each $\phi^2\Delta$ vertex (of which there are four) comes with a factor of $\frac{1}{n_\lm1}$, each \cc blob indicates a $j$ sum and hence a factor of $n_\lm1$, while a smaller blob representing a $2k$-point connected correlator of $(\l\*-\*1)$\,th-layer neurons comes with a factor of $\frac{1}{n_{\l\shortminus 2}^{k-1}}$. 
The end result is $\O\bigl(\frac{1}{n^3}\bigr)$ for all diagrams.

While it is straightforward to read off the expression of each diagram for general (nondegenerate) inputs $\x_1^{}, \dots, \x_8^{}$, the results are rather lengthy and not particularly illuminating. 
So for compactness we show final results in the degenerate input limit in the last column of Table~\ref{tab:8pt} and drop the input arguments. 
In each expression we write the symmetry factor in front; the remaining numerical factors come from combinatorics. 
The RG flow of connected eight-point correlator is given by the sum of all diagrams in Table~\ref{tab:8pt}, which can be used to recursively determine $V_8^{(\l)}$ from $V_8^{(\lm1)}$, $V_6^{(\lm1)}$, $V_4^{(\lm1)}$ and $\K_0^{(\lm1)}$.

\def\eightptA{
	\begin{tikzpicture}[baseline=(b.base)]
	\begin{feynman}
	\vertex[blob] (b) {\c};
	\vertex[left = 14pt of b] (l) {};
	\vertex[below = 14pt of l, dot, minimum size = 0pt] (v12) {};
	\vertex[below = 12pt of v12, dot, minimum size = 3pt] (x1) {};
	\vertex[left = 12pt of v12, dot, minimum size = 3pt] (x2) {};
	\vertex[above = 14pt of l, dot, minimum size = 0pt] (v34) {};
	\vertex[left = 12pt of v34, dot, minimum size = 3pt] (x3) {};
	\vertex[above = 12pt of v34, dot, minimum size = 3pt] (x4) {};
	\vertex[right = 14pt of b] (r) {};
	\vertex[above = 14pt of r, dot, minimum size = 0pt] (v56) {};
	\vertex[above = 12pt of v56, dot, minimum size = 3pt] (x5) {};
	\vertex[right = 12pt of v56, dot, minimum size = 3pt] (x6) {};
	\vertex[below = 14pt of r, dot, minimum size = 0pt] (v78) {};
	\vertex[right = 12pt of v78, dot, minimum size = 3pt] (x7) {};
	\vertex[below = 12pt of v78, dot, minimum size = 3pt] (x8) {};
	\diagram*{
		(x1) -- (v12) -- (x2),
		(x3) -- (v34) -- (x4),
		(x5) -- (v56) -- (x6),
		(x7) -- (v78) -- (x8),
		(v12) -- [photon] (b) -- [photon] (v34),
		(v56) -- [photon] (b) -- [photon] (v78)
	};
	\end{feynman}
	\end{tikzpicture}
}

\def\eightptAsub{
	\begin{tikzpicture}[baseline=(c.base)]
	\begin{feynman}
	\vertex[] (c) {};
	\vertex[above = 14pt of c, blob] (b1) {\c};
	\vertex[left = 20pt of b1, dot, minimum size = 0pt] (v12) {};
	\vertex[left = 8pt of v12] (v12l) {};
	\vertex[below = 8pt of v12l, dot, minimum size = 3pt] (x1) {};
	\vertex[above = 8pt of v12l, dot, minimum size = 3pt] (x2) {};
	\vertex[right = 20pt of b1, dot, minimum size = 0pt] (v34) {};
	\vertex[right = 8 pt of v34, dot, minimum size = 0pt] (v34r) {};
	\vertex[above = 8pt of v34r, dot, minimum size = 3pt] (x3) {};
	\vertex[below = 8pt of v34r, dot, minimum size = 3pt] (x4) {};
	\vertex[below = 14pt of c, blob] (b2) {\c};
	\vertex[left = 20pt of b2, dot, minimum size = 0pt] (v56) {};
	\vertex[left = 8pt of v56] (v56l) {};
	\vertex[below = 8pt of v56l, dot, minimum size = 3pt] (x5) {};
	\vertex[above = 8pt of v56l, dot, minimum size = 3pt] (x6) {};
	\vertex[right = 20pt of b2, dot, minimum size = 0pt] (v78) {};
	\vertex[right = 8 pt of v78, dot, minimum size = 0pt] (v78r) {};
	\vertex[above = 8pt of v78r, dot, minimum size = 3pt] (x7) {};
	\vertex[below = 8pt of v78r, dot, minimum size = 3pt] (x8) {};
	\diagram*{
		(x1) -- (v12) -- (x2),
		(x3) -- (v34) -- (x4),
		(x5) -- (v56) -- (x6),
		(x7) -- (v78) -- (x8),
		(v12) -- [photon] (b1) -- [photon] (v34),
		(v56) -- [photon] (b2) -- [photon] (v78)
	};
	\end{feynman}
	\end{tikzpicture}
}

\def\eightptBa{
\begin{tikzpicture}[baseline=(b12.base)]
\begin{feynman}
\vertex[blob, minimum size = 6pt] (V4) {};
\vertex[left = 3pt of V4, dot, minimum size = 0pt] (w12) {};
\vertex[left = 20pt of V4, blob] (b12) {\c};
\vertex[left = 20pt of b12, dot, minimum size = 0pt] (v12) {};
\vertex[left = 8pt of v12] (l) {};
\vertex[below = 8pt of l, dot, minimum size = 3pt] (x1) {};
\vertex[above = 8pt of l, dot, minimum size = 3pt] (x2) {};
\vertex[above = 20pt of b12, dot, minimum size = 0pt] (v34) {};
\vertex[above = 8pt of v34] (u) {};
\vertex[left = 8pt of u, dot, minimum size = 3pt] (x3) {};
\vertex[right = 8pt of u, dot, minimum size = 3pt] (x4) {};
\vertex[below = 20pt of b12, dot, minimum size = 0pt] (v78) {};
\vertex[below = 8pt of v78] (d) {};
\vertex[right = 8pt of d, dot, minimum size = 3pt] (x7) {};
\vertex[left = 8pt of d, dot, minimum size = 3pt] (x8) {};
\vertex[right = 3pt of V4, dot, minimum size = 0pt] (w56) {};
\vertex[right = 20pt of V4, blob] (b56) {\c};
\vertex[right = 20pt of b56, dot, minimum size = 0pt] (v56) {};
\vertex[right = 8 pt of v56, dot, minimum size = 0pt] (r) {};
\vertex[above = 8pt of r, dot, minimum size = 3pt] (x5) {};
\vertex[below = 8pt of r, dot, minimum size = 3pt] (x6) {};
\diagram*{
	(x1) -- (v12) -- (x2),
	(x3) -- (v34) -- (x4),
	(x5) -- (v56) -- (x6),
	(x7) -- (v78) -- (x8),
	(v12) -- [photon] (b12) -- [quarter left] (w12) -- [quarter left] (b12),
	(v56) -- [photon] (b56) -- [quarter left] (w56) -- [quarter left] (b56),
	(v34) -- [photon] (b12) -- [photon] (v78)
};
\end{feynman}
\end{tikzpicture}
}

\def\eightptBasub{
\begin{tikzpicture}[baseline=(c.base)]
\begin{feynman}
\vertex[] (c) {};
\vertex[above = 14pt of c, blob, minimum size = 6pt] (V4) {};
\vertex[left = 3pt of V4, dot, minimum size = 0pt] (w12) {};
\vertex[left = 20pt of V4, blob] (b12) {\c};
\vertex[left = 20pt of b12, dot, minimum size = 0pt] (v12) {};
\vertex[left = 8pt of v12] (l) {};
\vertex[below = 8pt of l, dot, minimum size = 3pt] (x1) {};
\vertex[above = 8pt of l, dot, minimum size = 3pt] (x2) {};
\vertex[right = 3pt of V4, dot, minimum size = 0pt] (w56) {};
\vertex[right = 20pt of V4, blob] (b56) {\c};
\vertex[right = 20pt of b56, dot, minimum size = 0pt] (v56) {};
\vertex[right = 8 pt of v56, dot, minimum size = 0pt] (r) {};
\vertex[above = 8pt of r, dot, minimum size = 3pt] (x5) {};
\vertex[below = 8pt of r, dot, minimum size = 3pt] (x6) {};
\vertex[below = 14pt of c, blob] (b) {\c};
\vertex[left = 20pt of b, dot, minimum size = 0pt] (v78) {};
\vertex[left = 8pt of v78] (v78l) {};
\vertex[below = 8pt of v78l, dot, minimum size = 3pt] (x7) {};
\vertex[above = 8pt of v78l, dot, minimum size = 3pt] (x8) {};
\vertex[right = 20pt of b, dot, minimum size = 0pt] (v34) {};
\vertex[right = 8 pt of v34, dot, minimum size = 0pt] (v34r) {};
\vertex[above = 8pt of v34r, dot, minimum size = 3pt] (x3) {};
\vertex[below = 8pt of v34r, dot, minimum size = 3pt] (x4) {};
\diagram*{
	(x1) -- (v12) -- (x2),
	(x3) -- (v34) -- (x4),
	(x5) -- (v56) -- (x6),
	(x7) -- (v78) -- (x8),
	(v12) -- [photon] (b12) -- [quarter left] (w12) -- [quarter left] (b12),
	(v56) -- [photon] (b56) -- [quarter left] (w56) -- [quarter left] (b56),
	(v34) -- [photon] (b) -- [photon] (v78)
};
\end{feynman}
\end{tikzpicture}
}

\def\eightptBb{
	\begin{tikzpicture}[baseline=(base.base)]
	\begin{feynman}
	\vertex[blob, minimum size = 6pt] (V4) {};
	\vertex[below = 2pt of V4] (base) {};
	\vertex[left = 3pt of V4, dot, minimum size = 0pt] (w12) {};
	\vertex[left = 20pt of V4, blob] (b12) {\c};
	\vertex[left = 14pt of b12] (l) {};
	\vertex[below = 14pt of l, dot, minimum size = 0pt] (v12) {};
	\vertex[left = 12pt of v12, dot, minimum size = 3pt] (x1) {};
	\vertex[below = 12pt of v12, dot, minimum size = 3pt] (x2) {};
	\vertex[above = 14pt of l, dot, minimum size = 0pt] (v34) {};
	\vertex[left = 12pt of v34, dot, minimum size = 3pt] (x3) {};
	\vertex[above = 12pt of v34, dot, minimum size = 3pt] (x4) {};
	\vertex[right = 3pt of V4, dot, minimum size = 0pt] (w56) {};
	\vertex[right = 20pt of V4, blob] (b56) {\c};
	\vertex[right = 14pt of  b56] (r) {};
	\vertex[above = 14pt of r, dot, minimum size = 0pt] (v56) {};
	\vertex[above = 12pt of v56, dot, minimum size = 3pt] (x5) {};
	\vertex[right = 12pt of v56, dot, minimum size = 3pt] (x6) {};
	\vertex[below = 14pt of r, dot, minimum size = 0pt] (v78) {};
	\vertex[right = 12pt of v78, dot, minimum size = 3pt] (x7) {};
	\vertex[below = 12pt of v78, dot, minimum size = 3pt] (x8) {};
	\diagram*{
		(x1) -- (v12) -- (x2),
		(x3) -- (v34) -- (x4),
		(x5) -- (v56) -- (x6),
		(x7) -- (v78) -- (x8),
		(v12) -- [photon] (b12) -- [quarter left] (w12) -- [quarter left] (b12),
		(v56) -- [photon] (b56) -- [quarter left] (w56) -- [quarter left] (b56),
		(v34) -- [photon] (b12),
		(b56) -- [photon] (v78)
	};
	\end{feynman}
	\end{tikzpicture}
}

\def\eightptCa{
	\begin{tikzpicture}[baseline=(base.base)]
	\begin{feynman}
	\vertex[blob, minimum size = 6pt] (d) {};
	\vertex[above = 14pt of d] (base) {};
	\vertex[left = 3pt of d, dot, minimum size = 0pt] (w12) {};
	\vertex[left = 20pt of d, blob] (b12) {\c};
	\vertex[left = 20pt of b12, dot, minimum size = 0pt] (v12) {};
	\vertex[left = 8pt of v12] (l) {};
	\vertex[below = 8pt of l, dot, minimum size = 3pt] (x1) {};
	\vertex[above = 8pt of l, dot, minimum size = 3pt] (x2) {};
	\vertex[above = 3pt of d, dot, minimum size = 0pt] (w34) {};
	\vertex[above = 20pt of d, blob] (b34) {\c};
	\vertex[above = 14pt of b34] (t) {};
	\vertex[left = 14pt of t, dot, minimum size = 0pt] (v34) {};
	\vertex[left = 12pt of v34, dot, minimum size = 3pt] (x3) {};
	\vertex[above = 12pt of v34, dot, minimum size = 3pt] (x4) {};
	\vertex[right = 14pt of t, dot, minimum size = 0pt] (v78) {};
	\vertex[right = 12pt of v78, dot, minimum size = 3pt] (x7) {};
	\vertex[above = 12pt of v78, dot, minimum size = 3pt] (x8) {};
	\vertex[right = 3pt of d, dot, minimum size = 0pt] (w56) {};
	\vertex[right = 20pt of d, blob] (b56) {\c};
	\vertex[right = 20pt of b56, dot, minimum size = 0pt] (v56) {};
	\vertex[right = 8 pt of v56, dot, minimum size = 0pt] (r) {};
	\vertex[above = 8pt of r, dot, minimum size = 3pt] (x5) {};
	\vertex[below = 8pt of r, dot, minimum size = 3pt] (x6) {};
	\diagram*{
		(x1) -- (v12) -- (x2),
		(x3) -- (v34) -- (x4),
		(x5) -- (v56) -- (x6),
		(x7) -- (v78) -- (x8),
		(v12) -- [photon] (b12) -- [quarter left] (w12) -- [quarter left] (b12),
		(v34) -- [photon] (b34) -- [quarter left] (w34) -- [quarter left] (b34) -- [photon] (v78),
		(v56) -- [photon] (b56) -- [quarter left] (w56) -- [quarter left] (b56)
	};
	\end{feynman}
	\end{tikzpicture}
}

\def\eightptCb{
\begin{tikzpicture}[baseline=(bc.base)]
\begin{feynman}
\vertex[blob] (bc) {\c};
\vertex[left = 20pt of bc, blob, minimum size = 6pt] (V4l) {};
\vertex[left = 3pt of V4l, dot, minimum size = 0pt] (V4ll) {};
\vertex[right = 3pt of V4l, dot, minimum size = 0pt] (V4lr) {};
\vertex[left = 20pt of V4l, blob] (bl) {\c};
\vertex[left = 20pt of bl, dot, minimum size = 0pt] (v12) {};
\vertex[left = 8pt of v12] (x12) {};
\vertex[below = 8pt of x12, dot, minimum size = 3pt] (x1) {};
\vertex[above = 8pt of x12, dot, minimum size = 3pt] (x2) {};
\vertex[right = 20pt of bc, blob, minimum size = 6pt] (V4r) {};
\vertex[left = 3pt of V4r, dot, minimum size = 0pt] (V4rl) {};
\vertex[right = 3pt of V4r, dot, minimum size = 0pt] (V4rr) {};
\vertex[right = 20pt of V4r, blob] (br) {\c};
\vertex[right = 20pt of br, dot, minimum size = 0pt] (v34) {};
\vertex[right = 8pt of v34] (x34) {};
\vertex[below = 8pt of x34, dot, minimum size = 3pt] (x3) {};
\vertex[above = 8pt of x34, dot, minimum size = 3pt] (x4) {};
\vertex[above = 20pt of bc, dot, minimum size = 0pt] (v56) {};
\vertex[above = 8pt of v56] (x56) {};
\vertex[left = 8pt of x56, dot, minimum size = 3pt] (x5) {};
\vertex[right = 8pt of x56, dot, minimum size = 3pt] (x6) {};
\vertex[below = 20pt of bc, dot, minimum size = 0pt] (v78) {};
\vertex[below = 8pt of v78] (x78) {};
\vertex[left = 8pt of x78, dot, minimum size = 3pt] (x7) {};
\vertex[right = 8pt of x78, dot, minimum size = 3pt] (x8) {};
\diagram*{
	(x1) -- (v12) -- (x2),
	(x3) -- (v34) -- (x4),
	(x5) -- (v56) -- (x6),
	(x7) -- (v78) -- (x8),
	(v12) -- [photon] (bl) -- [quarter left] (V4ll) -- [quarter left] (bl),
	(v34) -- [photon] (br) -- [quarter left] (V4rr) -- [quarter left] (br),
	(v56) -- [photon] (bc) -- [quarter left] (V4lr) -- [quarter left] (bc) -- [quarter left] (V4rl) -- [quarter left] (bc) -- [photon] (v78)
};
\end{feynman}
\end{tikzpicture}
}

\def\eightptCbsub{
\begin{tikzpicture}[baseline=(c.base)]
\begin{feynman}
\vertex[] (c) {};
\vertex[above = 14pt of c, blob, minimum size = 6pt] (V4t) {};
\vertex[left = 3pt of V4t, dot, minimum size = 0pt] (V4tl) {};
\vertex[right = 3pt of V4t, dot, minimum size = 0pt] (V4tr) {};
\vertex[left = 20pt of V4t, blob] (btl) {\c};
\vertex[left = 20pt of btl, dot, minimum size = 0pt] (v12) {};
\vertex[left = 8pt of v12] (x12) {};
\vertex[below = 8pt of x12, dot, minimum size = 3pt] (x1) {};
\vertex[above = 8pt of x12, dot, minimum size = 3pt] (x2) {};
\vertex[right = 20pt of V4t, blob] (btr) {\c};
\vertex[right = 20pt of btr, dot, minimum size = 0pt] (v34) {};
\vertex[right = 8pt of v34] (x34) {};
\vertex[below = 8pt of x34, dot, minimum size = 3pt] (x3) {};
\vertex[above = 8pt of x34, dot, minimum size = 3pt] (x4) {};
\vertex[below = 14pt of c, blob, minimum size = 6pt] (V4b) {};
\vertex[left = 3pt of V4b, dot, minimum size = 0pt] (V4bl) {};
\vertex[right = 3pt of V4b, dot, minimum size = 0pt] (V4br) {};
\vertex[left = 20pt of V4b, blob] (bbl) {\c};
\vertex[left = 20pt of bbl, dot, minimum size = 0pt] (v56) {};
\vertex[left = 8pt of v56] (x56) {};
\vertex[below = 8pt of x56, dot, minimum size = 3pt] (x5) {};
\vertex[above = 8pt of x56, dot, minimum size = 3pt] (x6) {};
\vertex[right = 20pt of V4b, blob] (bbr) {\c};
\vertex[right = 20pt of bbr, dot, minimum size = 0pt] (v78) {};
\vertex[right = 8pt of v78] (x78) {};
\vertex[below = 8pt of x78, dot, minimum size = 3pt] (x7) {};
\vertex[above = 8pt of x78, dot, minimum size = 3pt] (x8) {};
\diagram*{
	(x1) -- (v12) -- (x2),
	(x3) -- (v34) -- (x4),
	(x5) -- (v56) -- (x6),
	(x7) -- (v78) -- (x8),
	(v12) -- [photon] (btl) -- [quarter left] (V4tl) -- [quarter left] (btl),
	(v34) -- [photon] (btr) -- [quarter left] (V4tr) -- [quarter left] (btr),
	(v56) -- [photon] (bbl) -- [quarter left] (V4bl) -- [quarter left] (bbl),
	(v78) -- [photon] (bbr) -- [quarter left] (V4br) -- [quarter left] (bbr)
};
\end{feynman}
\end{tikzpicture}
}

\def\eightptCc{
	\begin{tikzpicture}[baseline=(bc.base)]
	\begin{feynman}
	\vertex[blob] (bc) {\c};
	\vertex[left = 20pt of bc, blob, minimum size = 6pt] (V4l) {};
	\vertex[left = 3pt of V4l, dot, minimum size = 0pt] (V4ll) {};
	\vertex[right = 3pt of V4l, dot, minimum size = 0pt] (V4lr) {};
	\vertex[left = 20pt of V4l, blob] (bl) {\c};
	\vertex[left = 14pt of bl] (l) {};
	\vertex[below = 14pt of l, dot, minimum size = 0pt] (v12) {};
	\vertex[below = 12pt of v12, dot, minimum size = 3pt] (x1) {};
	\vertex[left = 12pt of v12, dot, minimum size = 3pt] (x2) {};
	\vertex[above = 14pt of l, dot, minimum size = 0pt] (v34) {};
	\vertex[left = 12pt of v34, dot, minimum size = 3pt] (x3) {};
	\vertex[above = 12pt of v34, dot, minimum size = 3pt] (x4) {};
	\vertex[above = 20pt of bc, dot, minimum size = 0pt] (v56) {};
	\vertex[above = 8pt of v56] (x56);
	\vertex[left = 8pt of x56, dot, minimum size = 3pt] (x5) {};
	\vertex[right = 8pt of x56, dot, minimum size = 3pt] (x6) {};
	\vertex[right = 20pt of bc, blob, minimum size = 6pt] (V4r) {};
	\vertex[left = 3pt of V4r, dot, minimum size = 0pt] (V4rl) {};
	\vertex[right = 3pt of V4r, dot, minimum size = 0pt] (V4rr) {};
	\vertex[right = 20pt of V4r, blob] (br) {\c};
	\vertex[right = 20pt of br, dot, minimum size = 0pt] (v78) {};
	\vertex[right = 8pt of v78] (x78);
	\vertex[above = 8pt of x78, dot, minimum size = 3pt] (x7) {};
	\vertex[below = 8pt of x78, dot, minimum size = 3pt] (x8) {};
	\diagram*{
		(x1) -- (v12) -- (x2),
		(x3) -- (v34) -- (x4),
		(x5) -- (v56) -- (x6),
		(x7) -- (v78) -- (x8),
		(v12) -- [photon] (bl) -- [quarter left] (V4ll) -- [quarter left] (bl) -- [photon] (v34),
		(bc) -- [quarter left] (V4lr) -- [quarter left] (bc) -- [quarter left] (V4rl) -- [quarter left] (bc) -- [photon] (v56),
		(v78) -- [photon] (br) -- [quarter left] (V4rr) -- [quarter left] (br)
	};
	\end{feynman}
	\end{tikzpicture}
}

\def\eightptDa{
	\begin{tikzpicture}[baseline = (V8.base)]
	\begin{feynman}
	\vertex[blob, minimum size = 6pt] (V8) {};
	\vertex[left = 2pt of V8, dot, minimum size = 0pt] (V8l) {};
	\vertex[right = 2pt of V8, dot, minimum size = 0pt] (V8r) {};
	\vertex[above = 2pt of V8l, dot, minimum size = 0pt] (V8tl) {};
	\vertex[below = 2pt of V8l, dot, minimum size = 0pt] (V8bl) {};
	\vertex[above = 2pt of V8r, dot, minimum size = 0pt] (V8tr) {};
	\vertex[below = 2pt of V8r, dot, minimum size = 0pt] (V8br) {};
	\vertex[left = 14pt of V8, dot, minimum size = 0pt] (l) {};
	\vertex[above = 14pt of l, blob] (b12) {\c};
	\vertex[left = 14pt of b12] (tl) {};
	\vertex[above = 14pt of tl, dot, minimum size = 0pt] (v12) {};
	\vertex[left = 12pt of v12, dot, minimum size = 3pt] (x1) {};
	\vertex[above = 12pt of v12, dot, minimum size = 3pt] (x2) {};
	\vertex[right = 14pt of V8] (r) {};
	\vertex[above = 14pt of r, blob] (b34) {\c};
	\vertex[right = 14pt of b34] (tr) {};
	\vertex[above = 14pt of tr, dot, minimum size = 0pt] (v34) {};
	\vertex[right = 12pt of v34, dot, minimum size = 3pt] (x3) {};
	\vertex[above = 12pt of v34, dot, minimum size = 3pt] (x4) {};
	\vertex[below = 14pt of l, blob] (b56) {\c};
	\vertex[left = 14pt of b56] (bl) {};
	\vertex[below = 14pt of bl, dot, minimum size = 0pt] (v56) {};
	\vertex[left = 12pt of v56, dot, minimum size = 3pt] (x5) {};
	\vertex[below = 12pt of v56, dot, minimum size = 3pt] (x6) {};
	\vertex[below = 14pt of r, blob] (b78) {\c};
	\vertex[right = 14pt of b78] (br) {};
	\vertex[below = 14pt of br, dot, minimum size = 0pt] (v78) {};
	\vertex[right = 12pt of v78, dot, minimum size = 3pt] (x7) {};
	\vertex[below = 12pt of v78, dot, minimum size = 3pt] (x8) {};
	\diagram*{
		(x1) -- (v12) -- (x2),
		(x3) -- (v34) -- (x4),
		(x5) -- (v56) -- (x6),
		(x7) -- (v78) -- (x8),
		(v12) -- [photon] (b12) -- [quarter left] (V8tl) -- [quarter left] (b12),
		(v34) -- [photon] (b34) -- [quarter left] (V8tr) -- [quarter left] (b34),
		(v56) -- [photon] (b56) -- [quarter left] (V8bl) -- [quarter left] (b56),
		(v78) -- [photon] (b78) -- [quarter left] (V8br) -- [quarter left] (b78)
	};
	\end{feynman}
	\end{tikzpicture}
}

\def\eightptDb{
\begin{tikzpicture}[baseline=(bc.base)]
\begin{feynman}
\vertex[blob] (bc) {\c};
\vertex[left = 20pt of bc, blob, minimum size = 6pt] (V6) {};
\vertex[left = 2pt of V6] (V6l) {};
\vertex[above = 2pt of V6l, dot, minimum size = 0pt] (V6tl) {};
\vertex[below = 2pt of V6l, dot, minimum size = 0pt] (V6bl) {};
\vertex[right = 3pt of V6, dot, minimum size = 0pt] (V6r) {};
\vertex[left = 14pt of V6] (l) {};
\vertex[below = 14pt of l, blob] (b12) {\c};
\vertex[left = 14pt of b12] (bl) {};
\vertex[below = 14pt of bl, dot, minimum size = 0pt] (v12) {};
\vertex[below = 12pt of v12, dot, minimum size = 3pt] (x1) {};
\vertex[left = 12pt of v12, dot, minimum size = 3pt] (x2) {};
\vertex[above = 14pt of l, blob] (b34) {\c};
\vertex[left = 14pt of b34] (tl) {};
\vertex[above = 14pt of tl, dot, minimum size = 0pt] (v34) {};
\vertex[above = 12pt of v34, dot, minimum size = 3pt] (x3) {};
\vertex[left = 12pt of v34, dot, minimum size = 3pt] (x4) {};
\vertex[above = 20pt of bc, dot, minimum size = 0pt] (v56) {};
\vertex[above = 8pt of v56] (x56) {};
\vertex[left = 8pt of x56, dot, minimum size = 3pt] (x5) {};
\vertex[right = 8pt of x56, dot, minimum size = 3pt] (x6) {};
\vertex[right = 20pt of bc, blob, minimum size = 6pt] (V4) {};
\vertex[left = 3pt of V4, dot, minimum size = 0pt] (V4l) {};
\vertex[right = 3pt of V4, dot, minimum size = 0pt] (V4r) {};
\vertex[right = 20pt of V4, blob] (b78) {\c};
\vertex[right = 20pt of b78, dot, minimum size = 0pt] (v78) {};
\vertex[right = 8pt of v78] (x78) {};
\vertex[above = 8pt of x78, dot, minimum size = 3pt] (x7) {};
\vertex[below = 8pt of x78, dot, minimum size = 3pt] (x8) {};
\diagram*{
	(x1) -- (v12) -- (x2),
	(x3) -- (v34) -- (x4),
	(x5) -- (v56) -- (x6),
	(x7) -- (v78) -- (x8),
	(v12) -- [photon] (b12) -- [quarter left] (V6bl) -- [quarter left] (b12),
	(v34) -- [photon] (b34) -- [quarter left] (V6tl) -- [quarter left] (b34),
	(bc) -- [quarter left] (V6r) -- [quarter left] (bc) -- [quarter left] (V4l) -- [quarter left] (bc) -- [photon] (v56),
	(v78) -- [photon] (b78) -- [quarter left] (V4r) -- [quarter left] (b78)
};
\end{feynman}
\end{tikzpicture}
}

\def\eightptDc{
	\begin{tikzpicture}[baseline=(base.base)]
	\begin{feynman}
	\vertex[] (c) {};
	\vertex[above = 14pt of c] (base) {};
	\vertex[left = 20pt of c, blob, minimum size = 6pt] (V4l) {};
	\vertex[left = 3pt of V4l, dot, minimum size = 0pt] (V4ll) {};
	\vertex[right = 3pt of V4l, dot, minimum size = 0pt] (V4lr) {};
	\vertex[left = 20pt of V4l, blob] (bl) {\c};
	\vertex[left = 20pt of bl, dot, minimum size = 0pt] (v12) {};
	\vertex[left = 8pt of v12] (x12) {};
	\vertex[below = 8pt of x12, dot, minimum size = 3pt] (x1) {};
	\vertex[above = 8pt of x12, dot, minimum size = 3pt] (x2) {};
	\vertex[right = 20pt of c, blob, minimum size = 6pt] (V4r) {};
	\vertex[left = 3pt of V4r, dot, minimum size = 0pt] (V4rl) {};
	\vertex[right = 3pt of V4r, dot, minimum size = 0pt] (V4rr) {};
	\vertex[right = 20pt of V4r, blob] (br) {\c};
	\vertex[right = 20pt of br, dot, minimum size = 0pt] (v34) {};
	\vertex[right = 8pt of v34] (x34) {};
	\vertex[below = 8pt of x34, dot, minimum size = 3pt] (x3) {};
	\vertex[above = 8pt of x34, dot, minimum size = 3pt] (x4) {};
	\vertex[above = 20pt of c, blob, minimum size = 6pt] (V4t) {};
	\vertex[above = 3pt of V4t, dot, minimum size = 0pt] (V4tt) {};
	\vertex[below = 3pt of V4t, dot, minimum size = 0pt] (V4tb) {};
	\vertex[above = 20pt of V4t, blob] (bt) {\c};
	\vertex[above = 20pt of bt, dot, minimum size = 0pt] (v56) {};
	\vertex[above = 8pt of v56] (x56) {};
	\vertex[left = 8pt of x56, dot, minimum size = 3pt] (x5) {};
	\vertex[right = 8pt of x56, dot, minimum size = 3pt] (x6) {};
	\vertex[below = 20pt of c, dot, minimum size = 0pt] (v78) {};
	\vertex[below = 8pt of v78] (x78) {};
	\vertex[left = 8pt of x78, dot, minimum size = 3pt] (x7) {};
	\vertex[right = 8pt of x78, dot, minimum size = 3pt] (x8) {};
	\diagram*{
		(x1) -- (v12) -- (x2),
		(x3) -- (v34) -- (x4),
		(x5) -- (v56) -- (x6),
		(x7) -- (v78) -- (x8),
		(v12) -- [photon] (bl) -- [quarter left] (V4ll) -- [quarter left] (bl),
		(v34) -- [photon] (br) -- [quarter left] (V4rr) -- [quarter left] (br),
		(v56) -- [photon] (bt) -- [quarter left] (V4tt) -- [quarter left] (bt),
		(c) -- [quarter left] (V4lr) -- [quarter left] (c) -- [quarter left] (V4tb) -- [quarter left] (c) -- [quarter left] (V4rl) -- [quarter left] (c) -- [photon] (v78)
	};
	\vertex[above = 0pt of c, blob] (bc) {\c};
	\end{feynman}
	\end{tikzpicture}
}

\def\eightptDd{
\begin{tikzpicture}[baseline = (base.base)]
\begin{feynman}
\vertex[blob, minimum size = 6pt] (V4c) {};
\vertex[above = 6pt of V4c] (base) {};
\vertex[left = 3pt of V4c, dot, minimum size = 0pt] (V4cl) {};
\vertex[right = 3pt of V4c, dot, minimum size = 0pt] (V4cr) {};
\vertex[left = 20pt of V4c, blob] (b34) {\c};
\vertex[above = 20pt of b34, dot, minimum size = 0pt] (v34) {};
\vertex[above = 8pt of v34] (x34) {};
\vertex[left = 8pt of x34, dot, minimum size = 3pt] (x3) {};
\vertex[right = 8pt of x34, dot, minimum size = 3pt] (x4) {};
\vertex[left = 20pt of b34, blob, minimum size = 6pt] (V4l) {};
\vertex[left = 3pt of V4l, dot, minimum size = 0pt] (V4ll) {};
\vertex[right = 3pt of V4l, dot, minimum size = 0pt] (V4lr) {};
\vertex[left = 20pt of V4l, blob] (b12) {\c};
\vertex[left = 20pt of b12, dot, minimum size = 0pt] (v12) {};
\vertex[left = 8pt of v12] (x12) {};
\vertex[above = 8pt of x12, dot, minimum size = 3pt] (x1) {};
\vertex[below = 8pt of x12, dot, minimum size = 3pt] (x2) {};
\vertex[right = 20pt of V4c, blob] (b56) {\c};
\vertex[above = 20pt of b56, dot, minimum size = 0pt] (v56) {};
\vertex[above = 8pt of v56] (x56) {};
\vertex[left = 8pt of x56, dot, minimum size = 3pt] (x5) {};
\vertex[right = 8pt of x56, dot, minimum size = 3pt] (x6) {};
\vertex[right = 20pt of b56, blob, minimum size = 6pt] (V4r) {};
\vertex[left = 3pt of V4r, dot, minimum size = 0pt] (V4rl) {};
\vertex[right = 3pt of V4r, dot, minimum size = 0pt] (V4rr) {};
\vertex[right = 20pt of V4r, blob] (b78) {\c};
\vertex[right = 20pt of b78, dot, minimum size = 0pt] (v78) {};
\vertex[right = 8pt of v78] (x78) {};
\vertex[above = 8pt of x78, dot, minimum size = 3pt] (x7) {};
\vertex[below = 8pt of x78, dot, minimum size = 3pt] (x8) {};
\diagram*{
	(x1) -- (v12) -- (x2),
	(x3) -- (v34) -- (x4),
	(x5) -- (v56) -- (x6),
	(x7) -- (v78) -- (x8),
	(v12) -- [photon] (b12) -- [quarter left] (V4ll) -- [quarter left] (b12),
	(v78) -- [photon] (b78) -- [quarter left] (V4rr) -- [quarter left] (b78),
	(v34) -- [photon] (b34) -- [quarter left] (V4lr) -- [quarter left] (b34) -- [quarter left] (V4cl) -- [quarter left] (b34),
	(v56) -- [photon] (b56) -- [quarter left] (V4rl) -- [quarter left] (b56) -- [quarter left] (V4cr) -- [quarter left] (b56)
};
\end{feynman}
\end{tikzpicture}
}

\def\sc{0.68}
\begin{table}[htbp]
\begin{tabular}{c|c|c}
	\hline
	\# of \Tstrut & \multirow{2}{*}{diagrams} & \multirow{2}{*}{degenerate limit result $\bigr/\bigl(C_W^{(\l)}\bigr)^{\*4}$} \\
	\*$j$\;sums\Bstrut & & \\
	\hline
	1 & $\scalebox{\sc}[\sc]{\eightptA} \;-\,\left( \;\scalebox{\sc}[\sc]{\eightptAsub} \;\,+\,\text{perms.}\right)$ \rule[-25pt]{0pt}{55pt} & $\dfrac{1}{n_\lm1^3} \Bigl( \bigl\langle \Delta^4 \bigr\rangle - 3 \bigl\langle \Delta^2 \bigr\rangle^2 \Bigr)$\\
	\hline
	\multirow{2}{*}{2 \rule{0pt}{34pt}} & $\scalebox{\sc}[\sc]{\eightptBa} \;-\,\left( \;\scalebox{\sc}[\sc]{\eightptBasub} \;\,+\,\text{perms.}\right) \,+\,\text{perms.}$ \rule[-25pt]{0pt}{55pt} & $\,\dfrac{1}{4}\*\cdot\*\dfrac{V_4^{(\lm1)}}{n_\lm1^2 n_{\l\shortminus 2}^{}} \*\cdot 4\, \Bigl( \bigl\langle \partial^2\Delta^3 \bigr\rangle \bigl\langle \partial^2\Delta\bigr\rangle -3\, \bigl\langle \partial^2\Delta \bigr\rangle^2 \bigl\langle \Delta^2 \bigr\rangle \Bigr)\*$ \\
	\cline{2-3}
	& $\scalebox{\sc}[\sc]{\eightptBb} \;+\,\text{perms.}$ \rule[-25pt]{0pt}{55pt} & $\dfrac{1}{4}\*\cdot\*\dfrac{V_4^{(\lm1)}}{n_\lm1^2 n_{\l\shortminus 2}^{}} \*\cdot 3\, \bigl\langle \partial^2 \Delta^2 \bigr\rangle^2$ \\
	\hline
	& $\scalebox{\sc}[\sc]{\eightptCa}  \;+\,\text{perms.}$ \rule[-25pt]{0pt}{55pt} & $\dfrac{1}{8}\*\cdot\*\dfrac{V_6^{(\lm1)}}{n_\lm1^{} n_{\l\shortminus 2}^2} \*\cdot 6\, \bigl\langle \partial^2 \Delta^2 \bigr\rangle \bigl\langle \partial^2 \Delta \bigr\rangle^2$ \\
	\cline{2-3}
	3 & $\scalebox{\sc}[\sc]{\eightptCb} -\left( \,\scalebox{\sc}[\sc]{\eightptCbsub} \,+\text{perm.}\right) \*+\text{perms.}$ \rule[-25pt]{0pt}{55pt} & $\dfrac{1}{16}\*\cdot\*\dfrac{\bigl(V_4^{(\lm1)}\bigr)^2}{n_\lm1^{} n_{\l\shortminus 2}^2} \*\cdot 6\, \Bigl( \bigl\langle \partial^4 \Delta^2 \bigr\rangle \bigl\langle \partial^2 \Delta \bigr\rangle^2 -2\,\bigl\langle \partial^2 \Delta \bigr\rangle^4 \Bigr)$ \\
	\cline{2-3}
	& $\scalebox{\sc}[\sc]{\eightptCc} \; + \,\text{perms.}$ \rule[-25pt]{0pt}{55pt} & $\dfrac{1}{16}\*\cdot\*\dfrac{\bigl(V_4^{(\lm1)}\bigr)^2}{n_\lm1^{} n_{\l\shortminus 2}^2} \*\cdot 12\, \bigl\langle \partial^4 \Delta \bigr\rangle \bigl\langle \partial^2 \Delta^2 \bigr\rangle \bigl\langle \partial^2 \Delta \bigr\rangle$ \\
	\hline
	& $\scalebox{\sc}[\sc]{\eightptDa}$ \rule[-35pt]{0pt}{70pt} & $\dfrac{1}{16}\*\cdot\*\dfrac{V_8^{(\lm1)}}{n_{\l\shortminus 2}^3} \,\bigl\langle \partial^2 \Delta \bigr\rangle^4$ \\
	\cline{2-3}
	\multirow{2}{*}{4 \rule{0pt}{35pt}} & $\scalebox{\sc}[\sc]{\eightptDb} \; +\,\text{perms.}$ \rule[-30pt]{0pt}{65pt} & $\dfrac{1}{32}\*\cdot\*\dfrac{V_6^{(\lm1)} V_4^{(\lm1)}}{n_{\l\shortminus 2}^3} \*\cdot 12\, \bigl\langle \partial^4 \Delta \bigr\rangle \bigl\langle \partial^2 \Delta \bigr\rangle^3$ \\
	\cline{2-3}
	& $\scalebox{\sc}[\sc]{\eightptDc} \;+\,\text{perms.}$ \rule[-35pt]{0pt}{80pt} & $\dfrac{1}{64}\*\cdot\*\dfrac{\bigl( V_4^{(\lm1)} \bigr)^3}{n_{\l\shortminus 2}^3} \*\cdot 4\,\bigl\langle \partial^6 \Delta \bigr\rangle \bigl\langle \partial^2 \Delta \bigr\rangle^3$ \\
	\cline{2-3}
	& $\scalebox{\sc}[\sc]{\eightptDd} \;+\,\text{perms.}$ \rule[-20pt]{0pt}{45pt} & $\dfrac{1}{64}\*\cdot\*\dfrac{\bigl( V_4^{(\lm1)} \bigr)^3}{n_{\l\shortminus 2}^3} \*\cdot 12\,\bigl\langle \partial^4 \Delta \bigr\rangle^2 \bigl\langle \partial^2 \Delta \bigr\rangle^2$ \\
	\hline
\end{tabular}
\caption{\label{tab:8pt}
Diagrammatic calculation of the connected eight-point correlator.}
\end{table}

\subsection{Further discussion of criticality tuning}

In this section we briefly review the results on criticality tuning in Ref.~\cite{Roberts:2021fes} and explain its connection with our Eq.~\eqref{eq:crit}. 
In the nearby input limit, Ref.~\cite{Roberts:2021fes} identified the following criticality conditions:
\begin{eqnarray}
\chi_\parallel^{(\l)} (\K^\star) &\equiv& \frac{C_W^{(\l)}}{2\,\K^{\star 2}} \,\bigl\langle \Q^2 (\phi^2 -\K^\star) \bigr\rangle_{\K^\star} = \frac{C_W^{(\l)}}{2} \,\bigl\langle (\Q^2)'' \bigr\rangle_{\K^\star} = 1 \,,\label{seq:chi_parallel}\\[5pt]
\chi_\perp^{(\l)} (\K^\star) &\equiv& C_W^{(\l)} \,\bigl\langle \Q'^2 \bigr\rangle_{\K^\star} = 1 \,, \label{seq:chi_perp}
\end{eqnarray}
where a common input argument $\x$ is assumed for all functions involved, and $\K^\star (\x)$ is the fixed point of the kernel recursion Eq.~\eqref{eq:K_0_recursion} when $\x_1^{} = \x_2^{} \equiv \x$. 
The first condition Eq.~\eqref{seq:chi_parallel} ensures power-law scaling of the norm of preactivation for a given input $\x$, while the second condition Eq.~\eqref{seq:chi_perp} (also identified in earlier works Refs.~\cite{raghu2017expressive, poole2016exponential, schoenholz2016deep}) ensures power-law scaling of the distance between preactivations for infinitesimally-separated inputs with the same norm. 
These conditions imply the following tuning of initialization hyperparameters:
\begin{itemize}
	\item For scale-invariant activation functions, $\Q(\phi) = \begin{cases}
	a_- \phi & (\phi<0)\\
	a_+ \phi & (\phi\ge 0)
	\end{cases}$\, (including ReLU which corresponds to $a_-=0$, $a_+=1$), existence of a finite fixed point $\K^*$ for any input $\x$ requires $C_b^{(\l)}=0$. Meanwhile, $\chi_\parallel^{(\l)} (\K^\star) = \chi_\perp^{(\l)} (\K^\star) = \frac{C_W^{(\l)}}{2} (a_+^2 + a_-^2)$, independent of $\K^\star$. Therefore, setting $C_W^{(\l)} = \frac{2}{a_+^2 + a_-^2}$ ensures Eqs.~\eqref{seq:chi_parallel} and \eqref{seq:chi_perp} are satisfied for any input $\x$.
	\item For smooth activation functions that satisfy $\Q(0) = 0$, $\Q'(0)\ne 0$, we can Taylor-expand $\Q(\phi) = \Q_1 \phi + \frac{1}{2} \Q_2 \phi^2 +\dots$ and show that $\K^\star = 0$ is a fixed point for any $\x$ when we set $C_b^{(\l)}=0$. Meanwhile, $\chi_\parallel^{(\l)} (\K^\star) = \chi_\perp^{(\l)} (\K^\star) = C_W^{(\l)} \Q_1^2$ which means we should set $C_W^{(\l)}=\frac{1}{\Q_1^2}$ to satisfy Eqs.~\eqref{seq:chi_parallel} and \eqref{seq:chi_perp}.
\end{itemize}
It is interesting to note that although the criticality conditions Eqs.~\eqref{seq:chi_parallel} and \eqref{seq:chi_perp} may seem overconstraining since they should be applied to every input $\x$ while there are only two hyperparameters $C_W^{(\l)}, C_b^{(\l)}$ we can tune at each layer, it is actually possible to satisfy them at least for the two classes of activation functions above (referred to as scale-invariant and $\K^\star=0$ universality classes).

We now show that the hyperparameter tuning above in fact ensures the stronger condition Eq.~\eqref{eq:crit} is also satisfied at LO in $\frac{1}{n}$. 
Explicitly taking the functional derivatives in the definition of $\chi^{(\l)}\*(\x_1^{}, \x_2^{}; \y_1^{}, \y_2^{})$, Eq.~\eqref{eq:chi}, we find
\begin{eqnarray}
\chi^{(\l)}\*(\x_1^{}, \x_2^{}; \y_1^{}, \y_2^{}) &=&
\frac{C_W^{(\l)}}{2} \,\biggl\langle \frac{\delta^2 \bigl[\Q \bigl(\phi(\x_1^{})\bigr) \,\Q\bigl( \phi(\x_2^{}) \bigr)\bigr]}{\delta \phi(\y_1^{}) \,\delta \phi(\y_2^{})} \biggr\rangle_{\!\!\K^{(\lm1)}_0} +\O\Bigl(\frac{1}{n}\Bigr) \nonumber\\[5pt]
&=& \frac{C_W^{(\l)}}{2} \biggl\{
\Bigl[\delta(\x_1^{} - \y_1^{}) \,\delta(\x_2^{} - \y_2^{}) \*+\*\delta(\x_1^{} - \y_2^{}) \,\delta(\x_2^{} - \y_1^{}) \Bigr] \Bigl\langle \Q'(\phi \bigl(\x_1^{})\bigr) \,\Q'(\phi \bigl(\x_2^{})\bigr) \Bigr\rangle_{\!\K^{(\lm1)}_0}
\nonumber\\[2pt]
&&\qquad\quad +\,\delta(\x_1^{} - \y_1^{}) \,\delta(\x_1^{} - \y_2^{}) \,\Bigl\langle \Q''(\phi \bigl(\x_1^{})\bigr) \,\Q(\phi \bigl(\x_2^{})\bigr) \Bigr\rangle_{\!\K^{(\lm1)}_0} 
\nonumber\\[2pt]
&&\qquad\quad +\,\delta(\x_2^{} - \y_1^{}) \,\delta(\x_2^{} - \y_2^{}) \,\Bigl\langle \Q(\phi \bigl(\x_1^{})\bigr) \,\Q''(\phi \bigl(\x_2^{})\bigr) \Bigr\rangle_{\!\K^{(\lm1)}_0} 
\biggr\} 
+\O\Bigl(\frac{1}{n}\Bigr) \,.
\end{eqnarray}
Therefore, at LO in $\frac{1}{n}$, Eq.~\eqref{eq:crit} can be equivalently written as
\begin{equation}
C_W^{(\l)} \,\Bigl\langle \Q'(\phi \bigl(\x_1^{})\bigr) \,\Q'(\phi \bigl(\x_2^{})\bigr) \Bigr\rangle_{\!\K^\star} = 1 \,, \qquad
\Bigl\langle \Q''(\phi \bigl(\x_1^{})\bigr) \,\Q(\phi \bigl(\x_2^{})\bigr) \Bigr\rangle_{\!\K^\star} = 0 \qquad (\forall \x_1^{} , \, \x_2^{} ) \,,
\end{equation}
or
\begin{equation}
C_W^{(\l)} \,\Bigl\langle \Q'(\phi \bigl(\x_1^{})\bigr) \,\Q'(\phi \bigl(\x_2^{})\bigr) \Bigr\rangle_{\!\K^\star} 
= \frac{C_W^{(\l)}}{2} \,\Bigl\langle \Bigl[ \Q(\phi \bigl(\x_1^{})\bigr) \,\Q(\phi \bigl(\x_2^{})\bigr) \Bigr] '' \Bigr\rangle_{\!\K^\star} 
= 1
\qquad (\forall \x_1^{} , \, \x_2^{} ) \,.
\label{seq:crit}
\end{equation}
This latter form makes it clear that Eqs.~\eqref{seq:chi_parallel} and \eqref{seq:chi_perp} are the degenerate input limit of the generally stronger condition Eq.~\eqref{eq:crit}. 
Importantly, for both scale-invariant and $\K^\star=0$ activation functions, the expectation values in Eq.~\eqref{seq:crit} are independent of $\x_1^{}, \x_2^{}$, so the hyperparameter tuning derived from the nearby input analysis implies that Eq.~\eqref{eq:crit} is also satisfied.

Finally, we note that tuning hyperparameters to satisfy the criticality condition Eq.~\eqref{eq:crit} at LO in $\frac{1}{n}$ is sufficient to ensure power-law scaling of connected correlators, \ie\ $\delta\bigl\langle\G^{(\l)}\*(\x_1^{}, \x_2^{})\bigr\rangle \sim \l^\ce$ and $V_{2k}^{(\l)}\*(\x_1^{}, \dots \x_{2k}^{}) \sim \l^{\ce_k}$ (as opposed to $e^{\pm\l}$) from solving the RG equation, where $\ce, \ce_k$ are critical exponents.  
While $\O\bigl(\frac{1}{n}\bigr)$ corrections naively leads to $\delta\bigl\langle\G^{(\l)}\bigr\rangle \,, \, V_{2k}^{(\l)} \sim e^{\pm \frac{\l}{n}}$, such scaling is not really exponential when the network depth $L\ll n$. 
On the other hand, $\O\bigl(\frac{1}{n}\bigr)$ corrections can change the critical exponents, and finer tuning of $C_W^{(\l)}, C_b^{(\l)}$ by $\O\bigl(\frac{1}{n}\bigr)$ amounts can result in more favorable power-law scaling (\eg\ slower growth of fluctuations)~\cite{Roberts:2021fes}.

\subsection{Numerical verification of power-law scaling at criticality}
%

\begin{figure}
\includegraphics[scale=0.75]{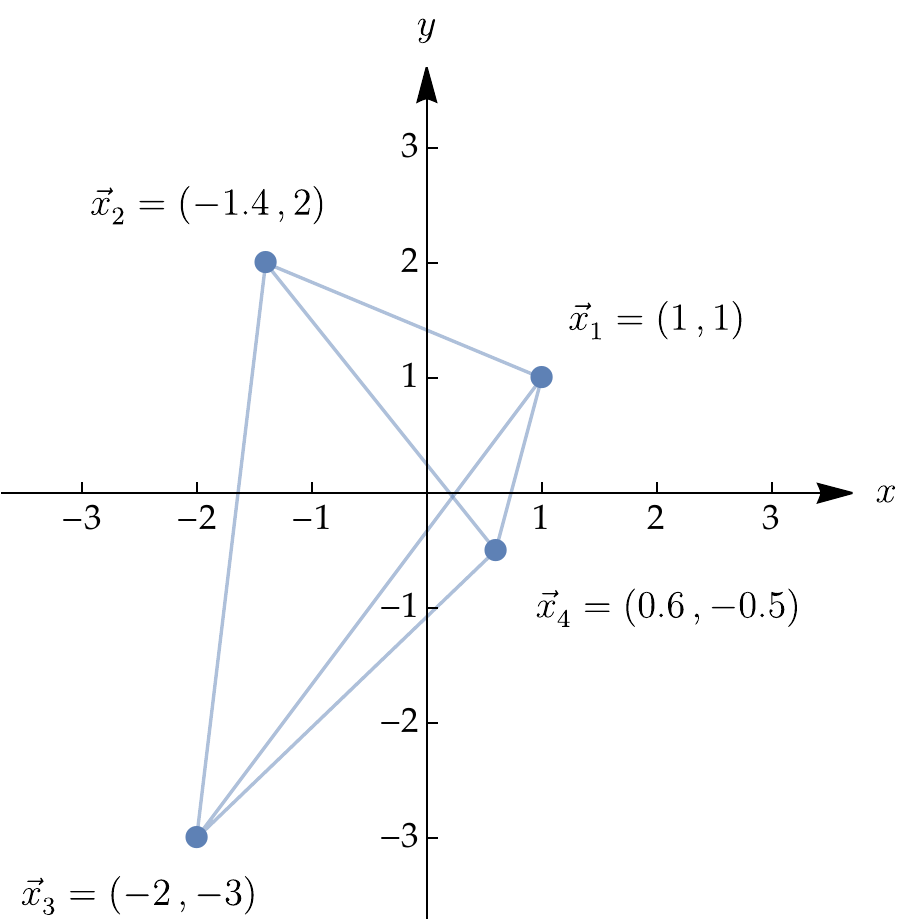}
\caption{\label{fig:inputs}
Input points chosen for illustration in our numerical experiments.}
\end{figure}

In this final section, we perform numerical experiments to verify power-law scaling of connected correlators when the hyperparameters $C_W^{(\l)}$, $C_b^{(\l)}$ are set to their critical values as discussed in the previous section. 
For each experiment, we generate an ensemble of $N_\text{net}=1,000$ neural networks, with hidden layer width $n=300$ and depth $L=30$, initialized according to Eq.~\eqref{seq:PWPb}. 
We consider an $n_0=2$ dimensional input space and pick four points $\x_1^{}, \x_2^{}, \x_3^{}, \x_4^{}$ as shown in Fig.~\ref{fig:inputs} for illustration. 
The activation function is chosen to be either ReLU or tanh. 
For each network, we compute neuron preactivations at every layer according to Eq.~\eqref{eq:operations}. 
The connected two-point and four-point correlators are then obtained from ensemble averaging:
\begin{eqnarray}
\bigl\langle \G^{(\l)} (\x_\alpha^{}, \x_\beta^{})\bigr\rangle &=& \frac{1}{N_\text{net}} \sum_{I=1}^{N_\text{net}} \,\frac{1}{n_\l}\, \sum_{i=1}^{n_\l} \phi_{I,i}^{(\l)} (\x_\alpha^{}) \,\phi_{I,i}^{(\l)} (\x_\beta^{}) \,,\\
V_4^{(\l)} (\x_\alpha^{}, \x_\beta^{} ; \x_\gamma^{}, \x_\delta^{}) &=& \frac{1}{N_\text{net}} \sum_{I=1}^{N_\text{net}} \,\frac{1}{n_\l} \sum_{i_1,i_2=1}^{n_\l} \phi_{I,i_1}^{(\l)} (\x_\alpha^{}) \,\phi_{I,i_1}^{(\l)} (\x_\beta^{}) \,\phi_{I,i_2}^{(\l)} (\x_\gamma^{}) \,\phi_{I,i_2}^{(\l)} (\x_\delta^{})
-n_\l\,\bigl\langle \G^{(\l)} (\x_\alpha^{}, \x_\beta^{})\bigr\rangle \bigl\langle \G^{(\l)} (\x_\gamma^{}, \x_\delta^{})\bigr\rangle
\nonumber\\
&& -\,\bigl\langle \G^{(\l)} (\x_\alpha^{}, \x_\gamma^{})\bigr\rangle \bigl\langle \G^{(\l)} (\x_\beta^{}, \x_\delta^{})\bigr\rangle
-\bigl\langle \G^{(\l)} (\x_\alpha^{}, \x_\delta^{})\bigr\rangle \bigl\langle \G^{(\l)} (\x_\beta^{}, \x_\gamma^{})\bigr\rangle 
+\O\Bigl(\frac{1}{n}\Bigr)\,,
\end{eqnarray}
where $I$ labels networks in the ensemble. 
We perform 10 experiments each for ReLU and tanh networks, and take the standard deviation across the 10 experiments to be the numerical uncertainty for each correlator computed, represented by vertical bars in the figures below.

\begin{figure}[t]
	\vspace{20pt}
	\includegraphics[width=\linewidth]{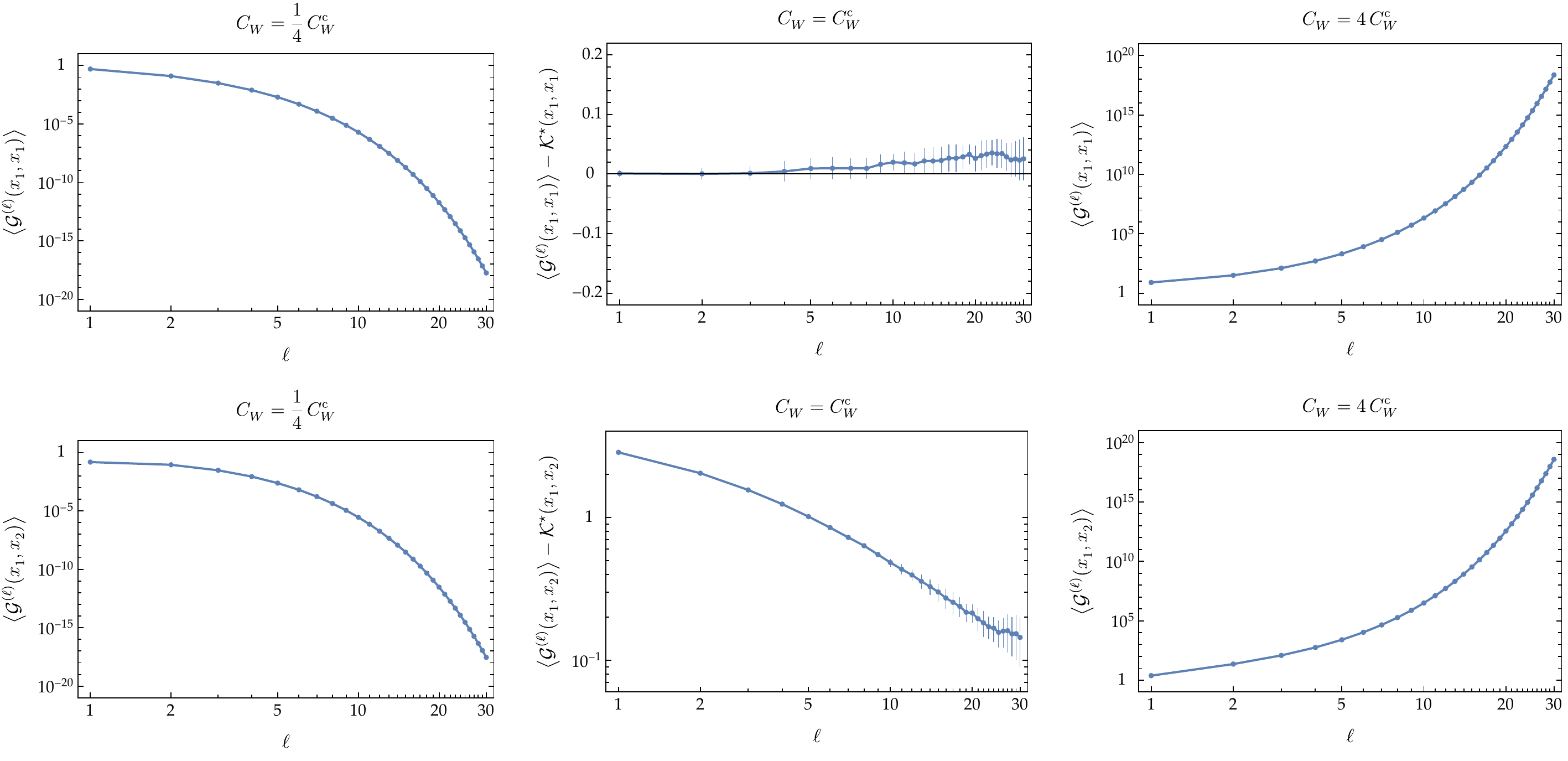}
	\caption{\label{fig:2-pt_ReLU}
		{\bf Two-point correlators for ReLU networks.} Asymptotic scaling toward fixed point is power-law at criticality (middle panels) and exponential away from criticality (left and right panels). The fixed points $\K^\star$ in the middle panels are given by Eq.~\eqref{seq:fp_relu}. See text for details.}
\end{figure}

\begin{figure}[h!]
	\vspace{25pt}
	\includegraphics[width=\linewidth]{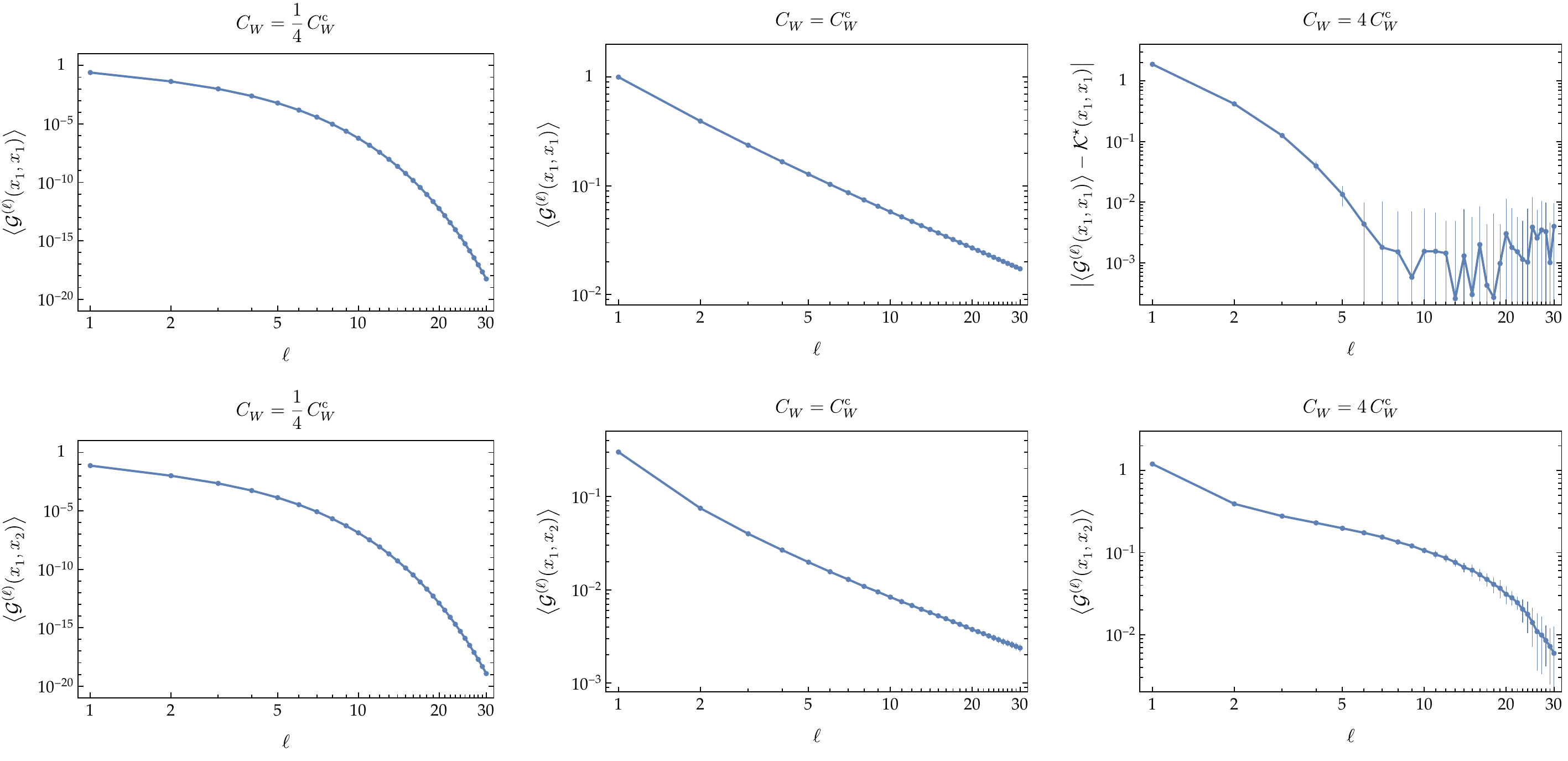}
	\caption{\label{fig:2-pt_tanh}
		{\bf Two-point correlators for tanh networks.} Asymptotic scaling toward fixed point is power-law at criticality (middle panels) and exponential away from criticality (left and right panels). The fixed point $\K^\star$ in the top-right panel is given by the nonzero solution to Eq.~\eqref{seq:fp_tanh}. See text for details.}
\end{figure}

In Figs.~\ref{fig:2-pt_ReLU} and \ref{fig:2-pt_tanh}, we show numerical results of two-point correlators for ReLU and tanh networks, respectively.
We choose to present $\langle \G^{(\l)} (\x_1^{}, \x_1^{})\rangle$ (top row) and $\langle \G^{(\l)} (\x_1^{}, \x_2^{})\rangle$ (bottom row) as examples of degenerate and nondegenerate inputs; we have checked that all other two-point correlators exhibit similar behaviors and are also consistent with theoretical expectations discussed below. 
In each figure, the middle panels are obtained from setting the hyperparameters to their critical values, $C_W^{(\l)} = C_W^\text{c}$, $C_b^{(\l)}=0$, while the left and right panels are obtained by setting $C_W^{(\l)}=\frac{1}{4}C_W^\text{c}$ and $4C_W^\text{c}$ (corresponding to the standard deviation of $W_{ij}^{(\l)}$ being half and twice the critical value), respectively, while keeping $C_b^{(\l)}=0$. 
The important takeaway here is that the scaling is power-law in all the middle panels, and exponential in all the left and right panels. 
Let us explain these plots in more detail:
\begin{itemize}
	\item For ReLU activation function (Fig.~\ref{fig:2-pt_ReLU}), we expect the two-point correlator for any inputs to flow exponentially to a trivial fixed point at $\K^\star=0$ ($\K^\star=\infty$) when $C_W^{(\l)} < C_W^\text{c}$ ($C_W^{(\l)} > C_W^\text{c}$); this is consistent with the left (right) panels of Fig.~\ref{fig:2-pt_ReLU}. 
	At criticality $C_W^{(\l)} = C_W^\text{c} = 2$, on the other hand, there is a nontrivial fixed point at~\cite{Williams1996,Roberts:2021fes,Halverson:2020trp}
	\begin{equation}
	\K^\star(\x_\alpha^{}, \x_\beta^{}) = \frac{C_W^\text{c}}{n_0} \sqrt{|\x_\alpha^{}||\x_\beta^{}|} = \sqrt{|\x_\alpha^{}||\x_\beta^{}|} \,.
	\label{seq:fp_relu}
	\end{equation}
	For degenerate inputs, the two-point correlator is already at this fixed point at the ultraviolet boundary $\ell=1$, and is expected to stay constant for all $\ell$ at LO in $\frac{1}{n}$. 
	The middle panel in the top row is consistent with this expectation. 
	For nondegenerate inputs, the middle panel in the bottom row confirms that RG flow to the fixed point is power-law as expected.
	\item For tanh activation function (Fig.~\ref{fig:2-pt_tanh}), first observe that at criticality $C_W^{(\l)} = C_W^\text{c} = 1$ (middle panels), RG flow to the fixed point $\K^\star=0$ is power-law as expected. 
	Next, for $C_W^{(\l)} < C_W^\text{c}$ (left panels), two-point correlators flow to the same fixed point $\K^\star=0$, but the scaling is exponential. 
	The final case $C_W^{(\l)} > C_W^\text{c}$ (right panels) is more subtle: for nondegenerate inputs, we again observe an exponential flow toward $\K^\star=0$; for degenerate inputs, the fixed point $\K^\star=0$ is repulsive and the two-point function actually flows to a different fixed point which solves
	\begin{equation}
	\K^\star = C_W \,\bigl\langle \tanh^2 \*\phi \bigr\rangle_{\K^\star} 
	= \frac{C_W}{\sqrt{2\pi\K^\star}} \int_{-\infty}^\infty d\phi \,\tanh^2\*\phi \,e^{-\frac{\phi^2}{2\K^\star}}\,,
	\label{seq:fp_tanh}
	\end{equation}
	where $C_W$ is the common value of $C_W^{(\l)}$. 
	In this latter case, the scaling is again exponential, as expected away from criticality.
\end{itemize}

\begin{figure}[t]
	\vspace{20pt}
	\includegraphics[width=\linewidth]{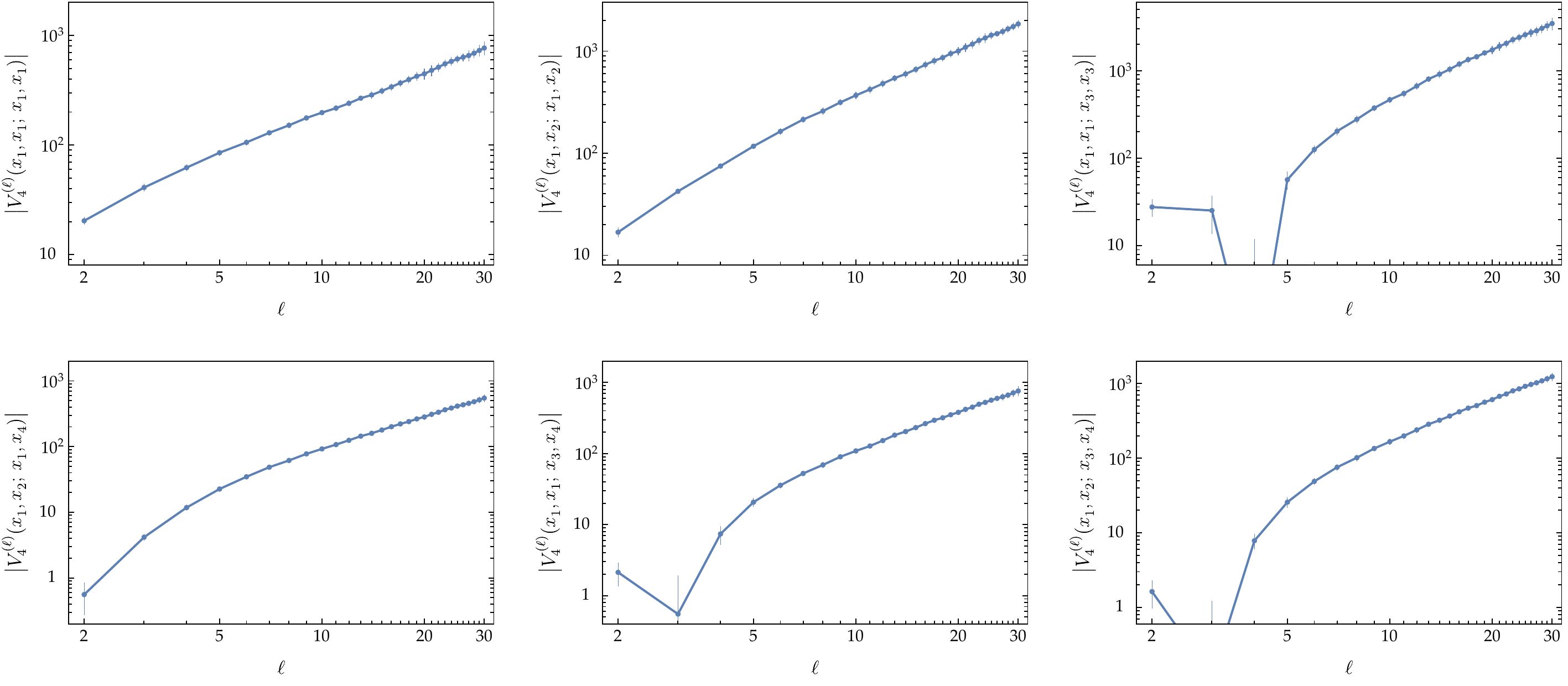}
	\caption{\label{fig:4-pt_ReLU}
		{\bf Connected four-point correlators for ReLU networks at criticality.} Asymptotic scaling is power-law for all input choices. See text for details.}
\end{figure}

\begin{figure}[h!]
	\vspace{25pt}
	\includegraphics[width=\linewidth]{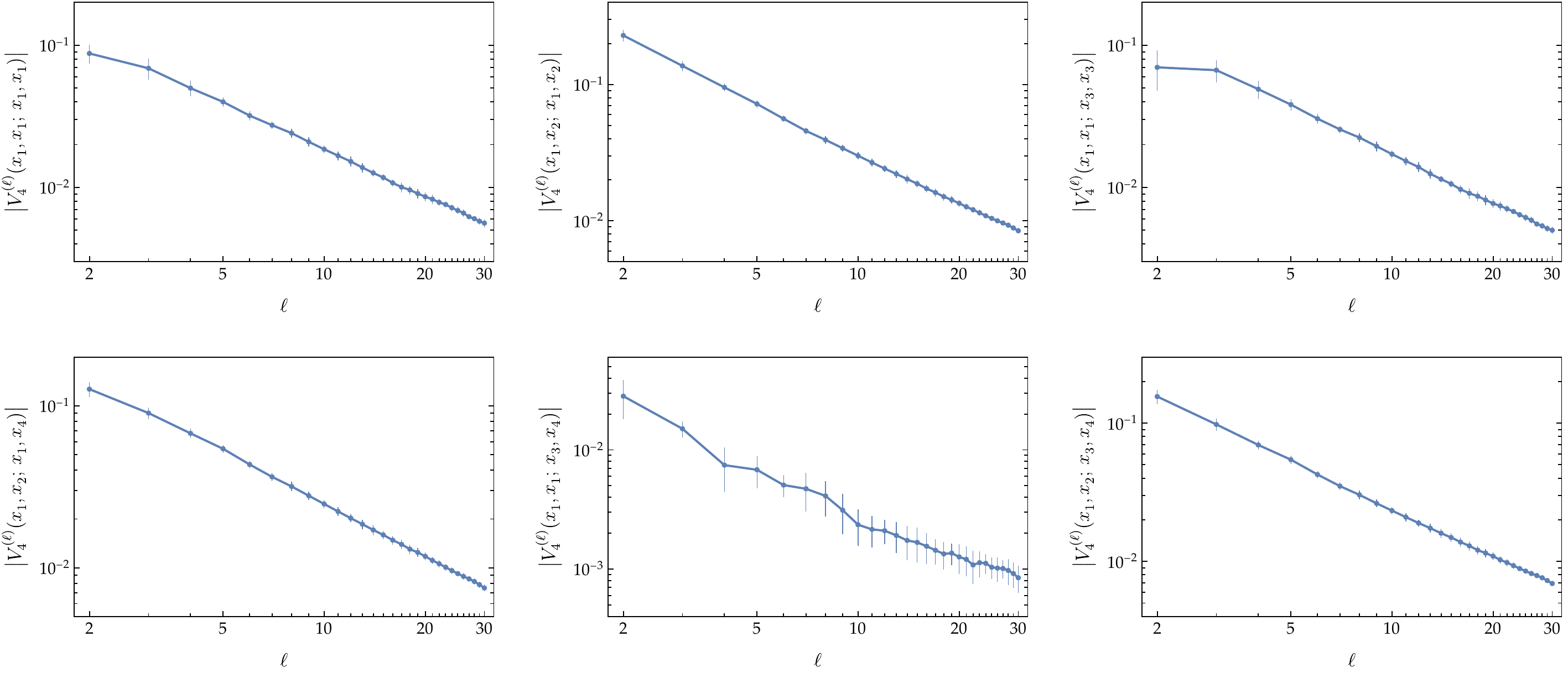}
	\caption{\label{fig:4-pt_tanh}
		{\bf Connected four-point correlators for tanh networks at criticality.} Asymptotic scaling is power-law for all input choices. See text for details.}
\end{figure}

Next, in Figs.~\ref{fig:4-pt_ReLU} and \ref{fig:4-pt_tanh}, we show numerical results for a representative set of connected four-point correlators at criticality for ReLU and tanh networks, respectively. 
These include choices of four inputs that are all identical, and take two, three and four distinct values. 
We plot $|V_4^{(\l)}|$ since $V_4^{(\l)}$ can have either sign, and in some cases actually changes sign along the RG flow. 
The plots start at $\ell=2$ since the connected four-point correlator vanishes at the first layer, and our numerical results are consistent with $V_4^{(1)}=0$. 
In all cases shown in Figs.~\ref{fig:4-pt_ReLU} and \ref{fig:4-pt_tanh}, we observe clear asymptotic power-law scaling. 
These results confirm that tuning $C_W^{(\l)}$, $C_b^{(\l)}$ to criticality ensures power-law scaling of all connected four-point correlators for general (nondegenerate) inputs. 
We have checked that other connected four-point correlators also exhibit power-law scaling at criticality. 
On the other hand, away from criticality we observe exponential scaling behaviors for connected four-point correlators similar to those of two-point correlators in the left and right panels of Figs.~\ref{fig:2-pt_ReLU} and \ref{fig:2-pt_tanh}. 

\newpage

\end{document}